\documentclass[12pt]{article}
\input{epsf}

\makeatletter
\def\Ddots{\mathinner{\mkern1mu\raise\p@
\vbox{\kern7\p@\hbox{.}}\mkern2mu
\raise4\p@\hbox{.}\mkern2mu\raise7\p@\hbox{.}\mkern1mu}}
\makeatother
\font\bm=cmmib10 at 10pt
\font\bms=cmmib10 at 7pt \textfont9=\bm \scriptfont9=\bms
\usepackage{amsfonts}
\usepackage{multirow}
\mathchardef\balpha= "790B
\mathchardef\bbeta= "790C
\mathchardef\bTheta= "7902
\mathchardef\bzeta= "7910
\mathchardef\bOmega= "790A
\mathchardef\bGamma= "7900
\mathchardef\bDelta= "7901
\mathchardef\bPhi= "7908
\mathchardef\bphi= "791E
\mathchardef\bomega= "7921
\mathchardef\bxi= "7918
\mathchardef\bet= "7911
\mathchardef\brho= "791A
\mathchardef\btau= "791C
\mathchardef\bmu= "7916
\mathchardef\bvarpi= "7924

\def \sgn {\hbox{ sgn}}

\def \infG {\,^{\infty}\!G}
\def \infF {\,^{\infty}\!F}

\def \lvec{(\kern-.26em(}
\def\pmb#1{\setbox0=\hbox{#1}
\kern-.025em\copy0\kern-\wd0
\kern.05em\copy0\kern-\wd0
\kern-.025em\raise.0433em\box0 }
\mathchardef\btheta= "7912
\usepackage{amsmath}
\usepackage{amssymb}
\usepackage{authblk}
\usepackage{url}

\usepackage{graphicx}
\begin{document}
\title{2n-Stream Thermal Emission from Clouds}
\author[1]{ W. A. van Wijngaarden}
\author[2] {W. Happer}
\affil[1]{Department of Physics and Astronomy, York University, Canada}
\affil[2]{Department of Physics, Princeton University, USA}
\renewcommand\Affilfont{\itshape\small}
\date{\today}
\maketitle
\begin{abstract}
We analyze thermal emission of radiation in homogeneous clouds.  The clouds have negligible horizontal variation, but the temperature can vary in the vertical direction. The radiation at the optical depth $\tau$ above the bottom of the cloud is characterized by the intensity values $I(\mu_i,\tau)$  at $2n$ Gauss-Legendre direction cosines $\mu_i$, the roots of the $2n$-th Legendre polynomial, $P_{2n}(\mu_i)=0$.  Scattering matrices are used to describe the fraction of intensity with incoming direction cosine $\mu_{i'}$ that is scattered to intensity with outgoing direction cosine $\mu_{i}$.  Green's-function matrices are used to describe radiation generated by  infinitesimally thin layers of cloud particulates, thermally emitting at the source optical depth $\tau'$.  For thin isothermal clouds of optical thickness $\tau_c\le1$, thermal emission is mainly determined by the single scattering albedo $\tilde \omega$. The emitted intensity of thin clouds is limb brightened and depends little on the  scattering phase function.  For optically thick isothermal clouds, with $\tau_c \gg 1$, purely absorbing clouds with $\tilde\omega = 0$ become perfect blackbodies and emit isotropic Planck intensity, and have neither limb brightening nor darkening. Thick isothermal  clouds with moderate single scattering albedos emit substantially less thermal radiation than black bodies, and the emission depends strongly on the anisotropy of the scattering phase function.  Clouds with strong forward scattering  are ``blackest" and emit the most thermal radiation. There is limb darkening, which is most pronounced for strongly forward-scattering phase functions. The limb darkening in thick, isothermal, scattering clouds is not due to temperature gradients, as it is for the Sun.  With scattering, nearly vertical radiation produced by thermal emission escapes through the surface more efficiently than nearly horizontal radiation. For isothermal clouds, the scattered and emitted intensities obey Kirchhoff's laws of thermal radiation. For clouds with hot interiors, there is limb darkening even for purely absorbing particulates. Radiative conduction of heat inside optically thick clouds is orders of magnitude faster than conduction due to molecular diffusion.
\end{abstract}
\newpage
\tableofcontents
\newpage
\section{Introduction\label{in}}
In a previous paper, {\it 2n-Stream Radiative Transfer}\,\cite{WH1} we outlined how to use operator and matrix methods, similar to those of quantum mechanics, to accurately and efficiently analyze radiative transfer with highly anisotropic scattering, like the scattering of sunlight by cloud particulates.  To facilitate subsequent discussions, we will refer to this paper as WH and to equation ($x$) in it as (WH-$x$). The basic ideas of the $2n$-stream method were first published in 1943 by G. C. Wick in connection with his analysis of neutron diffusion, {\it \"Uber ebene Diffusionsprobleme}\, \cite{Wick}.  In their paper, {\it The Penetration of Diffuse Ultraviolet Radiation into Stellar Clouds}, Flannery, Roberge and Rybicki [3], described one of the earliest applications of Wick’s methods to light scattering in clouds. 
The thermal emission of clouds was only briefly mentioned in Section 8 of WH. Here we discuss thermal emission in more detail.  
\section{The Intensity\label{in}}
As in WH, we assume that the intensity $I(\mu,\tau,\nu)$ is axially  symmetric and depends only on the cosine, $\mu=\cos\theta$, of the angle $\theta$ between direction of propagation and the zenith direction, and on the vertical optical depth, $\tau$, above the bottom of the cloud.  A representative unit of intensity is W m$^{-2}$ cm sr
$^{-1}$. The spatial frequency $\nu$ of the radiation has units of cm$^{-1}$ or wave numbers. Most of this paper is focussed on monochromatic radiation, so we will omit the argument $\nu$ of the intensity function, except in Section \ref{hc}, where we discuss the contribution of radiation to heat conduction in optically thick clouds. For axially symmetric radiation an increment of solid angle, in steradians (sr),  is $d\Omega = 2\pi d\mu$. The optical depth $\tau=\tau(z)$ at an altitude $z$ above the bottom of the cloud is 
\begin{equation}
\tau = \int_0^{z}dz' \,\alpha(z'),
\label{int0}
\end{equation}
where $\alpha =\alpha(z)$ is the attenuation rate, in units of $e$-foldings per altitude increment.   Both absorption and scattering contribute to $\alpha$.  We assume that the cloud is homogeneous in the sense that neither the single scattering albedo $\tilde\omega$, nor the scattering phase function $p(\mu,\mu')$ depend on $\tau$. The probability that  a collision with a cloud particulate scatters a photon, propagating with direction cosine $\mu'$, to a photon with a direction cosine between $\mu$ and $\mu+d\mu$ is $d\mu\, p(\mu,\mu')/2$.

\begin{figure}\centering
\includegraphics[height=80mm,width=1\columnwidth]{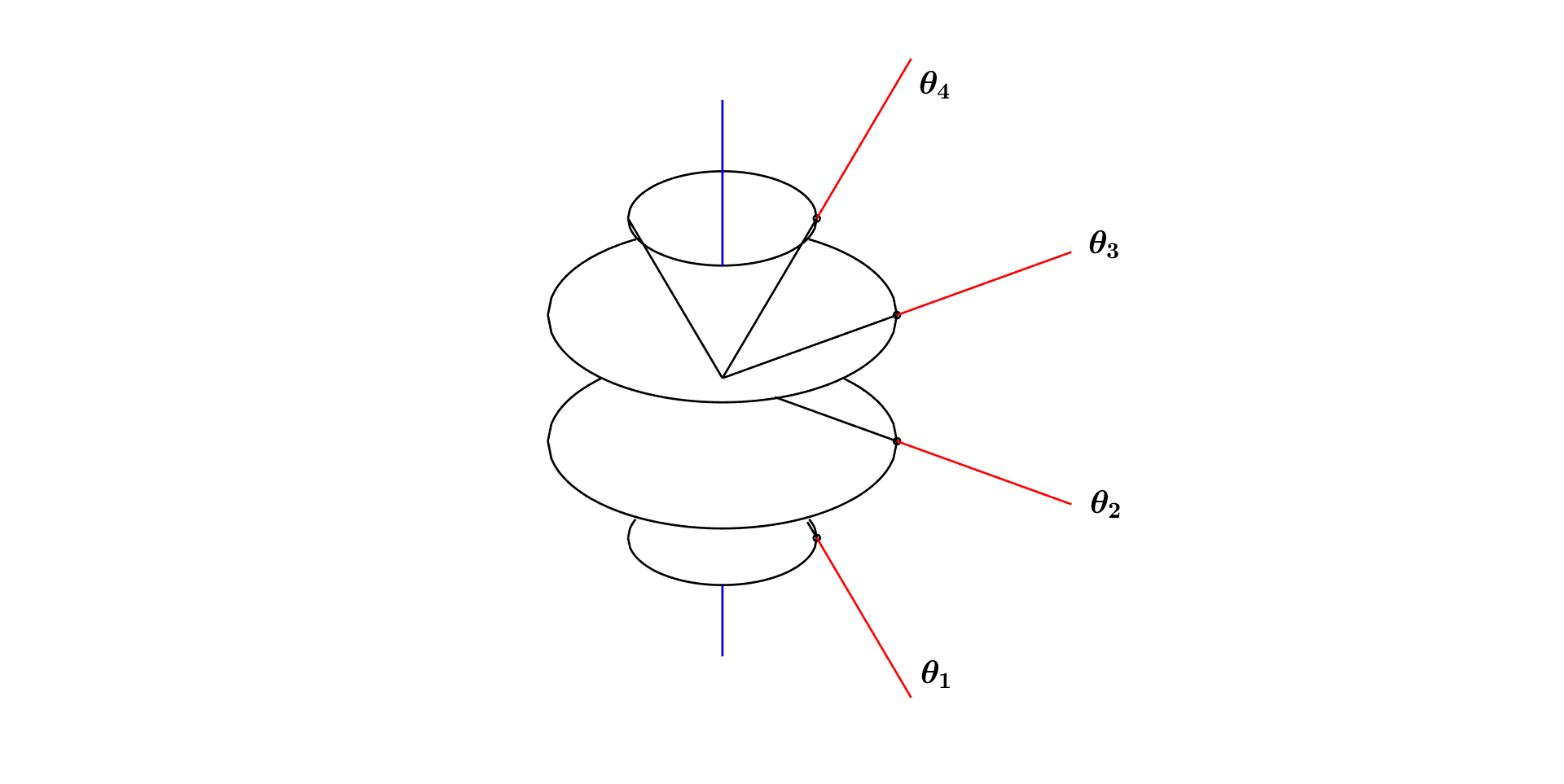}
\caption {Sample directions of $2n = 4$ streams of axially symmetric radiation. The radiation is centered on conical surfaces with opening angles $\theta_i$ to the zenith. The cosines of the opening angles, $\mu_i=\cos\theta_i$ are the roots of the Legendre polynomial $P_{2n}(\mu)=P_4(\mu)$, that is, $P_4(\mu_i)=0$ for $\mu_1, \mu_2, \mu_3, \mu_4 = -0.8611, -0.3400, \,0.3499,\,0.8611$.
\label{streams}}
\end{figure}

As in WH, we characterize angular dependence of  the  intensity with $2n$ sample values,
$I(\mu_i,\tau)\ge 0$, along the directions of the streams $i=1,2,3,\ldots, 2n$. As sketched in Fig. \ref{streams}, the $i$th stream makes an angle $\theta_i=\cos^{-1}\mu_i$ to the zenith.  
The {\it Gauss-Legendre} direction cosines, $\mu_i$,  are the zeros of the Legendre polynomial $P_{2n}$ of degree $2n$,  
\begin{equation}
P_{2n}(\mu_i)=0.
\label{int2}
\end{equation}
We will choose the indices $i=1,2,3,\cdots,2n$ such that
\begin{equation}
\mu_1<\mu_2<\mu_2<\cdots<\mu_{2n}.
\label{int4}
\end{equation}
Because the Legendre polynomial $P_{2n}$ is even, with $P_{2n}(\mu)=P_{2n}(-\mu)$,
the values of $\mu_i$ occur as equal and opposite pairs,
\begin{equation}
\mu_i=-\mu_{r(i)},
\label{int6}
\end{equation}
where the index reflection function is
\begin{equation}
r(i)=2n+1-i.
\label{int8}
\end{equation}
The weighted  sample values of the intensity, $w_i I(\mu_i,\tau)$, will be denoted with the symbol
 $\lvec\mu_i|I(\tau)\}$,
\begin{equation}
\lvec\mu_i|I(\tau)\}=w_i I(\mu_i,\tau)\ge 0.
\label{int10}
\end{equation}
A formula for the Gauss-Legendre weights, $w_i>0$, was given by (WH-72) as
\begin{eqnarray}
\frac{1}{w_i}=\sum_{l=0}^{2n-1}\frac{2l+1}{2}P_l^2(\mu_i).
\label{int12}
\end{eqnarray}
To simplify equations, we denote the intensity
as an abstract vector $|I\}=|I(\tau)\}$. We can represent the abstract vector $|I\}$ with a $2n\times 1$ column  vector
\begin{equation}
|I\}=\sum_{i=1}^{2n}|\mu_i)\lvec \mu_i|I\}=\left[\begin{array}{c}\lvec\mu_1|I\}\\\lvec\mu_2|I\}\\ \vdots \\ \lvec\mu_{2n}|I\}\end{array}\right].
\label{int14}
\end{equation}
We will call the column vector on the right of  (\ref{int14}) the {\it $\mu$-space} representation of  the abstract vector $|I\}$.  We will use other $2n\times 1$ arrays of numbers to represent the same abstract vector $|I\}$ in other bases. Examples discussed below  are the multipole basis $|l)$ or the penetration-length basis $|\lambda_i)$. 
%
%

The  {\it stream basis} vectors $|\mu_i)$ of (\ref{int14}) can be represented with the unit column vectors
\begin{equation}
|\mu_1)=\left[\begin{array}{c}1\\0\\ \vdots \\ 0\end{array}\right],\quad|\mu_2)=\left[\begin{array}{c}0\\1\\ \vdots \\ 0\end{array}\right],\quad \cdots,\quad|\mu_{2n})=\left[\begin{array}{c}0\\0\\ \vdots \\ 1\end{array}\right].
\label{int16}
\end{equation}
Corresponding left basis vectors can be represented as the unit row vectors
\begin{eqnarray}
\lvec\mu_1|&=&[1\quad 0\quad\cdots\quad 0],\nonumber\\
\lvec\mu_2|&=&[0\quad 1\quad\cdots\quad 0],\nonumber\\
&\vdots&\nonumber\\
\lvec\mu_{2n}|&=&[0\quad 0\quad\cdots\quad 1].\nonumber\\
\label{int18}
\end{eqnarray}
We use a double left parenthesis, $\lvec\mu_i|$, as a reminder that the row (or left)  basis vectors need not be Hermitian conjugates of the column (or right) basis vectors. The row  basis vectors $\lvec\mu_i|$ are like reciprocal lattice vectors of a crystal.  The column  basis vectors, $|\mu_i)$ are like direct lattice vectors. Just as low-symmetry crystals can have oblique, non-orthogonal lattice vectors, the right basis vectors for intensity need not be orthogonal \cite{Reciprocal}.

As discussed in connection with (WH-105), the stream basis vectors 
are right and left eigenvectors of the direction cosine matrix $\hat\mu$ 
\begin{equation}
\hat\mu|\mu_i)=\mu_i|\mu_i),\quad\hbox{and}\quad \lvec \mu_i|\hat\mu=\lvec \mu_i|\mu_i.
\label{int20}
\end{equation}
The eigenvectors are chosen to have the orthonormality property
\begin{equation}
\lvec \mu_i|\mu_{i'})=\delta_{i i'}.
\label{int22}
\end{equation}
They have the completeness property 
\begin{equation}
\sum_{i=1}^{2n}|\mu_i)\lvec \mu_i|=\hat 1.
\label{int24}
\end{equation}
In (\ref{int24}) and elsewhere, we will use the symbol $\hat 1$ to denote a square identity matrix with ones along the main diagonal and zeros elsewhere. It has the same dimensions as any square matrices that are added to, subtracted from,  or equated to it.

Multiplying (\ref{int24}) on the left or right by $\hat\mu$ and using (\ref{int20}) we find an expression for the direction-cosine operator,
\begin{equation}
\hat\mu =\sum_{i=1}^{2n}\mu_i|\mu_i)\lvec \mu_i|.
\label{int26}
\end{equation}
The direction secant matrix $\hat\varsigma$ is the inverse of the direction cosine matrix $\hat \mu$. We can use (\ref{int26}) to write
\begin{equation}
\hat\varsigma=\hat\mu^{-1} =\sum_{i=1}^{2n}\varsigma_i|\mu_i)\lvec \mu_i|.
\label{int28}
\end{equation}
The eigenvalues of the direction secant matrix are the inverses of the eigenvalues 
$\mu_i$ of the direction cosine matrix
\begin{equation}
\varsigma_i=\frac{1}{\mu_i}.
\label{int30}
\end{equation}
The direction secant matrix $\hat\varsigma$ has the same left and right eigenvectors as the direction cosine matrix $\hat\mu$,
\begin{equation}
\hat\varsigma|\mu_i)=\varsigma_i|\mu_i),\quad\hbox{and}\quad \lvec \mu_i|\hat\varsigma=\lvec \mu_i|\varsigma_i.
\label{int31}
\end{equation}

In accordance with  (WH-117) we define a reflection operator
\begin{equation}
\hat r = \sum_{i=1}^{2n}|\mu_i)\lvec \mu_{r(i)}|,
\label{int32}
\end{equation}
where the index reflection function $r(i)$ was defined by (\ref{int8}).  From (\ref{int32}) we see that the reflection operator transforms the stream basis vector $|\mu_i)$   to a basis vector for the opposite direction $|\mu_{r(i)})$, where $\mu_{r(i)}=-\mu_i$.
\begin{equation}
\hat r|\mu_i)=| \mu_{r(i)}),\quad\hbox{and}\quad \lvec \mu_i|\hat r = \lvec \mu_{r(i)}|.
\label{int33}
\end{equation}
In $\mu$-space the matrices representing $\hat r$ of (\ref{int32}) are unit anti-diagonal matrices. For  example, we can write the matrices $\lvec\mu_i|\hat r|\mu_{i'})$ representing $\hat r$ in $\mu$-space of $2n = 2$ or  $2n=4$ dimensions as
\begin{equation}
\hat r =\left[\begin{array}{cc}0&1\\1&0\end{array}\right],\quad\hbox{or}\quad\hat r =\left[\begin{array}{cccc}0&0&0&1
\\0&0&1&0 \\0&1&0&0\\1&0&0&0\end{array}\right].
\label{int34}
\end{equation}
The reflection matrices are their own inverses
\begin{equation}
\hat r^2=\hat 1.
\label{int36}
\end{equation}

In accordance with (WH-88), it is convenient to use the stream basis vectors to define a projection matrix $\mathcal{M}_{\bf d}$  for downward streams with indices $j\le n$ and $\mu_j<0$, and a projection matrix $\mathcal{M}_{\bf u}$ for upward streams with indices $k>n$ and $\mu_k>0$.
\begin{equation}
\mathcal{M}_{\bf d }=\sum_{j=1}^{n}|\mu_j)\lvec \mu_j|\quad\hbox{and}\quad
\mathcal{M}_{\bf u} =\sum_{k=n+1}^{2n}|\mu_k)\lvec \mu_k|.
\label{int38}
\end{equation}
The projection matrices  of (\ref{int38}) have the simple algebra
\begin{eqnarray}
\mathcal{M}_{\bf d }+\mathcal{M}_{\bf u} &=&\hat 1\label{int40}\\
\mathcal{M}^2_{\bf d }=\mathcal{M}_{\bf d}\quad&\hbox{and}&\quad
\mathcal{M}^2_{\bf u}=\mathcal{M}_{\bf u}\label{int42}\\
\mathcal{M}_{\bf d }\mathcal{M}_{\bf u}=\breve 0\quad&\hbox{and}&\quad
\mathcal{M}_{\bf u}\mathcal{M}_{\bf d} =\breve 0\label{int44}\\
\hat r\mathcal{M}_{\bf d}\hat r =\mathcal{M}_{\bf u}\quad&\hbox{and}&\quad
\hat r\mathcal{M}_{\bf u}\hat r =\mathcal{M}_{\bf d}.\label{int46}
\end{eqnarray}
Here and elsewhere, the symbol $\breve 0$ denotes a matrix, not necessarily square, for which all the elements are zero. The dimensions of $\breve 0$ are the same as the dimensions of other matrices to which it is added, subtracted or equated. 
We can use (\ref{int32}) to show that reflections change the signs of the direction-cosine operator of (\ref{int26}) and the direction secant operator of (\ref{int28})
\begin{equation}
\hat r \hat \mu\hat r = -\hat\mu, \quad\hbox{or}\quad\hat r \hat \mu=- \hat \mu\hat r, \label{int48}
\end{equation}
and
\begin{equation}
\hat r \hat \varsigma\hat r = -\hat \varsigma, \quad\hbox{or}\quad\hat r \hat \varsigma=- \hat \varsigma\hat r.\label{int50}
\end{equation}
In accordance with (WH-109) and (WH-110), we write the direction-cosine matrix 
$\hat\mu$ as the sum of a downward part $\hat\mu_{\bf d}$ and an upward part $\hat\mu_{\bf u}$ 
\begin{equation}
\hat \mu =\hat\mu_{\bf d} +\hat\mu_{\bf u}.
\label{int52}
\end{equation}
Expressions for the downward and upward parts are
\begin{eqnarray}
\hat\mu_{\bf d} &=&\mathcal{M}_{\bf d}\hat\mu =\hat\mu\mathcal{M}_{\bf d} =\sum_{j=1}^n\mu_j|\mu_j)\lvec \mu_j|,\label{int54}\\
 \hat\mu_{\bf u} &=&\mathcal{M}_{\bf u}\hat\mu =\hat\mu\mathcal{M}_{\bf u} =\sum_{k=n+1}^{2n}\mu_k|\mu_k)\lvec \mu_k|.
\label{int56}
\end{eqnarray}
Reflections transform the matrices of (\ref{int54}) and (\ref{int56}) into the negatives  each other 
\begin{equation}
\hat r\mu_{\bf d }\hat r=-\mu_{\bf u},\quad\hbox{and}\quad\hat r\mu_{\bf u }\hat r=-\mu_{\bf d}.
\label{int58}
\end{equation}
In like manner, we write the direction-secant matrix 
$\hat\varsigma$ as the sum of a downward part $\hat\varsigma_{\bf d}$ and an upward part $\hat\varsigma_{\bf u}$ 
\begin{equation}
\hat \varsigma =\hat\varsigma_{\bf d} +\hat\varsigma_{\bf u}.
\label{int60}
\end{equation}
Expressions for the downward and upward parts are
\begin{eqnarray}
\hat\varsigma_{\bf d} &=&\mathcal{M}_{\bf d}\hat\varsigma =\hat\varsigma\mathcal{M}_{\bf d} =\sum_{j=1}^n\varsigma_j|\mu_j)\lvec \mu_j|,\label{int62}\\
 \hat\varsigma_{\bf u} &=&\mathcal{M}_{\bf u}\hat\varsigma =\hat\varsigma\mathcal{M}_{\bf u} =\sum_{k=n+1}^{2n}\varsigma_k|\mu_k)\lvec \mu_k|.
\label{int64}
\end{eqnarray}
Reflections transform the matrices of (\ref{int62}) and (\ref{int64}) into the negatives of each other 
\begin{equation}
\hat r\varsigma_{\bf d }\hat r=-\varsigma_{\bf u}\quad\hbox{and}\quad\hat r\varsigma_{\bf u }\hat r=-\varsigma_{\bf d}.
\label{int66}
\end{equation}
\subsection{Pseudoinverse matrices\label{pim}}
To simplify notation for the analysis of cloud scattering it will be convenient to use pseudoinverse matrices. From (\ref{int22}), (\ref{int30}), (\ref{int54}) and (\ref{int62}) we find
\begin{eqnarray}
\hat\mu_{\bf d}\hat\varsigma_{\bf d} &=&\sum_{j=1}^n\mu_j|\mu_j)\lvec \mu_j|
\sum_{j'=1}^n\varsigma_{j'}|\mu_{j'})\lvec \mu_{j'}|\nonumber\\
&=&\sum_{j=1}^n\mu_j\varsigma_j|\mu_j)\lvec \mu_j|\nonumber\\
&=&\sum_{j=1}^n|\mu_j)\lvec \mu_j|\nonumber\\
\label{pim2}
\end{eqnarray}
or in view of (\ref{int38}),
\begin{equation}
\hat\mu_{\bf d}\hat\varsigma_{\bf d} =\mathcal{M}_{\bf d}.
\label{pim4}
\end{equation}
It is therefore convenient to write 
\begin{equation}
\hat\mu_{\bf d}^{-1}=\hat\varsigma_{\bf d} \quad\hbox{and}\quad\hat\varsigma_{\bf d} ^{-1}=\hat\mu_{\bf d}.
\label{pim6}
\end{equation}
The determinants of the $2n\times 2n$ matrices representing $\hat\mu_{\bf d}$ and $\hat\varsigma_{\bf d}$ are both zero, so neither has a true inverse matrix. But both have pseudoinverses in the sense that the product of the matrix with its pseudoinverse gives a projection operator $\mathcal{M}_{\bf d}$.  Using pseudoinverse notation simplifies several important equations in subsequent sections.
We will use the same superscript, $-1$, to denote both  true inverse matrices like $\hat r^{-1}=\hat r $, where $\det (\hat r)=(-1)^n\ne 0$, and pseudoinverse matrices like $\hat \mu_{\bf d}^{-1}$, where  $\det (\hat \mu_{\bf d})=0$. The context will normally make clear which type of inverse is meant.
\subsection{Multipoles}
Describing the angular distribution of the axially symmetric intensity $I(\mu, \tau)$ with the $2n$ samples
 values,  $I(\mu_i, \tau)$, at the  Gauss-Legendre direction cosines $\mu_i$ of  (\ref{int2}), is equivalent to approximating the intensity as a superposition of the first $2n$ Legendre polynomials,
\begin{equation}
I(\mu,\tau)= \sum_{l=0}^{2n-1}(2l+1)P_{l}(\mu)I_l(\tau).
\label {mm2}
\end{equation}
The intensity multipoles  are
\begin{eqnarray}
I_l(\tau)&=&\lvec l|I(\tau)\}\nonumber\\
&=&\sum_{i=1}^{2n}\lvec l|\mu_i)\lvec\mu_i|I(\tau)\}.
\label {mm4}
\end{eqnarray}
Projections $\lvec l|\mu_i)$ of  the left multipole basis vector $\lvec l|$ onto the right stream vector $|\mu_i)$, and vice versa, were given by (WH-84) and (WH-85) in terms of Legendre polynomials $P_l$, and weights $w_i$ of (\ref{int12}) as
\begin{equation}
 \lvec l|\mu_i) =\frac{1}{2}P_l(\mu_i),
\label {mm6}
\end{equation}
and
\begin{equation}
\lvec\mu_i|l) = w_i(2l+1)P_{l}(\mu_i).
\label {mm8}
\end{equation}
Substituting (\ref{mm6}) and (\ref{int10}) into (\ref{mm4}), and noting from (\ref{mm2}) that $I(\mu,\tau)$ can be written as a superpositon of the first $2n$ Legendre polynomials,  we can use the Gauss-Legendre quadrature\,\cite{Gauss} to write the $l$-th multipole moment of the intensity as
\begin{eqnarray}
I_l(\tau)
&=&\frac{1}{2}\sum_{i=1}^{2n}w_i P_l(\mu_i)I(\mu_i,\tau)\nonumber\\
&=&\frac{1}{2}\int_{-1}^1 d\mu P_l(\mu)I(\mu,\tau).
\label {mm9a}
\end{eqnarray}
In analogy to (\ref{int22}) the multipole basis vectors $|l')$ and $\lvec l|$ have been chosen to have the orthonormality property
\begin{equation}
\lvec l|l')=\delta_{l l'}.
\label{mm9b}
\end{equation}
In analogy to (\ref{int24}), they  have the completeness property 
\begin{equation}
\sum_{l=0}^{2n-1}|l)\lvec l|=\hat 1.
\label{mm9c}
\end{equation}
Using (\ref{mm9b}) we can write the intensity vector as
\begin{equation}
|I(\tau)\}=\sum_{l=0}^{2n-1}|l)\lvec l|I(\tau)\}=\sum_{l=0}^{2n-1}|l)I_l(\tau)
\label{mm9d}
\end{equation}
The expansion coefficients $\lvec l|I(\tau)\}=I_l(\tau)$ were given by (\ref{mm9a}).

From (\ref{mm8}) we see that the elements $\lvec \mu_i|0)$ of the right monopole basis vector  $|0)$ are the weights $w_i$ of (\ref{int12})
\begin{equation}
|0)=\sum_{i=1}^{2n}|\mu_i)\lvec \mu_i|0)=\left[\begin{array}{c}w_1\\ w_2\\ \vdots \\ w_{2n}\end{array}\right],
\label{mm10}
\end{equation}
From (\ref{mm6}) we see that the elements $\lvec 0|\mu_i)$ of the left monopole basis vector $\lvec 0|$  are all equal to 1/2,
\begin{equation}
\lvec 0|=\sum_{i=1}^{2n}\lvec 0|\mu_i)\lvec \mu_i|=\sum_{i=1}^{2n}\frac{1}{2}\lvec \mu_i|=\frac{1}{2}[1\quad 1\quad\cdots \quad 1].
\label{mm12}
\end{equation}
In multipole space ($l$-space) the reflection operator $\hat r$ of (\ref{int32}) becomes
\begin{equation}
\hat r=\sum_{l=0}^{2n-1}(-1)^l|l)\lvec l|.
\label{mm14}
\end{equation}
The matrices representing $\hat r$ have alternating 1's and -1's along the main diagonal and zeros elsewhere. For example, in $l$-space of $2n = 2$ or  $2n=4$ dimensions we can write the matrices $\lvec l|\hat r|l')$ representing $\hat r$ as
\begin{equation}
\hat r =\left[\begin{array}{rr}1&0\\0&-1\end{array}\right]\quad\hbox{or}\quad\hat r =\left[\begin{array}{rrrr}1&0&0&0
\\0&-1&0&0 \\0&0&1&0\\0&0&0&-1\end{array}\right].
\label{mm16}
\end{equation}
A simple consequence of (\ref{mm14}) is
\begin{equation}
\hat r|l)=(-1)^l|l),\quad\hbox{and}\quad \lvec l|\hat r = \lvec l|(-1)^l.
\label{mm18}
\end{equation}
Some useful identities for the multipole basis that follow from  (WH-17) are
\begin{eqnarray}
\lvec 0|\hat\mu=\lvec 1|\quad \hbox{and}\quad \hat\mu|0)=\frac{1}{3}|1),
\label{mm20}
\end{eqnarray}
and
\begin{eqnarray}
\lvec 1|\hat\mu=\frac{1}{3}\lvec 0|+\frac{2}{3}\lvec 2|\quad \hbox{and}\quad \hat\mu|1)=|0)+\frac{2}{5}|2).
\label{mm22}
\end{eqnarray}

Expansions like (\ref{mm9d}) of the intensity vector onto the basis functions $|l)$,  or  expansions of operators, like the reflection operator $\hat r$ of (\ref{mm14}) onto the basis dyadics $|l)\lvec l'|$  will be called {\it $l$-space} representations.
\subsection{Equation of radiative transfer\label{et}}
The change of the intensity vector $|I\}=|I(\tau)\}$  with optical depth $\tau$ in a cloud is  described by the equation of radiative transfer, given by (WH-62) as
\begin{equation}
\frac{d}{d\tau}|I\}=\hat \kappa\left(|B\}- |I\}\right).
\label{et2}
\end{equation}
The Planck intensity vector $|B\}=|B(\tau)\}$ of (WH-53) is
\begin{equation}
|B\}=|0)B.
\label{et4}
\end{equation}
The isotropic thermal emission of (\ref{et4}) is proportional to  $|0)$,  the right monopole basis vector of (\ref{mm10}).
For  monochromatic radiation of spatial frequency $\nu$ the Planck intensity $B=B(\nu,T)$ for the absolute temperature $T$ is
\begin{eqnarray}
B(\nu,T)&=&\frac{2h_{\rm P}c^2\nu^3}{e^{\nu c\, h_{\rm P}/(k_{\rm B}T)}-1}.
\label{et10}
\end{eqnarray}
Here $ h_{\rm P}$ is Planck's constant,  $c$ is the speed of light and $k_{\rm B}$ is Boltzmann's constant.
Integrating the upward flux, $\mu B$ over $2\pi$ solid-angle increments, $d\Omega = 2\pi d\mu$, gives a factor of $\pi$. The Stefan-Boltzmann flux, the frequency-integrated                                               and solid-angle-integrated flux is
\begin{eqnarray}
Z_{\rm SB}(T)=\int_0^{\infty}d\nu\, \pi B(\nu,T)=\sigma_{\rm SB} T^4,
\label{et11a}
\end{eqnarray}
The Stefan-Boltzmann constant is
\begin{equation}
\sigma_{\rm SB}=\frac{2\pi^5  k_{\rm B}^4}{15 h_{\rm P}^3 c^2} = 5.67\times 10^{-8} \hbox{W m$^{-2}$ K$^{-4}$}.
\label{et11b}
\end{equation}
If the temperature depends on optical depth, $T=T(\tau)$, the Planck intensity (\ref{et10}) will also be a function of optical depth, $B=B(\tau)$.

The exponentiation-rate matrix $\hat \kappa$ of (\ref{et2}) was given by (WH-63) as
\begin{equation}
\hat \kappa = \hat\varsigma\hat\eta,
\label{et12}
\end{equation}
the product of the direction-secant matrix $\hat\varsigma$ of  (\ref{int28}), and 
the efficiency matrix $\hat \eta$ of (WH-54),
\begin{equation}
\hat \eta =\hat 1-\frac{1}{2}\tilde\omega \hat p.
\label{et16}
\end{equation}
The single scattering albedo $\tilde \omega$ of (\ref{et16}) is the probability that a photon that collides with a cloud particulate is scattered, rather than being absorbed and converted to heat. Probabilities must be nonnegative and  no larger than 1. So $\tilde\omega$ must be  bounded by
\begin{equation}
0\le \tilde\omega \le 1.
\label{et18}
\end{equation}
The symbol $\hat p$ of (\ref{et16}) denotes the phase-function matrix (basically, the differential scattering cross section) of the particulates.
The matrix elements $\lvec \mu_i|\hat p|\mu_{i'})=w_i p(\mu_i,\mu_{i'})$  give the representation of $\hat p$ in $\mu$-space as a $2n\times 2n$ array of numbers. The matrix $\hat p$ is defined such  that 
a photon in the stream  $i'$ that is not absorbed in a collision with a cloud particulate 
has a probability
$\lvec \mu_i|\hat p|\mu_{i'})/2$ to be scattered into the stream $i$. Therefore we must have
\begin{equation}
\sum_{i=1}^{2n}\frac{1}{2}\lvec \mu_i|\hat p|\mu_{i'})=1\quad\hbox{and}\quad 
0\le\frac{1}{2}\lvec \mu_i|\hat p|\mu_{i'})\le 1.
\label{et20}
\end{equation}
In accordance with (WH-40), for a cloud of randomly oriented scattering particulates
the scattering operator can be written in terms of  the right and left multipole basis vectors $|l)$ and $\lvec l|$  of (\ref{mm6}) and (\ref{mm8}) as
\begin{equation}
\hat p = 2\sum_{l=0}^{2n-1}p_l|l)\lvec l|,
\label{et30}
\end{equation}
where the multipole scattering coefficients are $p_l$. 
Substituting (\ref{et30}) into (\ref{et16}) we see that the efficiency matrix is
\begin{equation}
\hat \eta =\sum_{l=0}^{2n-1}\eta_l|l)\lvec l|.
\label{et32}
\end{equation}
In view of (\ref{et30}) and (\ref{et16}), we see that
the coefficients $\eta_l$ (or the eigenvalues of $\hat \eta$) are
\begin{equation}
\eta_l =1-\tilde\omega p_l.
\label{et34}
\end{equation}
In accordance with the discussion in Section {\bf 3.2.1} of WH, for anisotropic intensity, with no dependence on optical depth $\tau$   in a spatially infinite cloud, the $l$th multipole moment of the intensity, $I_l=\lvec l|I\}$, decays with relative time $\theta$ as
\begin{equation}
I_l(\theta) = I_l(0)e^{-\theta \eta_l}.
\label{et35}
\end{equation}
One unit of relative time is the mean time between collisions of a photon with a cloud particulate.
The total number of photons must decay at the rate
\begin{equation}
\eta_0=1-\tilde\omega.
\label{et36}
\end{equation}
Comparing (\ref{et36}) with (\ref{et34}) we see that the monopole scattering coefficient must be unity,
\begin{equation}
p_0=1.
\label{et37}
\end{equation}
Even for hypothetical, non-absorbing particulates, with $\tilde\omega = 1$, successive 
scatterings must randomize anisotropies of the intensity. So (\ref{et35}) implies that we must have
\begin{equation}
\eta_l=1-p_l>0\quad\hbox{or}\quad p_l<1\quad\hbox{for}\quad l>0.
\label{et38}
\end{equation}
Multipole coefficients $p_l=\varpi^{\{p\}}_l$ for the maximum forward scattering phase functions  that can be constructed from the first $2p$ Legendre polynomials were listed in Table  1 of Section {\bf 5} of WH \cite{WH1}. 

In discussions of thermal emission, we will have occasion to use the identity
\begin{equation}
\hat \kappa|0) =(1-\tilde\omega) \hat\varsigma|0),
\label{et39}
\end{equation}
which follows from (\ref{et12}), (\ref{et32}) and (\ref{et36}).

The scattering-phase operator $\hat p$ of (\ref{et30}) and the efficiency operator $\hat\eta$  of (\ref{et32}) are diagonal in $l$-space, so they commute with the reflection operator $\hat r$ of (\ref{mm14}), which is also diagonal. Therefore  $\hat p$ and $\hat \eta$ are even under reflections,
\begin{equation}
\hat r\hat p \hat r = \hat p\quad\hbox{or}\quad \hat r\hat p =\hat p \hat r,\label{et40}
\end{equation}
and
\begin{equation}
\hat r\hat \eta \hat r = \hat p\quad\hbox{or}\quad \hat r\hat \eta =\hat \eta \hat r.\label{et42}
\end{equation}
We  can  use (\ref{et40}) and (\ref{et42}) together with the fact that $\hat r\hat r =\hat 1$ to show that the exponentiation-rate operator $\hat\kappa$ of (\ref{et12}) is odd under reflections,
\begin{eqnarray}
\hat r \hat \kappa\hat r &=& \hat r \hat\varsigma\hat r \hat r 
\hat\eta\hat r\nonumber\\
&=&-\hat\varsigma\hat\eta\nonumber\\
&=&-\hat\kappa\quad\hbox{or}\quad \hat r\hat \kappa =-\hat \kappa \hat r.
\label{et44}
\end{eqnarray}

\subsection{Penetration modes \label{lam}}
For homogeneous clouds the {\it penetration modes} are a convenient basis for representing the intensity vector $|I\}$. According to (WH-150), these are 
the left and right eigenvectors, $\lvec\lambda_i|$ and $|\lambda_i)$, of the exponentiation-rate matrix $\hat\kappa$ of (\ref{et12}),
\begin{equation}
\lvec \lambda_i|\hat \kappa =\lvec\lambda_i|\kappa_i\quad \hbox{and}\quad
\hat\kappa|\lambda_i)=\kappa_i|\lambda_i)\quad\hbox{where}\quad \lambda_i=1/\kappa_i.
\label{lam2}
\end{equation}
The eigenvalues of $\hat \kappa$ are 
\begin{equation}
\kappa_i=\frac{1}{\lambda_i}.
\label{lam3}
\end{equation}

Following (WH-154), we assume indices $i$ are chosen such that
\begin{equation}
\lambda_1<\lambda_2< \lambda_3<\cdots< \lambda_{2n}.
\label{lam4}
\end{equation}
In accordance with (\ref{et44}), the exponentiation-rate operator $\hat \kappa$  has odd reflection symmetry,
$\hat r\hat\kappa\hat r=-\hat\kappa$,  so the values of $\kappa_i$ or $\lambda_i$ occur as equal and opposite pairs,
\begin{equation}
\lambda_i=-\lambda_{r(i)}.
\label{lam5}
\end{equation}
The index reflection function $r(i)$ was given by (\ref{int8}).
A reflection transforms an upward penetration mode into a downward mode, and vice versa,
\begin{equation}
\hat r|\lambda_i)=|\lambda_{r(i)}),
\label{lam6}
\end{equation}
From (\ref{lam4}) and (\ref{int6}) it follows that the first half of the eigenvalues 
$\lambda_i$ are negative and the second half are positive,
\begin{eqnarray}
\lambda_i&<&0\quad\hbox{for}\quad i\le n,\nonumber\\
\lambda_i&>&0\quad\hbox{for}\quad i>n.
\label{lam8}
\end{eqnarray}
Following (WH-145) we assume that right eigenvectors are normalized such that
\begin{equation}
\lvec 0|\lambda_i) =\frac{1}{2}\quad\hbox{or}\quad \sum_{i'=1}^{2n}\lvec\mu_{i'}|\lambda_i) = 1.
\label{lam10}
\end{equation}
The elements of the $i$th column of the overlap matrix $\lvec \mu_{i'}|\lambda_i)$ sum to 1.
The left eigenvectors $\lvec\lambda_i|$ are chosen to ensure the orthonormality condition,
\begin{equation}
\lvec\lambda_i|\lambda_{i'}) =\delta_{i i'}.
\label{lam12}
\end{equation}
The left and right penetration-mode bases  have the completeness property 
\begin{equation}
\sum_{l=0}^{2n-1}|\lambda_i)\lvec \lambda_i|=\hat 1.
\label{lam13}
\end{equation}
Vectors expanded on  the basis vectors $|\lambda_i)$ or operators expanded on the basis dyadics $|\lambda_i)\lvec\lambda_{i'}|$  will be said to be represented in
 {\it $\lambda$-space}.

The projection operators for the lower and upper halves of $\lambda$-space are analogous to the projection operators (\ref{int38}) of $\mu$ space,
\begin{equation}
\mathcal{L}_{\bf d}=\sum_{j=1}^n|\lambda_j)\lvec \lambda_j|\quad\hbox{and}\quad\mathcal{L}_{\bf u}=\sum_{k=n+1}^{2n}|\lambda_k)\lvec \lambda_k|.
\label{lam14}
\end{equation}
The projection matrices  of (\ref{lam14}) have the simple algebra, analogous to (\ref{int40}) -- (\ref{int46}),
\begin{eqnarray}
\mathcal{L}_{\bf d }+\mathcal{L}_{\bf u} &=&\hat 1\label{lam16}\\
\mathcal{L}^2_{\bf d }=\mathcal{L}_{\bf d}\quad&\hbox{and}&\quad
\mathcal{L}^2_{\bf u}=\mathcal{L}_{\bf u}\label{lam18}\\
\mathcal{L}_{\bf d }\mathcal{L}_{\bf u}=\breve 0\quad&\hbox{and}&\quad
\mathcal{L}_{\bf u}\mathcal{L}_{\bf d} =\breve 0\label{lam20}\label{lam20a}\\
\hat r \mathcal{L}_{\bf d }\hat r =\mathcal{L}_{\bf u}\quad&\hbox{and}&\quad
\hat r \mathcal{L}_{\bf u}\hat r =\mathcal{L}_{\bf d}.\label{lam21}
\end{eqnarray}
In accordance with (WH-151), we write the exponentiation-rate matrix 
$\hat\kappa=\hat\kappa(\tau)$ as the sum of a downward part $\hat\kappa_{\bf d}$ and an upward part $\hat\kappa_{\bf u}$ 
\begin{equation}
\hat \kappa =\hat\kappa_{\bf d} +\hat\kappa_{\bf u}.
\label{lam22}
\end{equation}
Expressions for the downward and upward parts are
\begin{eqnarray}
\hat\kappa_{\bf d} &=&\mathcal{L}_{\bf d}\hat\kappa =\hat\kappa\mathcal{L}_{\bf d} =\sum_{j=1}^n\kappa_j|\lambda_j)\lvec \lambda_j|,\label{lam24a}\\
 \hat\kappa_{\bf u} &=&\mathcal{L}_{\bf u}\hat\kappa =\hat\kappa\mathcal{L}_{\bf u} =\sum_{k=n+1}^{2n}\kappa_k|\lambda_k)\lvec \lambda_k|.
\label{lam24b}
\end{eqnarray}
Reflections transform the matrices of (\ref{lam24a}) and (\ref{lam24b}) as
\begin{equation}
\hat r\hat\kappa_{\bf d }\hat r=-\hat \kappa_{\bf u}\quad\hbox{and}\quad\hat r\hat \kappa_{\bf u }\hat r=-\hat\kappa_{\bf d}.
\label{lam26}
\end{equation}

In like manner, we write the penetration-length matrix 
$\hat\lambda$ as the sum of a downward part $\hat\lambda_{\bf d}$ and an upward part $\hat\lambda_{\bf u}$ 
\begin{equation}
\hat \lambda =\hat\lambda_{\bf d} +\hat\lambda_{\bf u}.
\label{lam28}
\end{equation}
Expressions for the downward and upward parts are
\begin{eqnarray}
\hat\lambda_{\bf d} &=&\mathcal{L}_{\bf d}\hat\lambda =\hat\lambda\mathcal{L}_{\bf d} =\sum_{j=1}^n\lambda_j|\lambda_j)\lvec \lambda_j|,\label{lam30}\\
 \hat\lambda_{\bf u} &=&\mathcal{L}_{\bf u}\hat\lambda =\hat\lambda\mathcal{L}_{\bf u} =\sum_{k=n+1}^{2n}\lambda_k|\lambda_k)\lvec \lambda_k|.
\label{lam32}
\end{eqnarray}
Reflections transform the matrices of (\ref{lam30}) and (\ref{lam32}) into the negatives of each other 
\begin{equation}
\hat r\hat\lambda_{\bf d }\hat r=-\hat\lambda_{\bf u}\quad\hbox{and}\quad\hat r\hat\lambda_{\bf u }\hat r=-\hat\lambda_{\bf d}.
\label{lam34}
\end{equation}

In accord with (WH-162) we define overlap operators $\mathcal{C}_{\bf q q'}$ between the projection operators $\mathcal{M}_{\bf q}$ of (\ref{int38}) and the projection operators $\mathcal{L}_{\bf q'}$ of (\ref{lam14}) by
\begin{equation}
\mathcal{C}_{\bf q q'}=\mathcal{M}_{\bf q}\mathcal{L}_{\bf q'}.
\label{lam28a}
\end{equation}
The variable indices ${\bf q}$ and ${\bf q'}$ in (\ref{lam28a})  can take on the values ${\bf u}$ and ${\bf d}$.  In accord with (WH-163), when represented by matrices in $\mu$-space, $\lambda$-space or $l$-space, the overlap operators sum to the identity operator $\hat 1$,
\begin{eqnarray}
\hat 1 =\sum_{\bf q q'}\mathcal{C}_{\bf q q'}
= \mathcal{C}_{\bf d d}+ \mathcal{C}_{\bf u d}+ \mathcal{C}_{\bf d u}
+ \mathcal{C}_{\bf u u}.
\label{lam30a}
\end{eqnarray}

In the pure absorption limit, $\tilde\omega \to 0$,  limiting values of variables for $\lambda$-space are equal to the corresponding values for $\mu$-space. 
Examples are
\begin{eqnarray}
\hat \eta &\to& \hat 1\nonumber\\
\hat \lambda &\to& \hat \mu\nonumber\\
\hat \kappa &\to& \hat\varsigma\nonumber\\
|\lambda_i) &\to& |\mu_i)\nonumber\\
\lvec\lambda_i| &\to& \lvec \mu_i|\nonumber\\
\mathcal{L}_{\bf q}&\to& \mathcal {M}_{\bf q}.
\label{lam32a}
\end{eqnarray}
\subsection{External metrics\label{ex}}
One can remotely measure the radiation emitted from the top of a cloud with instruments on satellites,  or from the bottom of a cloud with ground-based instruments.
The upwelling intensity at the cloud top will consist of upward radiation that has been thermally generated in the cloud, plus  backscattered downward radiation from external sources above or transmittted radiation from external sources below.
Similarly,  the downwelling intensity at  the cloud bottom will consist of downward radiation that has been thermally generated in the cloud, plus backscattered upward radiation from external sources below or transmitted downward radiation
The boundary radiation  can be characterized  with the intensity vector $|I(\tau_c)\}$ at the cloud top and by  the intensity vector $|I(0)\}$  at the cloud bottom.
As discussed in (WH-174) and (WH-175), it is  convenient  to characterize radiation that can be observed externally with the outgoing intensity vector,
\begin{equation}
|I^{\{\rm out\}} \} =\mathcal{M}_{\bf u} |I(\tau_c)\}+
\mathcal{M}_{\bf d}|I(0)\} .
\label{ex2}
\end{equation}
The incoming intensity vector is
\begin{equation}
|I^{\{\rm in\}} \} =\mathcal{M}_{\bf d}|I(\tau_c)\}+\mathcal{M}_{\bf u}|I(0)\}.\label{ex4}
\end{equation}
The inverses of (\ref{ex2}) and  (\ref{ex4}) are
\begin{equation}
|I(0)\} =\mathcal{M}_{\bf u} |I^{\{\rm in\}}\}+
\mathcal{M}_{\bf d}|I^{\{\rm out\}} \}, \label{ex6}
\end{equation}
and
\begin{equation}
|I(\tau_c)\} =\mathcal{M}_{\bf d} |I^{\{\rm in\}}\}+
\mathcal{M}_{\bf u}|I^{\{\rm out\}} \}.
\label{ex8}
\end{equation}

The vertical flux vector is given by  (WH-210) as
\begin{equation}
|Z(\tau)\} =4\pi\hat\mu|I(\tau)\}.
\label{ex10}
\end{equation}
Unlike the vector intensity $|I\}$ , for which all elements in $\mu$ space are nonnegative, $\lvec\mu_i|I\}\ge 0$, the upward elements of $|Z\}$ are nonnegative, 
$\lvec\mu_k|Z\}=4\pi\mu_k\lvec\mu_k|I(\tau)\}\ge 0$ for $\mu_k>0$, and the downward elements are nonpositive 
$\lvec\mu_j|Z\}=4\pi\mu_j\lvec\mu_j|I(\tau)\}\le 0$ for $\mu_j<0$.
In analogy to (\ref{ex2})  we can write the outgoing flux vector as
\begin{eqnarray}
|Z^{\{\rm out\}} \} &=&\mathcal{M}_{\bf u} |Z(\tau_c)\}-
\mathcal{M}_{\bf d}|Z(0)\} \nonumber\\
&=&4\pi(\hat\mu_{\bf u}-\hat\mu_{\bf d})|I^{\{\rm out\}} \}.
\label{ex12}
\end{eqnarray}
In analogy to (\ref{ex4})  the incoming flux vector is
\begin{eqnarray}
|Z^{\{\rm in\}} \} &=&-\mathcal{M}_{\bf d}|Z(\tau_c)\}+\mathcal{M}_{\bf u}|Z(0)\}\nonumber\\
&=&4\pi(\hat\mu_{\bf u}-\hat\mu_{\bf d})|I^{\{\rm in\}} \}.
\label{ex14}
\end{eqnarray}
The negative signs of (\ref{ex12}) and (\ref{ex14}) ensure that all elements of the outgoing and incoming flux vectors are nonnegative,  $\lvec\mu_i|Z^{\{\rm out\}}\}\ge 0$ and $\lvec\mu_i|Z^{\{\rm in\}}\}\ge 0$
The inverses of (\ref{ex12}) and  (\ref{ex14}) are
\begin{equation}
|Z(0)\} =\mathcal{M}_{\bf u} |Z^{\{\rm in\}}\}-
\mathcal{M}_{\bf d}|Z^{\{\rm out\}} \}, \label{ex16}
\end{equation}
and
\begin{equation}
|Z(\tau_c)\} =-\mathcal{M}_{\bf d} |Z^{\{\rm in\}}\}+
\mathcal{M}_{\bf u}|Z^{\{\rm out\}} \}.
\label{ex18}
\end{equation}

\subsection{Externally incident and thermally emitted radiation \label{exin}}
The radiation intensity $|I\}=|I(\tau)\}$ of the equation of transfer (\ref{et2}) is the sum of a part $|\ddot I\}=|\ddot I(\tau)\}$ from external radiation, incident on the top and bottom of the cloud, and a part $|\dot I\}=|\dot I(\tau)\} $ that is created by  thermal emission of particulates inside the cloud.
\begin{equation}
|I\} =|\ddot I\}+|\dot I\}
\label{exin2}
\end{equation}
The dot and double dot above the intensity symbols $I$ should not be confused with Newton's notation for the first and second time derivatives.
Earth's atmosphere is too cool to emit visible light, so for optical frequencies,
$\nu\sim 20,\!000$ cm$^{-1}$, thermal emission intensity vanishes, $|\dot I\}=0$ and there is only externally generated intensity $|\ddot I\}$ from sunlight.  

Thermal radiation, with frequencies $\nu\sim 1,\!000$ cm$^{-1}$ is readily generated by cloud particulates, by greenhouse-gas molecules and by Earth's surface.
A cloud can contain externally generated thermal radiation $|\ddot I \}$, for example, the upward radiation from Earth's surface or from lower-altitude clouds, or downward radiation from higher-altitude clouds and the Sun. The Sun emits long-wave infrared radiation with frequencies, $\nu\sim 1,\!000$ cm$^{-1}$ similar to the thermal radiation of the Earth. But except for the strongly collimated long-wave infrared radiation coming directly from the Sun, locally generated radiation is much more intense.
 
Particulates inside a cloud emit thermal radiation $|\dot I \}$ in much the same way as incandescent soot particles emit yellow light from a hot candle flame.  Before escaping from the cloud, the emitted thermal radiation can be reabsorbed by either the particulates or greenhouse-gas molecules, and it can be scattered by particulates. A negligibly small fraction of radiation is scattered by greenhouse-gas molecules. The lack of scattering facilitates the calculation of thermal radiation transfer in cloud-free air \cite{WH0}.

The  intensity vector $|\ddot I\}$ for external radiation satisfies the cold-cloud  version of the equation of transfer (\ref{et2}), with vanishing Planck intensity $|B\}=\breve 0$,
\begin{equation}
\frac{d}{d\tau}|\ddot I\}=-\hat \kappa|\ddot I\}.
\label{exin4}
\end{equation}
To provide a unique solution to (\ref{exin4}), boundary conditions must be given. A commonly used boundary condition is   the value of the input  intensity $|\ddot I^{\{\rm in\}}\}$, given by (\ref{ex4}).  Then the solution of (\ref{exin4}) gives the scattering operator $\mathcal{S}$, which is defined by
\begin{equation}
|\ddot I^{\{\rm out\}}\}=\mathcal{S}|\ddot I^{\{\rm in\}}\}.
\label{exin6}
\end{equation}

The intensity vector $|\dot I\}=|\dot I(\tau)\}$  of thermally emitted radiation satisfies the full version of the equation of transfer (\ref{et2}) with a non-zero Planck intensity $|B\}$ of (\ref{et4}).
There can be no emitted radiation coming into the top or bottom of the cloud so the boundary condition that defines a unique solution of (\ref{et2}) is
\begin{equation}
|\dot I^{\{\rm in\}} \} =\breve 0.
\label{exin8}
\end{equation}
The intensity vector $|\dot I^{\{\rm out\}} \}$ of outgoing thermal radiation  has contributions from particulates in all optical depth intervals $d\tau'$ inside the cloud and can be written as
\begin{equation}
|\dot I^{\{\rm out\}} \} =\int_{0}^{\tau_c} d\tau' |G^{\{\rm out\}} (\tau')\} B(\tau').
\label{exin10}
\end{equation}
We will discuss the Green's-function vector  $|G^{\{\rm out\}}(\tau')\}$ for outgoing thermal radiation
in the following sections. For the special case of an isothermal cloud, where the Planck intensity $B$ is constant, $B(\tau') =B$, the outgoing thermal intensity (\ref{exin10}) simplifies to
\begin{equation}
|\dot I^{\{\rm out\}} \} = \mathcal{E}|0) B.
\label{exin12}
\end{equation}
In accordance with Kirchhoff's laws of thermal radiation, 
the isothermal emissivity operator $\mathcal{E}$ of (\ref{exin12}) and the scattering operator $\mathcal{S}$ of (\ref{exin6}) sum to the identity operator $\hat 1$,
\begin{equation}
\mathcal{S}+\mathcal{E}=\hat 1.
\label{exin14}
\end{equation}
We will prove (\ref{exin14}) in Section \ref{eic} below.
Multiplying both sides of (\ref{exin14}) on the left by $\lvec\mu_i|$ and on  the right by $|0)$, and recalling from (\ref{mm8}) that $\lvec \mu_i|0)=w_i$ we find
\begin{equation}
\lvec\mu_i|\mathcal{S}|0)+\lvec\mu_i|
\mathcal{E}|0)=w_i.
\label{exin16}
\end{equation}
Since both terms on the left of (\ref{exin16}) are nonnegative, it implies the bounds
\begin{equation}
0\le \lvec\mu_i|\mathcal{S}|0)\le w_i\quad\hbox{and}\quad 0\le \lvec\mu_i|\mathcal{E}|0)\le w_i.
\label{exin18}
\end{equation}
\section{Green's Functions for Intensity\label{gr}}
The intensity $|\dot I\}$ of thermally emitted radiation inside a cloud depends  on the optical depth profile of the Planck intensity, $B=B(\tau)$, or equivalently, on the temperature profile,  $T=T(\tau)$, of the cloud.
A convenient way to solve (\ref{et2})  for $|I\}=|\dot I\}$ is with a  Green's function, $|G(\tau,\tau')\}$. As discussed in (WH-256), the thermally emitted radiation can be written as
\begin{equation}
|\dot I(\tau)\}= \int_0^{\tau_c}d\tau'|G(\tau,\tau')\}B(\tau').
\label{gr2}
\end{equation}
The thermally emitting particulates in a optical-depth interval $d\tau'$, centered on the source optical depth $\tau'$, make a contribution $d\tau'|G(\tau,\tau')\}B(\tau')$ to the intensity $|\dot I(\tau)\}$ at the observation optical depth $\tau$.
To be consistent with the equation of transport  (\ref{et2})
the  Green's function must satisfy the differential equation
\begin{equation}
\left(\frac{\partial}{\partial\tau}+\hat\kappa\right)|G(\tau,\tau')\}=\delta(\tau-\tau')\,\hat \kappa |0).
\label{gr6}
\end{equation}
To fully determine the Green's function $|G(\tau,\tau')\}$ from the differential equation (\ref{gr6}), we use the boundary condition corresponding to (\ref{exin8}).
\begin{equation}
|G^{\{\rm in\}}(\tau')\}=\mathcal{M}_{\bf d}|G(\tau_c,\tau')\}+\mathcal{M}_{\bf u}|G(0,\tau')\}=\breve 0.
\label{gr8}
\end{equation}
It is convenient to write the Green's function as  the sum of a part $|\infG(\tau-\tau')\}$ for an unbounded cloud, and a part $|\Delta G(\tau,\tau')\}$ that accounts for the cloud particulates  that are missing for optical depths $\tau<0$ below the cloud bottom, and $\tau>\tau_c$ above the cloud top,
\begin{equation}
|G(\tau,\tau')\}=|\infG(\tau-\tau')\}+|\Delta G(\tau,\tau')\}.
\label{gr10}
\end{equation}
\begin{figure}\centering
\includegraphics[height=80mm,width=1\columnwidth]{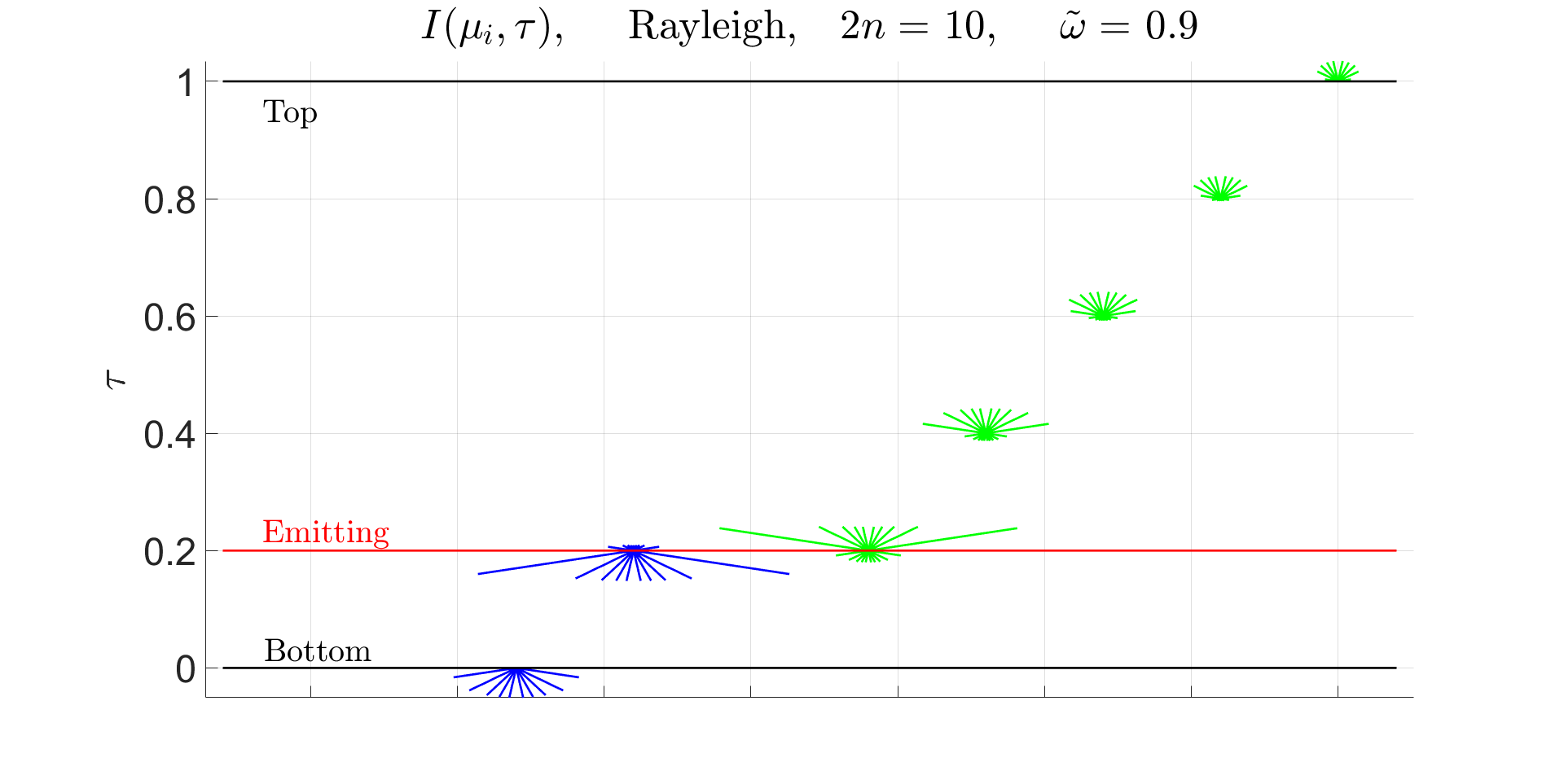}
\caption{ The radiation intensity $\dot I(\mu_i,\tau)$ of (\ref{gr26}) at the observation optical depths $\tau$ in a cloud of vertical optical depth $\tau_c = 1$. At the source optical depth $\tau'=0.2$ above the bottom, there is an infinitesimally thin emitting layer with the Planck intensity $B(\tau'')=\delta(\tau''-\tau')$. See the text  following (\ref{gr26}) for details.
\label{Gr1}}
\end{figure}

%

\begin{figure}\centering
\includegraphics[height=80mm,width=1\columnwidth]{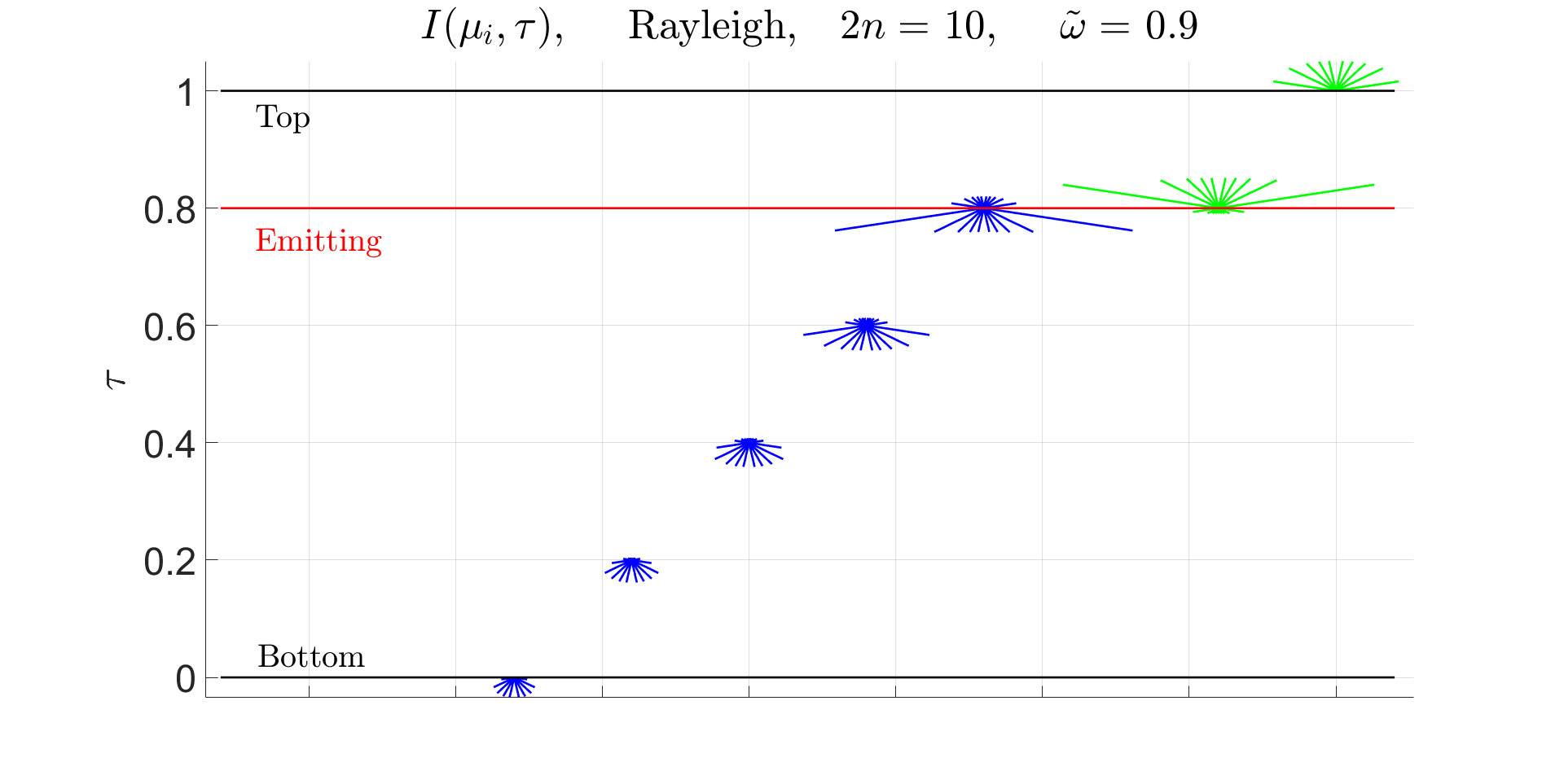}
\caption{Like Fig. \ref{Gr1} but with the source optical depth $\tau'=0.8$ of the emitting layer an equal distance below the cloud top as the emitting layer of Fig. \ref{Gr1} was above the bottom. In accordance with the reflection symmetry (\ref{rs34})  Fig, \ref{Gr1} and Fig. \ref{Gr2} are  mirror images with respect to midcloud optical depth, $\tau=\tau_c/2=0.5$.
\label{Gr2}}
\end{figure}

%
According to (WH-260), the Green's function $|\infG(\tau)\}$ for an unbounded, homogeneous cloud can be written as
\begin{equation}
|\infG(\tau)\} =\left[H(\tau)e^{-\hat \kappa_{\bf u}\tau}
-H(-\tau)e^{-\hat \kappa_{\bf d}\tau}\right]\hat\kappa|0).
\label{gr12}
\end{equation}
The Heaviside unit step function is
\begin{equation}
H(\tau)=\left \{\begin{array}{rl}0, &\mbox{if $\tau<0$, }\\
1/2, &\mbox{if $\tau=0$,}\\
1,&\mbox{if $\tau>0$.} \end{array}\right . 
\label{gr14}
\end{equation}
The upward and downward parts $\hat \kappa_{\bf u}$ and $\hat \kappa_{\bf d}$ of the exponentiation-rate matrix $\hat\kappa$ were given by (\ref{lam24a}) and (\ref{lam24b}).

According to (WH-273) the boundary  contribution  $|\Delta G(\tau,\tau')\}$ of (\ref{gr10}) to the Green's function is 
\begin{equation}
|\Delta G(\tau,\tau')\} =\mathcal{U}(\tau)\mathcal{I}^{-1} \mathcal{R}(\tau')\hat\kappa|0).
\label{gr16}
\end{equation}
The first factor on the right of (\ref{gr16}) is the  {\it propagation operator} $\mathcal{U}(\tau)$, which is given by (WH-193) as
\begin{eqnarray}
\mathcal{U}(\tau)
&=&e^{-\hat\kappa_{\bf d} (\tau-\tau_c)}+e^{-\hat\kappa_{\bf u} \tau}.
\label{gr18}
\end{eqnarray}
The propagation operator $\mathcal{U}(\tau)$, which is diagonal in $\lambda$-space, gives the intensity at the optical depth $\tau$ within a cloud. The intensity consists of
downward penetration modes that start at the top of the cloud where the optical depth is $\tau = \tau_c$, and upward modes that start at the bottom where the optical depth is $\tau = 0$.
The second factor $\mathcal{I}^{-1}$  of (\ref{gr16}) is the inverse of the {\it  incoming operator} $\mathcal{I}$, which is given by (WH-199) as
\begin{eqnarray}
 \mathcal{I} &=&
\mathcal{M}_{\bf u}\mathcal{U}(0)+\mathcal{M}_{\bf d}\mathcal{U}(\tau_c)\nonumber\\
&=&\mathcal{C}_{\bf u  d}e^{\hat\kappa_{\bf d}\tau_c}+\mathcal{C}_{\bf u u} + \mathcal{C}_{\bf d d}+\mathcal{C}_{\bf d u}e^{-\hat\kappa_{\bf u} \tau_c}.
\label{gr20}
\end{eqnarray}
The overlap operators $\mathcal{C}_{\bf q  q'}$ were given by (\ref{lam28a}).
The third factor $\mathcal{R}(\tau')$  on the right of (\ref{gr16}) is the {\it retro operator}, given by (WH-270) as
\begin{eqnarray}
\mathcal{R}(\tau')&=&\mathcal{M}_{\bf u}e^{\hat\kappa_{\bf d}\tau'}-\mathcal{M}_{\bf d}e^{-\hat\kappa_{\bf u}(\tau_c-\tau')}\nonumber\\
&=&\mathcal{C}_{\bf u d}e^{\hat\kappa_{\bf d}\tau'}-\mathcal{C}_{\bf d u}e^{-\hat\kappa_{\bf u}(\tau_c-\tau')}.
\label{gr24}
\end{eqnarray}

In Fig. \ref{Gr1} and Fig. \ref{Gr2} we show the intensities
\begin{eqnarray}
\dot I(\mu_i,\tau) 
&=& w_i^{-1}\lvec \mu_i|G(\tau,\tau')\}
\label{gr26}
\end{eqnarray}
that follow from (\ref{gr2})  and from an infinitesimal emission layer at the optical depth, $\tau'$, with the Planck intensity
\begin{equation}
B(\tau'')=\delta(\tau''-\tau').
\label{gr28}
\end{equation}
Here it is understood that the dummy variable of integration $\tau'$ of (\ref{gr2}) is to be replaced by the dummy variable $\tau''$, since we use $\tau'$ in (\ref{gr28}) to denote the source optical depth of the  emitting particulates in the cloud.
In both figures the emitted radiation is
Rayleigh scattered, with a single scattering albedo of $\tilde\omega = 0.9$, until it is absorbed or leaves the top or bottom of the cloud. The intensities for $\tau>\tau'$ are proportional to the lengths of the green rays, and the intensities for $\tau<\tau'$ are proportional to the lengths of the blue rays. In Fig. \ref{Gr1} the infinitesimal emission layer at $\tau'= 0.2$ is closer to the bottom, where $\tau =0$, than at the top where $\tau=\tau_c=1$. So more radiation comes out of the bottom than out of the top.  For both Fig. \ref{Gr1} and Fig. \ref{Gr2} there are only upward streams at the cloud top, and only downward streams at the cloud bottom. Because of scattering from cloud particulates, there are both upward and  downward streams for optical depths between the top and bottom of the cloud. Close to the source optical depth $\tau'$, the nearly horizontal streams with long slant paths are more intense than  the  near-vertical streams with short slant paths. The intensity is limb-brightened near the source. At the cloud top, the nearly horizontal streams with longer slant paths from the source are more attenuated and weaker than the intensities of nearly vertical streams with shorter slant paths and less attenuation. The intensity is increasingly limb-darkened for increasing distances from the source layer. 
\subsection{Pure absorption\label{pag}}
An important special case is a cloud with absorption and emission but no scattering.  A layer of clear air  with only greenhouse gases but no condensed water droplets, ice crystallites or other particulates is an example of a purely absorbing ``cloud'' for Earth's thermal radiation. It would not be a purely absorbing cloud for visible or ultraviolet sunlight, which is Rayleigh scattered by nitrogen and oxygen, the main constituents of the atmosphere.

For a purely absorbing cloud, the single scattering albedo vanishes, $\tilde\omega \to 0$, and we can use the limiting values of (\ref{lam32a}) together with (\ref{gr24}) and  (\ref{gr16}) to show that
\begin{equation}
R(\tau')\to \breve 0,\quad\hbox{and}\quad |\Delta G(\tau,\tau')\}\to \breve 0.
\label{pag0}
\end{equation}
From (\ref{pag0}) and (\ref{gr10})  we see that for the pure absorption limit
\begin{equation}
|G(\tau,\tau')\}\to |\infG(\tau-\tau')\},
\label{pag2}
\end{equation}
where  (\ref{lam32a}) and (\ref{gr12}) imply that
\begin{equation}
|\infG(\tau)\} =\left[H(\tau)e^{-\hat \varsigma_{\bf u}\tau}
-H(-\tau)e^{-\hat \varsigma_{\bf d}\tau}\right]\hat\varsigma|0).
\label{pag4}
\end{equation}
\subsection{Scattering operator $\mathcal{S}$\label{so} }
Although this paper is focussed on radiation intensity generated by thermal emission of particulates inside clouds, it turns out that the internal emission is closely related to the scattering of external intensity.  
The scattering operator $\mathcal{S}$ was defined by  (\ref{exin6}) as the ``ratio'' of outgoing external intensity $|\ddot I^{\{\rm out\}}\}$ to incoming external intensity $|\ddot I^{\{\rm in\}}\}$ .  For a homogeneous cloud,   (WH-206) showed that the scattering operator can be written as
\begin{equation}
\mathcal{S}=\mathcal{O}\mathcal{I}^{-1}.
\label{so2}
\end{equation}
We already discussed the incoming operator, $\mathcal{I}$ of (\ref{gr20}). We can use (\ref{gr18}) and (\ref{lam28a}) to write 
the outgoing operator of (WH-201) as
\begin{eqnarray}
\mathcal{O}&=&\mathcal{M}_{\bf u}\mathcal{U}(\tau_c) +\mathcal{M}_{\bf d} \mathcal{U}(0))\nonumber\\
&=&\mathcal{M}_{\bf u}\left[\mathcal{L}_{\bf d}+e^{-\hat\kappa_{\bf u}\tau_c}\right]+\mathcal{M}_{\bf d}\left[e^{\hat\kappa_{\bf d}\tau_c} +\mathcal{L}_{\bf u}\right] \nonumber\\
&=&\mathcal{C}_{\bf u d}+\mathcal{C}_{\bf u u}e^{-\hat\kappa_{\bf u}\tau_c}+\mathcal{C}_{\bf d d}e^{\hat\kappa_{\bf d}\tau_c}
+\mathcal{C}_{\bf d u}.
\label{so4}
\end{eqnarray}
Of particular interest is the scattering operator for a purely absorbing cloud, with $\tilde \omega \to 0$. The we can use the limiting expressions of (\ref{lam32a}) to write the limiting values of the input operator $\mathcal{I}$ of (\ref{gr20}) and the output operator $\mathcal{O}$ of (\ref{so4}) as
\begin{equation}
\mathcal{I} \to \hat 1\quad\hbox{and}\quad \mathcal{O}\to e^{-\hat \varsigma_{\bf u}\tau_c}+e^{\hat\varsigma_{\bf d}\tau_c}.
\label{so6}
\end{equation}
Using  (\ref{so2}) and (\ref{so4}) we write  the scattering matrix for a purely absorbing cloud as
\begin{equation}
\mathcal{S}= e^{-\hat \varsigma_{\bf u}\tau_c}+e^{\hat\varsigma_{\bf d}\tau_c},\quad\hbox{if}\quad \tilde \omega = 0.
\label{so8}
\end{equation}

\subsection{Reflection symmetries\label{rs}}
A reflection transforms the Green's function (\ref{gr10}) for a homogeneous cloud  in a relatively simple way that can facilitate numerical calculations and give  physical insight.
Multiplying the Green's function (\ref{gr12}) for an unbounded cloud on the left  by $\hat r$, and recalling that $\hat r \hat r = \hat 1$, we can use (\ref{mm18}), (\ref{et44}) and (\ref{lam26}) to write
\begin{eqnarray}
\hat r \,|\infG (\tau)\}&=&\hat r\left[H(\tau) e^{-\hat \kappa_{\bf u}\tau}
-H(-\tau)\hat e^{-\hat \kappa_{\bf d}\tau}\right]\hat r\hat r\hat\kappa\hat r\hat r|0)\nonumber\\
&=&\left[H(\tau) e^{-(\hat r\hat \kappa_{\bf u}\hat r)\tau}
-H(-\tau)\hat e^{-(\hat r\hat \kappa_{\bf d}\hat r)\tau}\right]\left[-\hat\kappa\right]|0)\nonumber\\
&=&\left[H(-\tau) e^{\hat \kappa_{\bf u}\tau}-H(\tau) e^{\hat \kappa_{\bf d}\tau} \right]\hat\kappa|0)
\nonumber\\
&=& |\infG (-\tau)\}.
\label{rs14}
\end{eqnarray}
Multiplying  (\ref{gr18}) on the left and right by $\hat r$ and using (\ref{lam26}) we find
\begin{eqnarray}
\hat r \mathcal{U}(\tau)\hat r&=&\hat r\left[e^{-\hat\kappa_{\bf d} (\tau-\tau_c)}+e^{-\hat\kappa_{\bf u} \tau}\right]\hat r \nonumber\\
&=&e^{-\hat r\hat\kappa_{\bf d}\hat r (\tau-\tau_c)}+e^{-\hat r\hat\kappa_{\bf u} \hat r\tau} \nonumber\\
&=&e^{\hat\kappa_{\bf u} (\tau-\tau_c)}+e^{\hat\kappa_{\bf d} \tau}\nonumber\\
&=&\mathcal{U}(\tau_c-\tau).
\label{rs18}
\end{eqnarray}
Multiplying  (\ref{gr20}) on the left and right by $\hat r$, inserting factors of $\hat r \hat r = \hat 1$,  and using (\ref{int46}) with  (\ref{rs18}), we find 
\begin{eqnarray}
 \hat r \mathcal{I} \hat r &=&\hat r\left[\mathcal{M}_{\bf u}\hat r\hat r\mathcal{U}(0)+\mathcal{M}_{\bf d}\hat r \hat r\mathcal{U}(\tau_c)\right]\hat r\nonumber\\
 &=&\mathcal{M}_{\bf d}\mathcal{U}(\tau_c)+\mathcal{M}_{\bf u} \mathcal{U}(0)\nonumber\\
 &=&\mathcal{I}.
\label{rs20}
\end{eqnarray}
and therefore
\begin{equation}
\hat r \mathcal{I}^{-1}\hat r=\mathcal{I}^{-1}.
\label{rs22}
\end{equation}
Multiplying  (\ref{so4}) on the left and right by $\hat r$, inserting factors of $\hat r \hat r = \hat 1$,  and using (\ref{int46}) with  (\ref{rs18}), we find 
\begin{eqnarray}
 \hat r \mathcal{O} \hat r &=&\hat r\left[\mathcal{M}_{\bf u}\hat r\hat r\mathcal{U}(\tau_c)+\mathcal{M}_{\bf d}\hat r \hat r\mathcal{U}(0)\right]\hat r\nonumber\\
 &=&\mathcal{M}_{\bf d}\mathcal{U}(0)+\mathcal{M}_{\bf u} \mathcal{U}(\tau_c)\nonumber\\
 &=&\mathcal{O}.
\label{rs23}
\end{eqnarray}
Multiplying  (\ref{gr24}) on the left and right by $\hat r$ and using (\ref{lam26})  with (\ref{int46}) we find
\begin{eqnarray}
\hat r \mathcal{R}(\tau')\hat r&=&\hat r\left[\mathcal{M}_{\bf u}\hat r\hat r e^{\hat\kappa_{\bf d}\tau'}-\mathcal{M}_{\bf d}\hat r\hat r e^{-\hat\kappa_{\bf u}(\tau_c-\tau')}\right]\hat r\nonumber\\
&=&\mathcal{M}_{\bf d}e^{-\hat\kappa_{\bf u}\tau'}-\mathcal{M}_{\bf u}e^{\hat\kappa_{\bf d}(\tau_c-\tau')}\nonumber\\
&=&-\mathcal{R}(\tau_c-\tau').
\label{rs24}
\end{eqnarray}
From (\ref{mm18}), (\ref{et44}),  (\ref{rs18}), (\ref{rs22}) and (\ref{rs24}) we see that a reflection transforms the boundary contribution (\ref{gr16}) to the matrix Green's function as
\begin{eqnarray}
\hat r|\Delta G(\tau,\tau')\} &=&\hat r\mathcal{U}(\tau)\hat r\hat r\mathcal{I}^{-1}\hat r\hat r \mathcal{R}(\tau')\hat r\hat r\hat\kappa\hat r\hat r |0)\nonumber\\
&=&-\mathcal{U}(\tau_c-\tau)\mathcal{I}^{-1}\mathcal{R}(\tau_c-\tau')\hat\kappa|0)\nonumber\\
&=&|\Delta G(\tau_c-\tau,\tau_c-\tau')\}.
\label{rs26}
\end{eqnarray}
Adding (\ref{rs14}) and (\ref{rs26})  we see that a reflection transforms  the total Green's function  (\ref{gr10}) into
\begin{eqnarray}
\hat r|G(\tau,\tau')\} =|G(\tau_c-\tau,\tau_c-\tau')\}.
\label{rs34}
\end{eqnarray}
Fig. \ref{Gr1} and Fig. \ref{Gr2} give one example of the reflection symmetry (\ref{rs34}) of Green's functions.

Multiplying  (\ref{so2}) on the left and right by $\hat r$, inserting factors of $\hat r \hat r = \hat 1$,  and using (\ref{rs22}) with  (\ref{rs23}), we find
\begin{eqnarray}
\hat r\mathcal{S}\hat r&=&\hat r\mathcal{O}\hat r\hat r\mathcal{I}^{-1}\hat r\nonumber\\
&=&\mathcal{O}\mathcal{I}^{-1}\nonumber\\
&=&S.
\label{rs36}
\end{eqnarray}
The scattering operator of a homogeneous cloud is invariant to reflections. Inhomogeous clouds, where the single scattering albedo $\tilde\omega(\tau)$ and scattering phase operator $\hat p(\tau) $ vary with optical depth $\tau$ in the cloud, do not have the reflection symmetry (\ref{rs36}).

\section{Outgoing Intensity}
The outgoing thermal intensity  can be written as 
\begin{equation}
|\dot I^{\{\rm out\}} \} =\int_0^{\tau_c} d\tau' |G^{\{\rm out\}}(\tau')\}B(\tau'). 
\label{go2}
\end{equation}
From (\ref{ex2}), (\ref{gr2}) and (\ref{gr10}) we see that the Green's function for outgoing intensity in  (\ref{go2}) is

\begin{eqnarray}
|G^{\{\rm out\}} (\tau')\} &=&
\mathcal{M}_{\bf u}|G(\tau_c,\tau')\} +\mathcal{M}_{\bf d} |G(0,\tau')\}\nonumber\\
&=&|\infG^{\{\rm out\}} (\tau')\} +|\Delta G^{\{\rm out\}} (\tau')\}. 
\label{go4}
\end{eqnarray}

Using (\ref{gr12}) we write the contribution to (\ref{go4}) from an infinite cloud as
\begin{eqnarray}
|\infG^{\{\rm out\}} (\tau')\} &=&
\mathcal{M}_{\bf u}|\infG(\tau_c-\tau')\} +\mathcal{M}_{\bf d} |\infG(-\tau')\}\nonumber\\
&=&\mathcal{Q}(\tau') \hat\kappa|0)\nonumber\\
&=&\frac{\partial}{\partial \tau'}\mathcal{Q}(\tau') |0).
\label{go6}
\end{eqnarray}
Here the matrix $\mathcal{Q}(\tau') $ is
\begin{eqnarray}
\mathcal{Q}(\tau') 
&=&\mathcal{M}_{\bf u}e^{-\hat\kappa_{\bf u}(\tau_c-\tau')} -\mathcal{M}_{\bf d} e^{\hat\kappa_{\bf d}\tau'}\nonumber\\
&=&\mathcal{C}_{\bf u u}e^{-\hat\kappa_{\bf u}(\tau_c-\tau')} -\mathcal{C}_{\bf d d} e^{\hat\kappa_{\bf d}\tau'}.
\label{go8}
\end{eqnarray}

Using (\ref{gr16}) with (\ref{so2}) and (\ref{so4}), and noting from (\ref{gr24}) that $\partial \mathcal{R}(\tau')/\partial \tau' = \mathcal{R}(\tau')\hat\kappa$, we write the contribution from the cloud boundaries as
\begin{eqnarray}
|\Delta G^{\{\rm out\}} (\tau')\}&=&
\mathcal{M}_{\bf u}|\Delta G(\tau_c,\tau')\} +\mathcal{M}_{\bf d} |\Delta G(0,\tau')\}\nonumber\\
&=&\mathcal{O}\mathcal{I}^{-1} \mathcal{R}(\tau')\hat\kappa|0)\nonumber\\
&=&\mathcal{S} \mathcal{R}(\tau')\hat\kappa|0)\nonumber\\
&=&\frac{\partial}{\partial \tau'}\mathcal{S} \mathcal{R}(\tau')|0).
\label{go10}
\end{eqnarray}
Adding (\ref{go6}) and (\ref{go10}) we find
\begin{eqnarray}
|G^{\{\rm out\}} (\tau')\} &=&\left[\mathcal{Q}(\tau')+\mathcal{S} \mathcal{R}(\tau')\right]\hat\kappa |0)\nonumber\\
&=&\frac{\partial}{\partial \tau'}\left[\mathcal{Q}(\tau')+\mathcal{S} \mathcal{R}(\tau')\right] 
|0). \label{go18}
\end{eqnarray}
We can use (\ref{gr24}) and  (\ref{go8}) to factor out the $\tau'$-independent {\it thermal emission operator}
\begin{eqnarray}
\mathcal{T} &=&(\mathcal{M}_{\bf u}-\mathcal{S}\mathcal{M}_{\bf d})\mathcal{L}_{\bf u}
+(\mathcal{M}_{\bf d}-\mathcal{S}\mathcal{M}_{\bf u})\mathcal{L}_{\bf d}.
 \label{go20}
\end{eqnarray}
and to write (\ref{go18}) as 
\begin{eqnarray}
|G^{\{\rm out\}} (\tau')\} &=&\mathcal{T}\mathcal{X}(\tau')\hat\kappa |0)\nonumber\\
 &=&\mathcal{T} \frac{\partial}{\partial \tau'} \mathcal{X}(\tau') |0).
 \label{go22}
\end{eqnarray}
The $\tau'$-dependent operator $\mathcal{X}(\tau')$ of (\ref{go22}) is diagonal in $\lambda$-space, and it can be written as
\begin{eqnarray}
\mathcal{X}(\tau') = e^{-\hat \kappa_{\bf u}(\tau_c-\tau') }-e^{\hat \kappa_{\bf d}\tau'}
 \label{go24}
\end{eqnarray}

Using (\ref{go22}) in (\ref{go2}) we see that the outgoing thermal radiation from a cloud can be written as
\begin{eqnarray}
|\dot I^{\{\rm out\}} \} =\mathcal{T}|\tilde B\}.
 \label{go26}
\end{eqnarray}
The {\it Planck source vector} can be written as
\begin{eqnarray}
 |\tilde B\}&=&\int_0^{\tau_c}d\tau'\mathcal{X}(\tau')\hat \kappa |B(\tau')\}\nonumber\\
&=&\int_0^{\tau_c}d\tau'  \frac{\partial}{\partial \tau'} \mathcal{X}(\tau') |B(\tau')\}\nonumber\\
&=& |\tilde B_{\bf u}\}+ |\tilde B_{\bf d}\}.
 \label{go28}
\end{eqnarray}
As indicated in the last line of (\ref{go28}) we
 write $|\tilde B\}$ as the sum of upward and downward parts in $\lambda$-space. The upward part of $|\tilde B\}$ is
\begin{eqnarray}
 |\tilde B_{\bf u}\}&=&\int_{0}^{\tau_c}d\tau'  e^{-\hat\kappa_{\bf u}(\tau_c-\tau')}
\hat\kappa_{\bf u}|B(\tau')\}\nonumber\\
&=&\sum_{k=n+1}^{2n}|\lambda_k)\lvec\lambda_k|\tilde B\}.
 \label{go30}
\end{eqnarray}
The upward coefficients are
\begin{eqnarray}
\lvec\lambda_k|\tilde B\} &=&\lvec\lambda_k|0)\int_{0}^{\tau_c}d\tau' B(\tau') \kappa_k e^{-\kappa_{k}(\tau_c-\tau')}.
 \label{go32}
\end{eqnarray}
The downward part of $|\tilde B\}$ is
\begin{eqnarray}
 |\tilde B_{\bf d}\}&=&-\int_{0}^{\tau_c}d\tau' B(\tau') e^{\hat\kappa_{\bf d}\tau'}
\hat\kappa_{\bf d}|0)\nonumber\\
&=&\sum_{j=1}^{n}|\lambda_j)\lvec\lambda_j|\tilde B_{\bf d}\}.
 \label{go34}
\end{eqnarray}
The downward coefficients are 
\begin{equation}
\lvec\lambda_j|\tilde B\} =-\lvec\lambda_j|0)\int_{0}^{\tau_c}d\tau' B(\tau')\kappa_j e^{\kappa_{j}\tau'}.
 \label{go36}
\end{equation}
As shown in Fig. 7 of WH, the least attenuated intensity modes have penetration lengths $\lambda_{2n}=-\lambda_1$. From inspection of (\ref{go32}) and (\ref{go36}) we see that for source optical depths $\tau'$ deep inside optically thick clouds, $\tau' \gg \lambda_{2n}$ and $\tau_c-\tau' \gg \lambda_{2n}$, the Planck intensity $B(\tau')$  will make negligible contributions to the Planck source vector $|\tilde B\}$. The emitted radiation is absorbed before it can escape from the top or bottom surface. For most scattering conditions $\lambda_{2n}\approx 1$. Then only regions within about 1 optical depth from the top or bottom contribute to the thermal emission $|\dot I^{\{\rm out\}}\}$ from the cloud.

 For nearly conservative scattering, with $\tilde\omega \to 1$, the quasi-isotropic penetration length can be many optical depths in magnitude, $\lambda_{2n}\gg 1$, as shown in Fig. 7 of WH\,\cite{WH1} . Earth's clouds can have nearly conservative scattering for visible sunlight.   Modifications to the theory of WH needed to account for exact conservative scattering, with $\tilde \omega = 1$, are discussed in the paper {\it $2n$-Stream Conservative Scattering} \cite{WH2}. For nearly conservative scattering, a photon thermally generated many optical depths from the top or bottom of a cloud can have a good chance of escape, since it can undergo many scattering collisions and make a random walk to the nearest  surface before it is lost in an absorptive collision. But negligibly few visible photons are generated inside Earth's clouds. For the photons of thermal radiation that are copiously generated inside clouds, representative values of the single scattering albedo are $\tilde \omega\approx 0.5$. So a thermal-radiation photon has a good chance of being absorbed and converted to heat on each collision with a cloud particulate.
\subsection{Isothermal clouds\label{eic}}
For an isothermal cloud, with a $\tau'$-independent Planck intensity, $B$, we can take the factor, $|B(\tau')\}=|0)B$, outside the integral and use (\ref{go24}) to write the second line of (\ref{go28}) as
\begin{eqnarray}
 |\tilde B\}
&=&\bigg[\int_0^{\tau_c}d\tau'  \frac{\partial}{\partial \tau'} \mathcal{X}(\tau')\bigg] |B\}\nonumber\\
&=&\left[\mathcal{X}(\tau_c)-\mathcal{X}(0)\right] |B\}\nonumber\\
&=& [\hat 1- e^{\hat\kappa_{\bf d}\tau_c}- e^{-\hat\kappa_{\bf u}\tau_c}]|B\}
 \label{eic2}
\end{eqnarray}
Using (\ref{eic2}) and (\ref{go20}) in (\ref{go26}) we find
\begin{eqnarray}
|\dot I^{\rm out}\} &=&\left[(\mathcal{M}_{\bf u}-\mathcal{S}\mathcal{M}_{\bf d})\mathcal{L}_{\bf u}
+(\mathcal{M}_{\bf d}-\mathcal{S}\mathcal{M}_{\bf u})\mathcal{L}_{\bf d}\right] [\hat 1- e^{\hat\kappa_{\bf d}\tau_c}- e^{-\hat\kappa_{\bf u}\tau_c}]|B\}\nonumber\\
&=&\left[(\mathcal{M}_{\bf u}-\mathcal{S}\mathcal{M}_{\bf d})(\mathcal{L}_{\bf u}- e^{-\hat\kappa_{\bf u}\tau_c})
+(\mathcal{M}_{\bf d}-\mathcal{S}\mathcal{M}_{\bf u})(\mathcal{L}_{\bf d}-e^{\hat\kappa_{\bf d}\tau_c})\right] |B\}\nonumber\\
&=&\mathcal{C}_{\bf u u}-\mathcal{C}_{\bf u u}e^{-\hat\kappa_{\bf u}\tau_c}
+\mathcal{C}_{\bf d d}-\mathcal{C}_{\bf d d}e^{\hat\kappa_{\bf d}\tau_c}\nonumber\\
&&-\mathcal{S}\left[\mathcal{C}_{\bf d u}-\mathcal{C}_{\bf d u}e^{-\hat\kappa_{\bf u}\tau_c}
+\mathcal{C}_{\bf u d}-\mathcal{C}_{\bf u d}e^{\hat\kappa_{\bf d}\tau_c}\right]|B\}\nonumber\\
&=&\left[\hat 1-\mathcal{O}-\mathcal{S}(\hat 1-\mathcal{I})\right]|B\},
\label{eic4}
\end{eqnarray}
The last line of (\ref{eic4}) follows from (\ref{lam30a}), (\ref{gr20}) and (\ref{so4}). 
Since $\mathcal{S}\mathcal{I}=\mathcal{O}$ according to (\ref{so2}), we can write (\ref{eic4}) as
\begin{eqnarray}
|\dot I^{\rm out}\} =\mathcal{E}|B\}.
\label{eic6}
\end{eqnarray}
The isothermal emissivity operator of (\ref{eic6}) is
\begin{equation}
\mathcal{E} =\hat 1 -\mathcal{S}.
\label{eic14}
\end{equation}
Eq. (\ref{eic14}) is one version of Kirchhoff's law of thermal radiation\,\cite{Kirchhoff}
.
The total outgoing intensity from an isothermal cloud, the sum of internally emitted intensity $|\dot I^{\rm out}\}$ and reflected and transmitted external intensity $|\ddot I^{\rm out}\}$, is therefore
\begin{eqnarray}
|I^{\{\rm out\}} \}&=&|\dot I^{\{\rm out\}} \}+|\ddot I^{\{\rm out\}} \}\nonumber\\
&=&\mathcal{E} |B\} +\mathcal{S}|\ddot I^{\{\rm in \}}\}.
\label{eic16}
\end{eqnarray}
The last line of (\ref{eic16}) follows from (\ref{eic6}) and  (\ref{exin6}).

In accordance with (\ref{exin8}), thermal emission makes no contribution to the incoming radiation, so the total incoming radiation is 
\begin{equation}
|I^{\{\rm in\}} \}=|\ddot I^{\{\rm in\}} \}
\label{eic18}
\end{equation}
The excess of outgoing over incoming intensity vectors  for an isothermal cloud, the difference between (\ref{eic16}) and (\ref{eic18}), is
\begin{eqnarray}
|I^{\{\rm out\}} \}-|I^{\{\rm in \}} \}
&=&\mathcal{E} |B\} +(\mathcal{S}-\hat 1)|\ddot I^{\{\rm in\}}\}\nonumber\\
&=&\mathcal{E}\left( |B\} -|\ddot I^{\{\rm in\}}\}\right).
\label{eic20}
\end{eqnarray}
We  used (\ref{eic14}) to get the last line of (\ref{eic20}) from the previous one.
From (\ref{eic20}) we see that there will be thermal equilibrium, with equal outgoing and incoming intensities and no heating or cooling of the isothermal cloud if the incoming intensity is equal to the $\tau'$-independent Planck intensity $|B\}=|0)B$ of the cloud,
\begin{equation} 
|I^{\{\rm in\}} \}=|\ddot I^{\{\rm in\}} \}=|B\}.
\label{eic22}
\end{equation}
Substituting (\ref{eic22}) into (\ref{eic16}) and using (\ref{eic14}) we see that the total outgoing intensity from the isothermal cloud, the sum of thermally emitted and reflected or transmitted external intensity, is
\begin{eqnarray}
|I^{\{\rm out\}} \}
&=&\left(\mathcal{E}  +\mathcal{S}\right)|B\}\nonumber\\
&=&|B\}.
\label{eic24}
\end{eqnarray}
The incoming intensity will equal the Planck intensity as in (\ref{eic22}) if the cloud of interest is irradiated from above and below by Planck radiation of the same temperature, for example, from optically thick, purely absorbing clouds of the same temperature below and above the cloud of interest. 

In the conservative scattering limit there will be no thermal emission, $|\dot I^{\{\rm out\}} \}=\breve 0$, so (\ref{eic6}) and (\ref{eic14}) imply that
\begin{eqnarray}
\mathcal{E}|0)\to \breve 0,\quad\hbox{and}\quad \mathcal{S}|0)\to|0),\quad\hbox{as}\quad \tilde\omega \to 1. 
\label{eic26}
\end{eqnarray}
For a purely absorptive cloud with $\tilde\omega =0$ we can use (\ref{so8}) with (\ref{eic14}) to write
\begin{equation}
\mathcal{E} =\hat 1 - e^{-\hat \varsigma_{\bf u}\tau_c}-e^{\hat\varsigma_{\bf d}\tau_c},\quad\hbox{for}\quad \tilde \omega = 0.
\label{eic28}
\end{equation}

\begin{figure}\centering
\includegraphics[height=80mm,width=1\columnwidth]{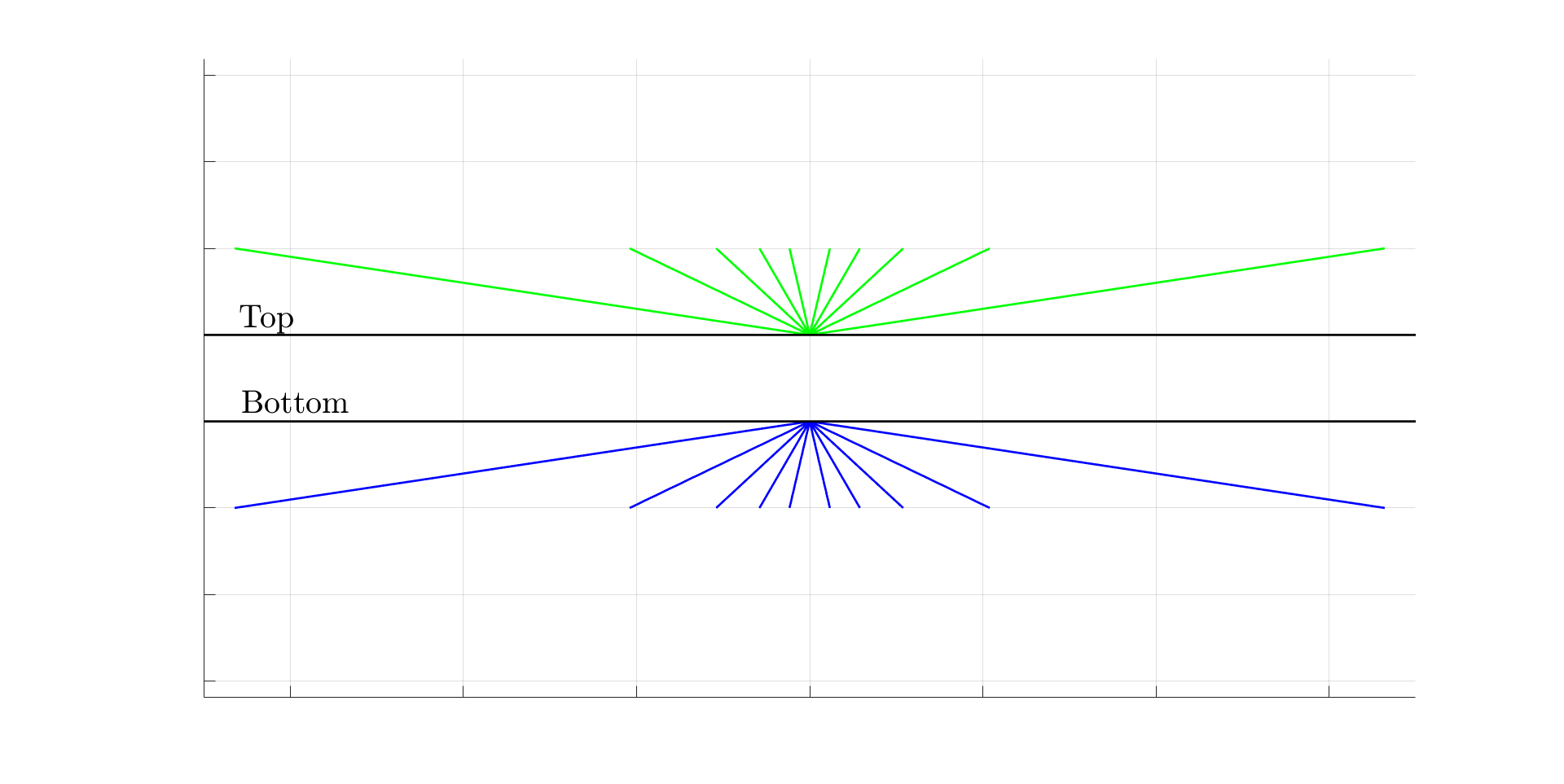}
\caption{ Upward (green) and  downward (blue) relative outgoing intensities $\dot I^{\{\rm out\}}(\mu_i,\tau)$ of (\ref{otn12}) from an optically thin cloud. The direction secant of the stream is $\varsigma_i$. The intensity of the $i$th stream is proportional to $|\varsigma_i|$  or to the slant distance through the thin cloud.   For $\tau_c\ll 1$, the pattern is independent of $\tau_c$, $\tilde\omega$, $\hat p$ or $\overline B$.
\label{Gr6}}
\end{figure}

%
\subsection{Optically thin clouds}
For an optically thin cloud, the total optical depth $\tau_c$ is small enough that
\begin{equation}
|\kappa_i|\tau_c\ll 1 \quad\hbox{for}\quad i=1,2,3,\ldots, 2n.
\label{otn2}
\end{equation}
We recall from (WH-184) that the scattering matrix for an optically thin cloud can be written, to order $\tau_c\ll 1$, as
\begin{equation}
\mathcal{S} = \hat 1 -\tau_c (\hat\varsigma_{\bf u}-\hat\varsigma_{\bf d})\hat\eta. 
\label{otn4}
\end{equation}
From (\ref{otn4}) and (\ref{eic14}) we see that the isothermal emissivity operator of the thin cloud is
\begin{equation}
\mathcal{E} = \hat 1-\mathcal{S}  =\tau_c (\hat\varsigma_{\bf u}-\hat\varsigma_{\bf d})\hat\eta. 
\label{otn5}
\end{equation}
For the thin-cloud limit we can write the operator $\mathcal{X}$ of (\ref{go24})  to order $\tau_c\ll 1$ as
\begin{eqnarray}
\mathcal{X}(\tau') =\mathcal{L}_{\bf u}-\mathcal {L}_{\bf d} -\hat \kappa_{\bf u}(\tau_c-\tau') -\hat \kappa_{\bf d}\tau'.
 \label{otn5a}
\end{eqnarray}
We can use (\ref{et39}) and (\ref{otn5a}) 
to write the Planck source vector (\ref{go28}), to order $\tau_c\ll 1$ as
\begin{eqnarray}
 |\tilde B\}&=&\overline B \tau_c(\mathcal{L}_{\bf u}-\mathcal{L}_{\bf d})\hat\kappa|0)\nonumber\\
 &=&(1-\tilde\omega)\overline B \tau_c(\mathcal{L}_{\bf u}-\mathcal{L}_{\bf d})\hat\zeta|0),
 \label{otn6}
\end{eqnarray}
where the average Planck intensity of the thin cloud is
\begin{equation}
\overline B=\frac{1}{\tau_c}\int_0^{\tau_c}d\tau' B(\tau').
 \label{otn8}
\end{equation}
We can substitute (\ref{go20}) and (\ref{otn6}) into (\ref{go26}) to find that the outgoing intensity for a thin cloud is
\begin{eqnarray}
|\dot I^{\{\rm out\}} \} &=&(1-\tilde\omega)\overline B \tau_c\left[(\mathcal{M}_{\bf u}-\mathcal{S}\mathcal{M}_{\bf d})\mathcal{L}_{\bf u}
+(\mathcal{M}_{\bf d}-\mathcal{S}\mathcal{M}_{\bf u})\mathcal{L}_{\bf d}\right](\mathcal{L}_{\bf u}-\mathcal{L}_{\bf d})\hat\zeta|0)\nonumber\\
 &=&(1-\tilde\omega)\overline B \tau_c\left[(\mathcal{M}_{\bf u}-\mathcal{M}_{\bf d})\mathcal{L}_{\bf u}
-(\mathcal{M}_{\bf d}-\mathcal{M}_{\bf u})\mathcal{L}_{\bf d}\right]\hat\zeta|0)\nonumber\\
&=&(1-\tilde\omega)\overline B \tau_c\left[\mathcal{M}_{\bf u}(\mathcal{L}_{\bf u}
+\mathcal{L}_{\bf d}) - \mathcal{M}_{\bf d}(\mathcal{L}_{\bf u}
+\mathcal{L}_{\bf d})\right]\hat\zeta|0)\nonumber\\
&=&(1-\tilde\omega)\overline B \tau_c\left[\mathcal{M}_{\bf u} - \mathcal{M}_{\bf d}\right]\hat\zeta|0),
 \label{otn10}
\end{eqnarray}
To get the second line from the first, we noted  from (\ref{otn4}) that $\,\mathcal {S}\mathcal{M}_{\bf q}= \mathcal{M}_{\bf q}$ to order $\tau_c\ll 1$, and we used (\ref{lam18}) and (\ref{lam20a}). We used (\ref{lam16}) to get the final line.
Using (\ref{int62}) and (\ref{int64}) we write (\ref{otn10}) in final form as
\begin{equation}
|\dot I^{\{\rm out\}} \}=(1-\tilde\omega)\overline B \tau_c\left[\hat\zeta_{\bf u} - \hat\zeta_{\bf d}\right]|0).
 \label{otn12}
\end{equation}
The intensity of the $i$th outgoing, thermally emitted stream is
\begin{equation}
\dot I^{\{\rm out\}}(\mu_i) = w_i^{-1}\lvec\mu_i|\dot I^{\{\rm out\}} \}
=(1-\tilde\omega)\overline B \tau_c|\varsigma_i|.
 \label{otn12}
\end{equation}
Not surprisingly, in view of Kirchhoff's laws of thermal radiation, the radiation intensity
$\dot I^{\{\rm out\}}(\mu_i)$  of the $i$th stream is proportional to the single scattering absorption probability,  $1-\tilde\omega$, to the mean Planck intensity $\overline B$ of (\ref{otn8}), and to the slant optical thickness,  $\tau_c|\varsigma_i| \ll 1$.
The relative intensities $\dot I^{\{\rm out\}}(\mu_i)$ of (\ref{otn12}) are illustrated in Fig. \ref{Gr6}.  Nearly horizontal streams have much larger intensities than nearly vertical streams.
\subsection{Optically thick clouds}
Consider a homogeneous cloud with some absorption and $\tilde\omega < 1$. As the optical thickness approaches infinity, $\tau_c\to\infty$, the magnitudes of all of the mode attenuation factors  also  approach infinity, $|\kappa_i\tau_c|\to \infty$. Therefore
\begin{equation}
e^{-\hat \kappa_{\bf u}\tau}\to \breve 0\quad\hbox{and}\quad e^{\hat \kappa_{\bf d}\tau}\to \breve 0\quad\hbox{as}\quad \tau_c\to\infty.
\label{ot2}
\end{equation}
Then the incoming operator (\ref{gr20}) becomes a block diagonal matrix in
{\it $\mu\lambda$-space}. That is, for a $2n\times 2n$ matrix with elements $\lvec\mu_i|\mathcal{I}|\lambda_{i'})$,
\begin{equation}
 \mathcal{I} =\left[\begin{array}{ll}\mathcal{I}_{\bf dd}&\mathcal{I}_{\bf du}\\\mathcal{I}_{\bf ud}&\mathcal{I}_{\bf uu}\end{array}\right]
=\left[\begin{array}{ll}\mathcal{C}_{\bf dd}&\breve 0\\ \breve 0&\mathcal{C}_{\bf uu}\end{array}\right].
\label{ot4}
\end{equation}
We can therefore write the inverse of the incoming operator as a block diagonal matrix in {\it $\lambda\mu$-space}. That is, for a $2n\times 2n$ matrix with elements $\lvec\lambda_{i'}|\mathcal{I}^{-1}|\mu_{i})$,
\begin{equation}
 \mathcal{I}^{-1} =\left[\begin{array}{ll}\left(\mathcal{I}^{-1}\right)_{\bf dd}&\left(\mathcal{I}^{-1}\right)_{\bf du}\\ \left(\mathcal{I}^{-1}\right)_{\bf ud}&\left(\mathcal{I}^{-1}\right)_{\bf uu}\end{array}\right]
=\left[\begin{array}{ll}\left(\mathcal{C}_{\bf dd}\right)^{-1}&\breve 0\\ \breve 0&\left(\mathcal{C}_{\bf uu}\right)^{-1}\end{array}\right].
\label{ot6}
\end{equation}
Here the $n\times n$ inverse matrix $\left(\mathcal{C}_{\bf u u}\right)^{-1}$, which can also be thought of as a  $2n\times 2n$ pseudoinverse matrix analogous to those of Section \ref{pim}, has the properties
\begin{eqnarray}
\mathcal{C}_{\bf u u}\left(\mathcal{C}_{\bf u u}\right)^{-1}&=&\mathcal{M}_{\bf u},\quad\hbox{and}\quad
\left(\mathcal{C}_{\bf u u}\right)^{-1}\mathcal{C}_{\bf u u}=\mathcal{L}_{\bf u}.
\label{ot8}
\end{eqnarray}
Similarly, the $n\times n$ inverse operator $\left(\mathcal{C}_{\bf d d}\right)^{-1}$ has the properties
\begin{eqnarray}
\mathcal{C}_{\bf d d }\left(\mathcal{C}_{\bf d d }\right)^{-1}&=&\mathcal{M}_{\bf d},\quad\hbox{and}\quad
\left(\mathcal{C}_{\bf d d}\right)^{-1}\mathcal{C}_{\bf d d}=\mathcal{L}_{\bf d}.
\label{ot10}
\end{eqnarray}
For the optically thick limit (\ref{ot2}), the outgoing operator (\ref{so4}) becomes the block anti-diagonal matrix in $\mu \lambda$-space,
\begin{equation}
 \mathcal{O} =\left[\begin{array}{ll}\mathcal{O}_{\bf dd}&\mathcal{O}_{\bf du}\\\mathcal{O}_{\bf ud}&\mathcal{O}_{\bf uu}\end{array}\right]
=\left[\begin{array}{ll}\breve 0&\mathcal{C}_{\bf du}\\ \mathcal{C}_{\bf ud}&\breve 0\end{array}\right].
\label{ot12}
\end{equation}
Substituting (\ref{ot12}) and (\ref{ot6}) into (\ref{so2}) we see that the scattering matrix for an optically thick cloud can be written as a  block anti-diagonal matrix in $\mu$ space
\begin{equation}
\mathcal{S}=\mathcal{O} \mathcal{I}^{-1} =\left[\begin{array}{rr}\mathcal{S}_{\bf dd}&\mathcal{S}_{\bf du}\\\mathcal{S}_{\bf ud}&\mathcal{S}_{\bf uu}\end{array}\right]=\left[\begin{array}{ll}\breve 0&\mathcal{C}_{\bf d u} \left(\mathcal{C}_{\bf u u}\right)^{-1}\\ \mathcal{C}_{\bf u d} \left(\mathcal{C}_{\bf d d}\right)^{-1}&\breve 0\end{array}\right].
\label{ot16}
\end{equation}
We recall from (\ref {int44}),  (\ref{lam28a}) and (\ref{lam32a}) that $\mathcal{C}_{\bf ud}\to \breve 0$ and $\mathcal{C}_{\bf  du}\to \breve 0$ as the scattering vanishes and $\tilde\omega \to 0$.  The scattering matrix of a homogeneous, optically thick cloud vanishes as $\tilde \omega\to 0$, and the cloud becomes purely absorbing,
\begin{equation}
\mathcal{S}\to  \breve 0,\quad\hbox{as}\quad \tilde\omega \to 0,\quad\hbox{and}\quad \tau_c\to \infty.
\label{ot18}
\end{equation}
From (\ref{ot18}) and (\ref{otn5}) we see that the emission operator $\mathcal{E}$  of a homogeneous, optically thick, purely absorbing cloud becomes the identity operator
\begin{equation}
\mathcal{E} = \hat 1.
\label{ot20}
\end{equation}
Such a cloud is an  ideal blackbody which absorbs all incident external radiation and scatters none. As we shall show in the next section, an isothermal, optically thick, purely absorbing cloud also becomes an ideal Lambertian emitter of blackbody radiation.
\begin{figure}\centering
\includegraphics[height=80mm,width=1\columnwidth]{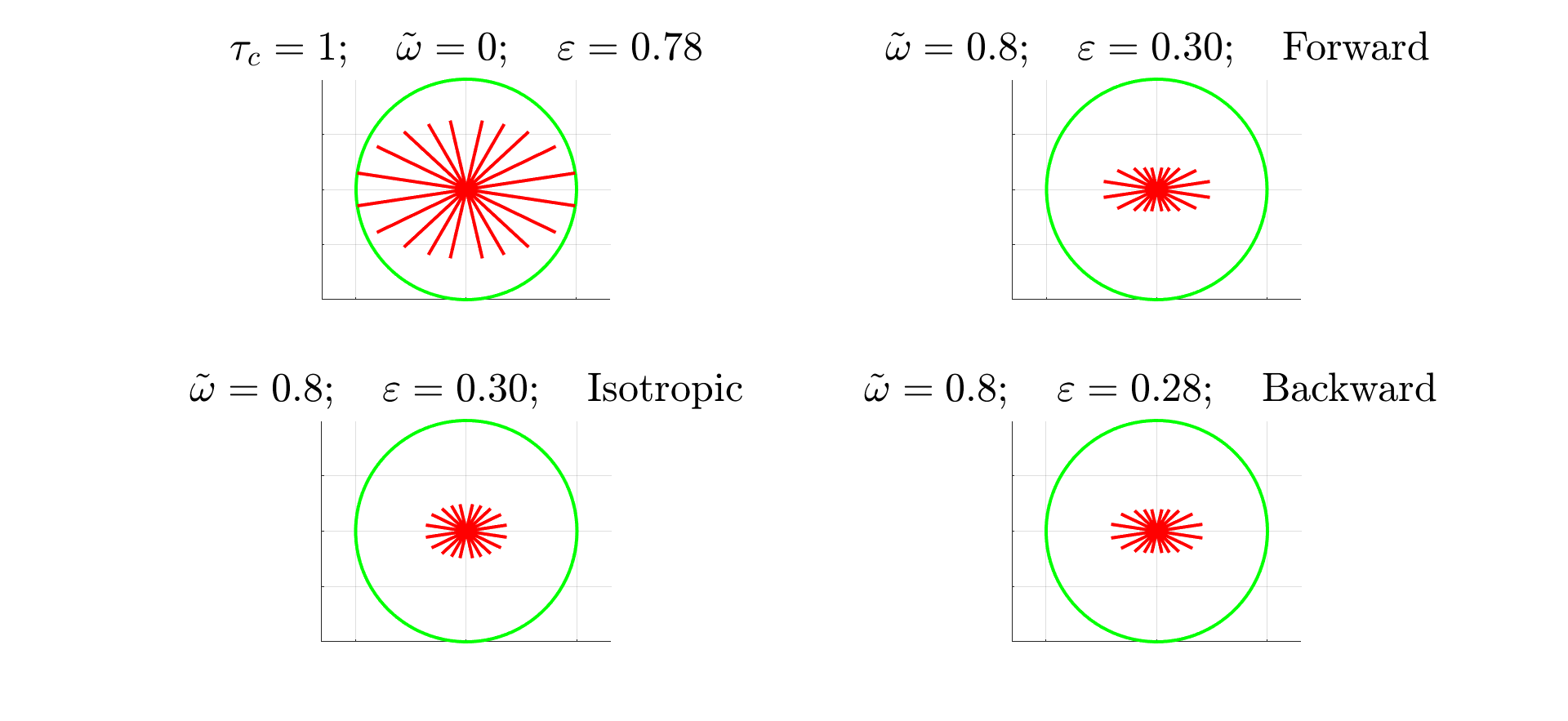}
\caption{ Angular dependence of emission from isothermal clouds of optical thickness $\tau_c = 1$, modelled with $2n=10$ streams.  The green circle is the blackbody limit, when all of the sample intensities $\dot I^{\{\rm out\}}(\mu_i)$ would be equal to the Planck intensity $B$ of (\ref{et10}). Upward rays come from the cloud top and downward rays from the cloud bottom.  For the moderate optical depth of these examples, all of the intensities are limb-brightened. Nearly horizontal outgoing streams have intensity contributions from longer slant paths through the cloud than nearly vertical streams. The upper right panel, labeled ``Forward,'' has the scattering phase function $p(\mu)=\varpi^{\{5\}}(\mu)$ of (WH-134) and the lower right panel labeled ``Backward'' has the scattering phase function $p(\mu)=\varpi^{\{5\}}(-\mu)$.  The scalar emissivities $\varepsilon$ of (\ref{se6}) for  all clouds are much less than the maximum possible value, $\varepsilon = 1$. The maximum penetration lengths for radiation in the clouds with pure absorption, forward, isotropic, and backward scattering are $\lambda_{2n} =$ 0.9739, 2.6649, 1.4076, and 1.2627.
\label{em1}}
\end{figure}

%
%
\begin{figure}\centering
\includegraphics[height=80mm,width=1\columnwidth]{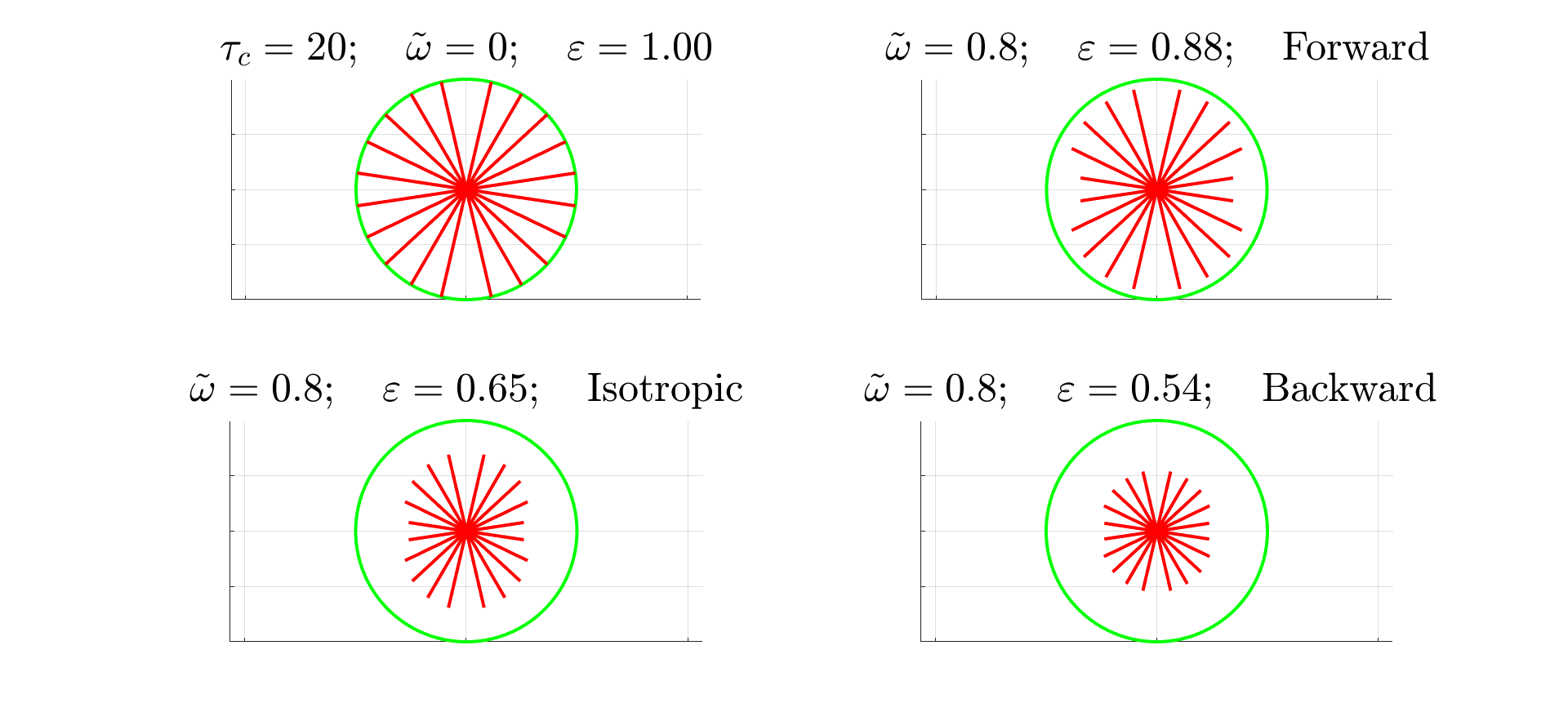}
\caption{ Like Fig. \ref{em1} but for optically thick, isothermal clouds with optical depths $\tau_c=20$.  The purely absorbing cloud with $\tilde\omega=0$ shown in the upper left panel has become an ideal blackbody with a Lambertian (uniform) angular distribution of intensities,  $\dot I^{\{\rm out\}}(\mu_i)$, all equal to the Planck intensity $B$ of (\ref{et10}).  The emission intensities of the three scattering clouds with $\tilde\omega = 0.8$ are limb-darkened because of preferential attenuation of more nearly horizontal rays as they emerge from the bottoms and tops of the clouds. For the scattering clouds, the scalar emissivities $\varepsilon$ of (\ref{se6}) remain substantially less than the maximum possible value, $\varepsilon = 1$.
\label{em20}}
\end{figure}

%
\subsection {Directions of thermal emission}
According to (\ref{exin12}) the intensity of the $i$th stream of outgoing radiation from an isothermal cloud is
\begin{eqnarray}
\dot I^{\{\rm out\}}(\mu_i)&=&w_i^{-1}\lvec \mu_i|\dot I^{\{\rm out\}}\}\nonumber\\
&=&w_i^{-1}\lvec \mu_i|\mathcal{E}|0)B\nonumber\\
&=&w_i^{-1}\lvec \mu_i|\hat 1-\mathcal{S}|0)B\nonumber\\
&=&B\left[1-w_i^{-1}\lvec \mu_i|\mathcal{S}|0) \right].
\label{dir2}
\end{eqnarray}
Using (\ref{exin18}) with (\ref{dir2}) we conclude that the outgoing intensity from an isothermal cloud must have values bounded by
\begin{equation}
0\le \dot I^{\{\rm out\}}(\mu_i)\le B.
\label{dir4}
\end{equation}
For the blackbody limit of an optically thick cloud with $\tilde\omega = 0$, the outgoing thermal intensity is equal to the Planck intensity $B$ of (\ref{et10}) for all emission angles,
\begin{equation}
\dot I^{\{\rm out\}}(\mu_i)= B,\quad\hbox{if}\quad \mathcal{S}=\breve 0.
\label{dir6}
\end{equation}
Equal emitted intensity for all outward directions from the surface is called Lambertian emission\, \cite{Lambert}.

For a transparent cloud with $\mathcal{S}=\hat 1$, or for the conservative scattering limit, $\tilde\omega \to 1$, when $\mathcal{S}|0)\to|0)$ but $\mathcal{S}\ne \hat 1$,  the outgoing thermal intensity vanishes
\begin{equation}
\dot I^{\{\rm out\}}(\mu_i)= \breve 0,\quad\hbox{if}\quad \mathcal{S}|0)=|0).
\label{dir8}
\end{equation}

Some representative outgoing thermal intensities $\dot I^{\{\rm out\}}(\mu_i)$ calculated with (\ref{dir2}) for clouds with a moderate optical depth, $\tau_c=1$, are shown in Fig. \ref{em1}. The intensities of the purely absorbing cloud, with $\tilde\omega = 0$, are moderately close to the Planck blackbody limit $B$, denoted by the green circle. For the scattering clouds with $\tilde\omega = 0.8$ the intensities are much less than $B$. Limb brightening similar to that of the infinitesimally thin cloud of Fig. \ref{Gr6} can be seen.  Also shown in the figure are the scalar emissivities $\varepsilon$ of (\ref{se6}) , which we discuss in Section \ref{em}.  All of the emissivities are less than the blackbody limit, $\varepsilon = 1$.

Fig \ref{em20} shows what happens when the isothermal clouds of  Fig. \ref{em1} become optically thick with $\tau_c = 20$.  The purely absorbing cloud becomes an ideal blackbody with $\dot I^{\{\rm out\}}(\mu_i)=B$ and a Lambertian angular distribution of intensities.  For the optically thick scattering clouds with $\tilde \omega =0.8$, the outgoing thermal intensities $\dot I^{\{\rm out\}}(\mu_i)$  remain smaller than  the blackbody limit $B$, and the scalar emissivities $\varepsilon$ remain  less than the maximum possible value of $\varepsilon =1$.  The cloud with forward scattering comes closest to the blackbody limit.  There is limb darkening  for all three scattering clouds.

\begin{figure}\centering
\includegraphics[height=80mm,width=1\columnwidth]{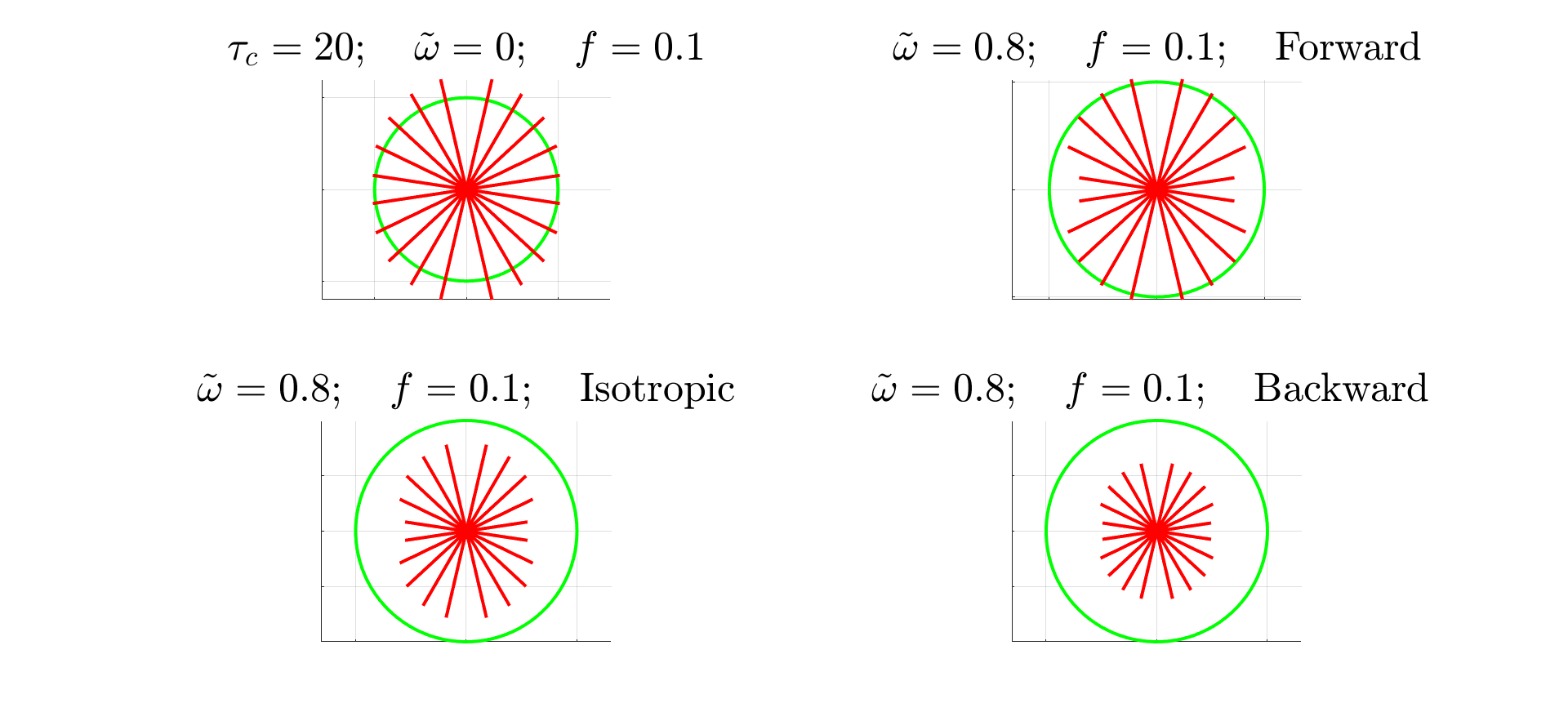}
\caption{Like the optically thick isothermal clouds of Fig. \ref{em20} but for clouds with a hot interiors, described by (\ref{chi2}) with $f = 0.1$. The green circle is the Planck intensity $B_0$ at the cool surfaces of the cloud.  There is limb darkening for the purely absorbing cloud with $\tilde\omega = 0$, as well as for  scattering clouds with $\tilde\omega = 0.8$. 
\label{em20h}}
\end{figure}

%
%
\subsection{Clouds with hot interiors\label{chi}}
There is little scattering of light in the surface layers of the Sun. Most of the opacity is absorptive\cite{Photosphere} and due to photodetachment of electrons from negative hydrogen ions, H$^{-}$. To good approximation, the photosphere of the Sun is a purely absorptive cloud, like that modeled on the upper left panel of Fig. \ref{em20}. But unlike the example of Fig. \ref{em20}, the Sun is strongly limb darkened due to the increase of temperature below the emitting surface.

A simple model for the Planck intensity in a cloud with a hot interior and cooler surfaces at the top and bottom is
\begin{equation}
B(\tau') =\left \{\begin{array}{ll}B_0 e^{f\tau'/\lambda_{2n}} &\mbox{if $\tau'<\tau_c/2$, }\\B_0 e^{f(\tau_c-\tau')/\lambda_{2n}} &\mbox{if $\tau'>\tau_c/2$.} \end{array}\right . 
\label{chi2}
\end{equation}
The {\it heating parameter} $f$,  determines the exponential growth rate (for $f>0$) or attenuation rate (for $f<0$) of the Planck intensity (\ref{chi2}) toward the hot center of the cloud, at the optical depth $\tau=\tau_c/2$.
Using (\ref{chi2}) to evaluate (\ref{go32}) for upward streams, we find
\begin{eqnarray}
\lvec\lambda_k|\tilde B\} &=&\lvec\lambda_k|0)B_0\kappa_k\int_{-\infty}^{\tau_c}d\tau'  e^{-(\kappa_{k}-f/\lambda_{2n})(\tau_c-\tau')}\nonumber\\
&=&\frac{\lvec\lambda_k|0)B_0}{1-f\lambda_k/\lambda_{2n}}.
 \label{chi8}
\end{eqnarray}
In (\ref{chi8})  we have assumed that $f<1$ and that that the cloud is so optically thick that 
\begin{equation}
\frac{|\lambda_{i}|}{1-f}\le\frac{\lambda_{2n}}{1-f}\ll \tau_c.
 \label{chi6}
\end{equation}
If (\ref{chi6}) is true, we can replace the lower limit $0$ of the integral of (\ref{go32}) by $-\infty$. In like manner we use  (\ref{chi2}) to evaluate  (\ref{go36}) for downward streams to  find
\begin{eqnarray}
\lvec\lambda_j|\tilde B\} &=&-\lvec\lambda_j|0)B_0\kappa_j\int_{0}^{\infty}d\tau' e^{(\kappa_{j}+f/\lambda_{2n})\tau'}\nonumber\\
&=&\frac{\lvec\lambda_k|0)B_0}{1+f\lambda_j/\lambda_{2n}}.
 \label{chi10}
\end{eqnarray}

We use (\ref{chi8}) and (\ref{chi10}) to write the Planck source vector of (\ref{go28}) as
\begin{eqnarray}
|\tilde B\}&=&\sum_k|\lambda_k)\lvec\lambda_k|\tilde B\} +\sum_j|\lambda_j)\lvec\lambda_j|\tilde B\} \nonumber\\
&=&\left(\sum_k\frac{|\lambda_k)\lvec\lambda_k|}{1-f\lambda_k/\lambda_{2n}}+
\sum_j\frac{|\lambda_j)\lvec\lambda_j|}{1+f\lambda_j/\lambda_{2n}}\right)
|0)B_0\nonumber\\
&=&\left[(\mathcal{L}_{\bf u}-f\hat\lambda_{\bf u}/\lambda_{2n})^{-1}+
(\mathcal{L}_{\bf d}+f\hat\lambda_{\bf d}/\lambda_{2n})^{-1}\right]|0)B_0.
 \label{chi12}
\end{eqnarray}
The projection matrices $\mathcal{L}_{\bf d}$ and $\mathcal{L}_{\bf u}$ were defined by (\ref{lam14}) and the upward and downward penetration operators  $\hat\lambda_{\bf d}$ and $\hat\lambda_{\bf u}$ were given by (\ref{lam30}) and (\ref{lam32}).
The meaning of the pseudoinverse matrices $(\mathcal{L}_{\bf d}+f\hat\lambda_{\bf d}/\lambda_{2n})^{-1}$ and
$(\mathcal{L}_{\bf u}-f\hat\lambda_{\bf u}/\lambda_{2n})^{-1}$ of (\ref{chi12}) was discussed in Section \ref{pim}. 
Using (\ref{go20}) and (\ref{chi12}), we write the
outgoing radiation of (\ref{go26}) as
\begin{eqnarray}
|\dot I^{\{\rm out\}} \} &=&(\mathcal{M}_{\bf u}-\mathcal{S}\mathcal{M}_{\bf d})(\mathcal{L}_{\bf u}-f\hat\lambda_{\bf u}/\lambda_{2n})^{-1}|0)B_0+\nonumber\\
&&(\mathcal{M}_{\bf d}-\mathcal{S}\mathcal{M}_{\bf u})
(\mathcal{L}_{\bf d}+f\hat\lambda_{\bf d}/\lambda_{2n})^{-1}|0)B_0.
 \label{chi14}
\end{eqnarray}

Representative evaluations of (\ref{chi14}) for clouds with hot interiors are shown in Fig. \ref{em20h} for a heating parameter $f=0.1$.  As mentioned in the caption of Fig. \ref{em1}, for $2n=10$ streams the maximum penetration length is that for forward scattering, with  $\lambda_{2n} = 2.66$. The condition (\ref{chi6}) that the cloud be optically thick is satisfied for forward scattering, where $\lambda_{2n}/(1-f) =2.96 \ll \tau_c = 20$. For the  examples of Fig. \ref{em20} the clouds are even more optically thick for isotropic, backward, or purely  absorptive  scattering, where the penetration lengths,  $\lambda_{2n}= 1.4076,\, 1.2627$ and $0.9739$, are smaller.  The most striking change between Fig. \ref{em20} and Fig. \ref{em20h}  is limb darkening of emission from the cloud with pure absorption. For  heating parameters $f$ much larger than the  value $f=0.1$ of Fig. \ref{em20h}, the vertically emitted intensities of all of the clouds with hot interiors, with or without scattering,  exceed the surface Planck intensity $B_0$.

\section{Flux \label{sf}}
To facilitate discussions of vertical heat transport by radiation,  we use the upward flux vector $|Z(\tau)\}$ of (WH-210).
The  flux vector $|Z(\tau)\}$ at an observation optical depth $\tau$ in the cloud is related to the intensity
vector $|I(\tau)\}$ by
\begin{eqnarray}
|Z(\tau)\}&=&4\pi \hat\mu|I(\tau)\}\nonumber\\
&=&\int_0^{\tau_c}d\tau'|F(\tau,\tau')\}B(\tau').
\label{sf2}
\end{eqnarray}
The Green's function $|F(\tau,\tau')\}$ for the flux is related to the Green's function $|G(\tau,\tau')\}$ of (\ref{gr2})  by
\begin{equation}
|F(\tau,\tau')\}
=4\pi \hat\mu|G(\tau,\tau')\}.
\label{sf4}
\end{equation}
In analogy to (\ref{gr10}), we write $|F(\tau,\tau')\}$ as a part $|\infF(\tau-\tau')\}$ from an unbounded  cloud, and a part $|\Delta F(\tau, \tau')\}$ that accounts for the cloud boundaries
\begin{equation}
|F(\tau,\tau')\}
=|\infF(\tau-\tau')\}+|\Delta F(\tau, \tau')\}.
\label{sf5}
\end{equation}
The contribution for an infinite cloud follows from (\ref{gr12}) and is
\begin{eqnarray}
|\infF(\tau)\}&=&4\pi\hat\mu|\infG(\tau)\}\nonumber\\
&=&4\pi\hat\mu\left[H(\tau)e^{-\hat \kappa_{\bf u}\tau}
-H(-\tau)e^{-\hat \kappa_{\bf d}\tau}\right]\hat\kappa|0).
\label{sf6}
\end{eqnarray}
The contribution from the cloud boundaries follows from  (\ref{gr16})  and is
\begin{eqnarray}
|\Delta F(\tau,\tau')\}&=&4\pi\hat\mu|\Delta G(\tau,\tau')\}\nonumber\\
&=&4\pi\hat\mu \,\mathcal{U}(\tau)\mathcal{I}^{-1} \mathcal{R}(\tau')\hat\kappa|0).
\label{sf6a}
\end{eqnarray}

\subsection{Scalar flux \label{sf}}
In accordance with (WH-209), we can write the scalar flux as the scalar 
\begin{eqnarray}
Z(\tau)&=&\lvec 0|Z(\tau)\}\nonumber\\
&=&\int_0^{\tau_c}d\tau' F(\tau,\tau')B(\tau').
\label{sf8}
\end{eqnarray}
Unlike the vector flux $|Z(\tau)\}$, which must be specified by  $2n$ real numbers,  the scalar flux $Z(\tau)$ of (\ref{sf8})  is specified by  a single real number, which can be positive, negative or zero. Representative units of the (spectral) scalar flux are W m$^{-2}$ cm. In  (\ref{sf8})
we have used (\ref{sf2}) to write the scalar flux as an integral  transform of the Planck intensity profile $B(\tau)$. The kernel of the transform is the scalar Green's function
\begin{eqnarray}
 F(\tau,\tau')&=&\lvec 0|F(\tau,\tau')\}\nonumber\\
&=& \infF(\tau-\tau')+\Delta F(\tau, \tau').
\label{sf12}
\end{eqnarray}
We can use (\ref{sf6}) and (\ref{et39}) to write the part of (\ref{sf12}) for an infinite cloud as
\begin{eqnarray}
\infF(\tau)&=&\lvec 0|\infF(\tau)\}\nonumber\\
&=&4\pi \lvec 0|\hat\mu |\infG(\tau)\}\nonumber\\
&=&4\pi (1-\tilde\omega)\lvec 0|\hat\mu\left[ H(\tau)e^{-\hat \kappa_{\bf u}\tau}
-H(-\tau)e^{-\hat \kappa_{\bf d}\tau}\right]  \hat\varsigma|0).
\label{sf14}
\end{eqnarray}
Inserting factors of $\hat r\hat r =1$ into the second line of (\ref{sf14}) and using (\ref{int48}), (\ref{mm18}) and (\ref{rs14}) we find
\begin{eqnarray}
 \infF(\tau)
&=&4\pi \lvec 0|\hat r\hat r\hat\mu \hat r\hat r\,|\!\infG(\tau)\}\nonumber\\
&=&-4\pi \lvec 0|\hat\mu\, |\!\infG(-\tau)\}\nonumber\\
&=& - \infF(-\tau).
\label{sf16}
\end{eqnarray}
For an infinite cloud the Green's function $\infF(\tau)$ for the scalar flux is antisymmetric in the difference $\tau$ between the observation and source optical depths. 

\begin{figure}\centering
\includegraphics[height=80mm,width=1\columnwidth]{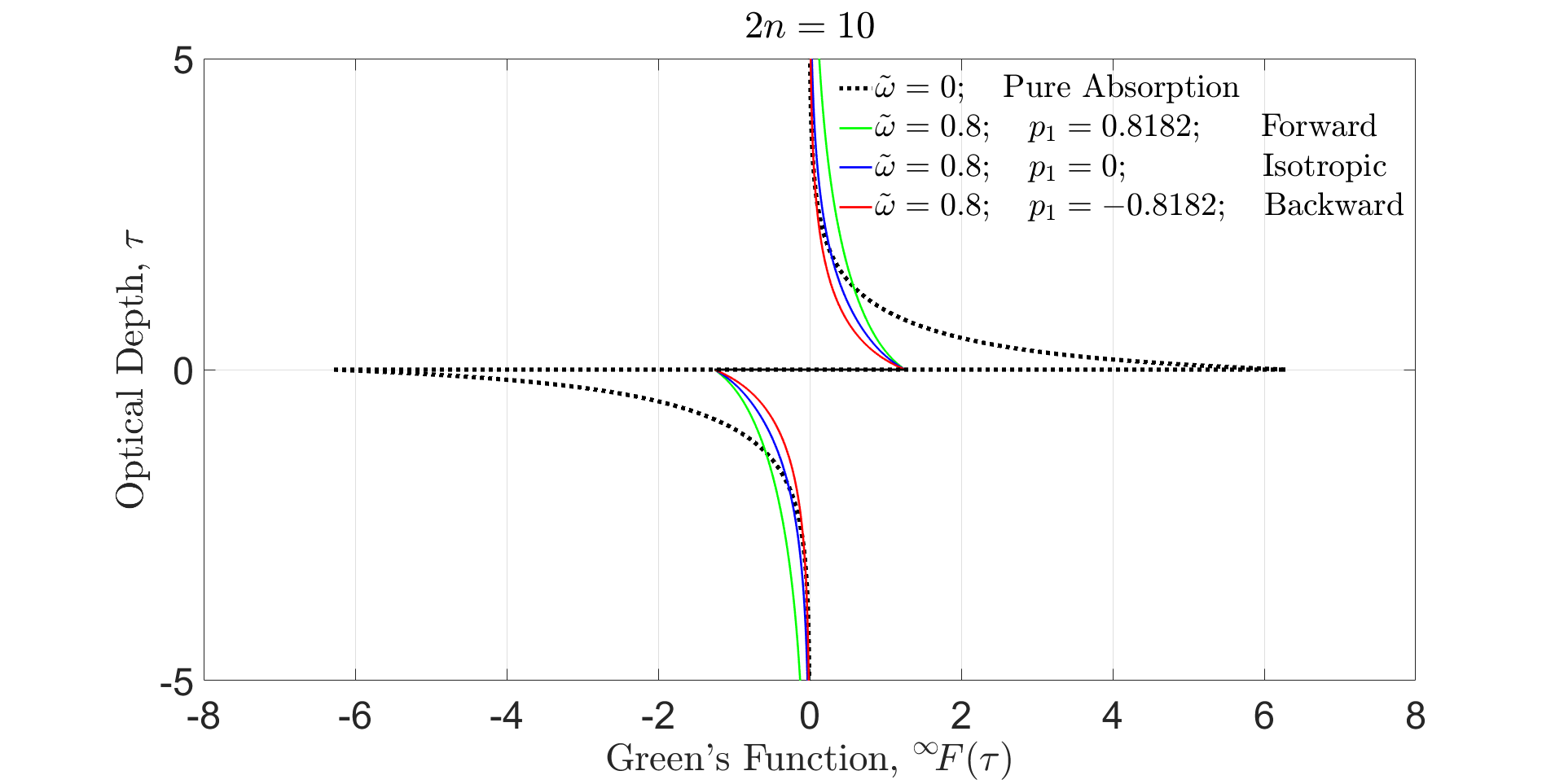}
\caption{Green's function of (\ref{sf18}) for the scalar flux in an infinite cloud in which the dipole scattering phase coefficient $p_1 =\varpi^{\{n\}}_1= 0.8182=(2n-1)/(2n+1)$ corresponds to the maximum possible forward scattering phase function, $p(\mu) =\varpi^{\{n\}}(\mu)$ of (WH-134), that can be constructed from the first $2n=10$ Legendre polynomials. For this phase function the forward-scattering amplitude is $p(1)=\varpi^{\{5\}}(1)=n(n+1) = 30$.  The curve with $p_1 = -0.8182$ corresponds to the maximum possible backward scattering, and the curve with $p_1 = 0$ corresponds to isotropic scattering.  
\label{Gr4}}
\end{figure}

%

In view of the symmetry (\ref{sf16}), we can write 
the last line of (\ref{sf14}) as
\begin{eqnarray}
\infF(\tau)
&=&4\pi (1-\tilde\omega)\sgn(\tau)\lvec 0|\hat\mu e^{-\hat \kappa_{\bf u}|\tau|} \hat\varsigma|0)\nonumber\\
&=&4\pi (1-\tilde\omega)\sgn(\tau) \sum_{k=n+1}^{2n}\lvec 0|\hat\mu|\lambda_k)e^{-\hat \kappa_k|\tau|}\lvec \lambda_k| \hat\varsigma|0).
\label{sf18}
\end{eqnarray}
The summation indices of the upward modes with $\kappa_k>0$  are $k=n+1, n+2,\ldots, 2n$. The sign function $\sgn(\tau)$ is related to the Heaviside function 
$H(\tau)$ of (\ref{gr14}) by
\begin{equation}
\sgn(\tau) = H(\tau)-H(-\tau)=\left \{\begin{array}{rl}-1, &\mbox{if $\tau<0$, }\\
0, &\mbox{if $\tau=0$,}\\
1,&\mbox{if $\tau>0$.} \end{array}\right . 
\label{sf20}
\end{equation}
At an infinitesimal optical depth $\tau=0^+$ above the emitting layer, (\ref{sf18}) simplifies to
\begin{eqnarray}
 \infF(0^+)&=&4\pi (1-\tilde\omega) \sum_{k=n+1}^{2n}\lvec 0|\hat\mu|\lambda_k)\lvec \lambda_k| \hat\varsigma|0)\nonumber\\
 &=&2\pi (1-\tilde\omega).
\label{sf22}
\end{eqnarray}
To prove (\ref{sf22}) we use (\ref{lam14}), (\ref{lam21}), (\ref{int48}), (\ref{int58}) and (\ref{mm18}) to write the sum as
\begin{eqnarray}
\sum_{k=n+1}^{2n}\lvec 0|\hat\mu|\lambda_k)\lvec \lambda_k| \hat\varsigma|0)&=&
\lvec 0|\hat\mu \mathcal{L}_{\bf u}\hat\varsigma|0)\nonumber\\
&=&\lvec 0|\hat\mu \hat r\mathcal{L}_{\bf d}\hat r\hat\varsigma|0)\nonumber\\
 &=&\lvec 0|(-\hat r\hat\mu) \mathcal{L}_{\bf d}(- \hat\varsigma \hat r)|0)\nonumber\\
 &=&\lvec 0|\hat\mu\mathcal{L}_{\bf d}  \hat\varsigma|0).
\label{sf24}
\end{eqnarray}
Averaging the two equivalent expressions on the right of (\ref{sf24}) and using (\ref{lam16}) and (\ref{int28})  we find
\begin{eqnarray}
\sum_{k=n+1}^{2n}\lvec 0|\hat\mu|\lambda_k)\lvec \lambda_k| \hat\varsigma|0)
 &=&\frac{1}{2}\lvec 0|\hat\mu\left[ \mathcal{L}_{\bf u}+ \mathcal{L}_{\bf d}\right]\hat\varsigma|0)\nonumber\\
&=&\frac{1}{2} \lvec 0|\hat\mu \hat\varsigma|0)\nonumber\\
&=&\frac{1}{2}.
\label{sf26}
\end{eqnarray}
This  completes the proof of (\ref{sf22}).

Some representative Green's functions $\infF(\tau) $ of (\ref{sf18}) for the scalar flux in an infinite cloud are shown in Fig. \ref{Gr4}.  For pure absorption and a vanishing single scattering albedo, $\tilde\omega=0$, the Green's function just above the scattering layer has the maximum possible value $\infF(0^+)=2\pi$,  in accordance with 
(\ref{sf22}). For non-zero scattering, in this example $\tilde \omega = 0.8$, the peak values are reduced to $\infF(0^+)=2\pi(1-\tilde\omega) = 0.4 \pi$, and the Green's function is broadened. At a given value of $\tilde\omega$, forward scattering ($p_1>0$), allows the Green's function to extend to greater optical depths from the emission layer at $\tau =0$ than for phase functons with no forward-backward asymmetry and with a vanishing dipole coefficient, $p_1=0$, like isotropic or Rayleigh scattering. Backward scattering ($p_1<0$), decreases the extent of the Green's function. As shown in Section \ref{mo} below, the value of the important first moment $\langle\tau\rangle = 4\pi [3(1-\tilde\omega p_1)]^{-1}$  of  $\infF(\tau) $ depends only on the single scattering albedo $\tilde \omega$ and on the dipole coefficient $p_1$ of the scattering phase matrix (\ref{et30}). The higher multpole moments $p_2, p_3, p_4,\ldots$ have no influence on $\langle\tau\rangle$.
\subsection{Exponential integral functions $E^{\{n\}}_q$ for pure absorption}
An important limiting case  is the scalar flux for pure absorption, discussed in Section \ref{pag}, when 
$\tilde\omega \to 0$ and the other radiation  transfer parameters take on limiting values of (\ref{lam32a}). Then according to (\ref{sf12}), (\ref{sf6a}) and (\ref{pag0}), 
\begin{eqnarray}
\Delta F(\tau,\tau')&=&\lvec 0|\Delta F(\tau,\tau')\}\nonumber\\
&=&4\pi \lvec 0|\hat\mu|\Delta G(\tau,\tau')\}\nonumber\\
&\to& 0.
\label{pa10}
\end{eqnarray}
Then (\ref{sf12}) implies that
\begin{equation}
F(\tau,\tau')\to \infF(\tau-\tau'),\quad\hbox{as}\quad \tilde\omega \to 0.
\label{pa12}
\end{equation}
We can use (\ref{mm6}) and (\ref{mm8}) to write (\ref{sf18}) as
\begin{eqnarray}
\infF(\tau)
&=&4\pi \sgn(\tau) \sum_{k}
\lvec 0|\mu_k)e^{-\hat \varsigma_k|\tau|}\lvec \mu_k|0)\nonumber\\
&=&2\pi \sgn(\tau) \sum_{k}w_k e^{-\hat \varsigma_k|\tau|}\nonumber\\
&=&2\pi \sgn(\tau)E^{\{n\}}_2(|\tau|).
\label{pa14}
\end{eqnarray}
Here we have introduced the $n$-exponential approximation
\begin{eqnarray}
E_{q}^{\{n\}}(\tau)&=&\sum_{k=n+1}^{2n}w_k\mu_k^{q-2}e^{-\tau/\mu_k}\nonumber\\
&=&2\lvec 0|\hat\mu_{\bf u}^{q-2}e^{-\hat \varsigma_{\bf u}\tau}|0).
\label{pa16}
\end{eqnarray}
to the exponential integral function
\begin{equation}
E_{q}(\tau)=\int_0^1 d\mu\,\mu^{q-2}e^{-\tau/\mu}.
\label{pa18}
\end{equation}
\begin{figure}\centering
\includegraphics[height=80mm,width=1\columnwidth]{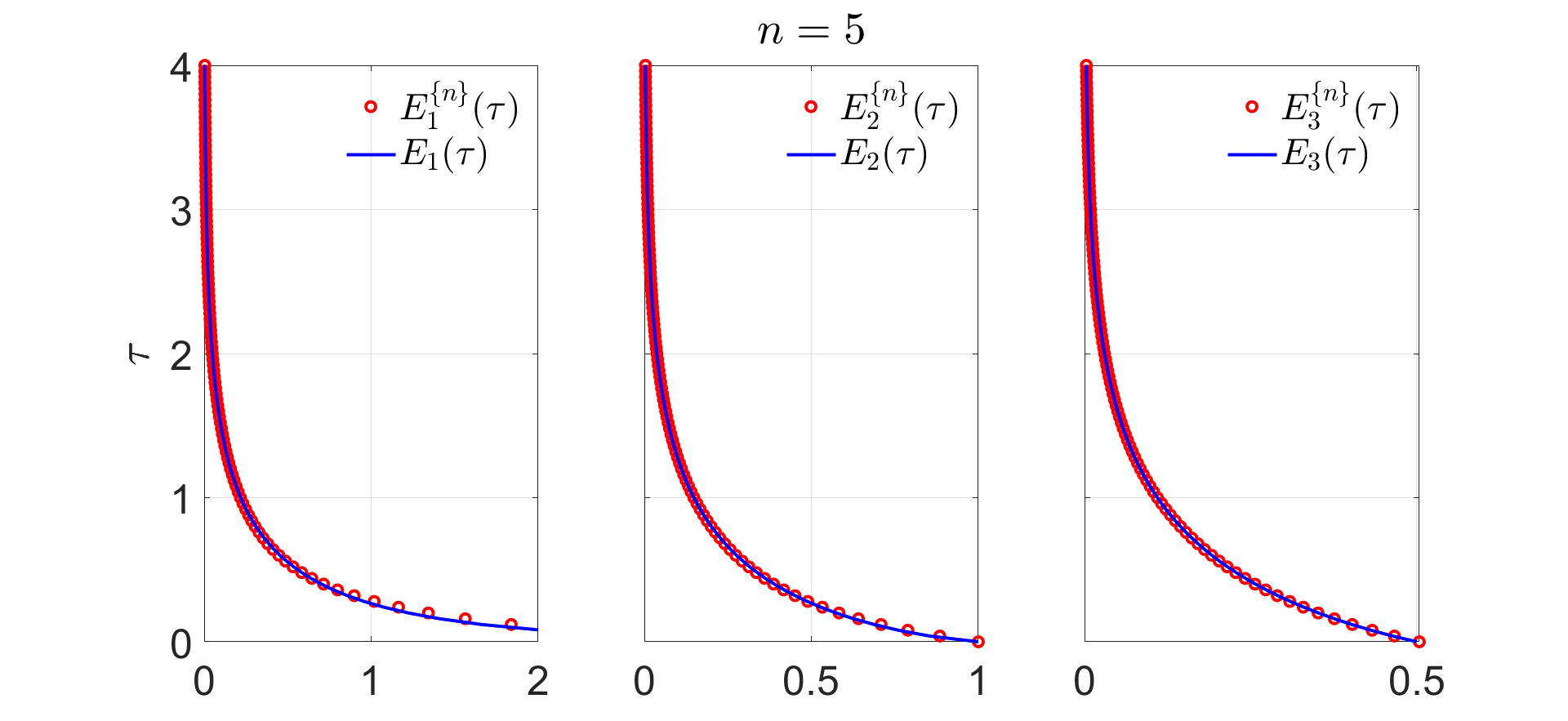}
\caption{A comparison of the exact exponential-integral functions $E_q(\tau)$ of (\ref{pa18})  with the $n$-exponential approximations $E^{\{n\}}_q(\tau)$ of  (\ref{pa16}). Even for the modest  number of stream pairs of this example, $n=5$, the difference between the two functions is hardly perceptible.
\label{Ex5}}
\end{figure}

%
%
\begin{table}[t]
\begin{center}
\begin{tabular}{|r|l|l|l|l|l|l|l|l|l|l|}
 \hline
$ n $ & 1 & 2 & 3 & 4 & 5 & 6 & 7 & 8 & 9 &10\\ [0.5ex]
 \hline\hline 
 $E^{\{n\}}_1 (0) $ &1.7321 &2.3321  & 2.6903 & 2.9587 & 3.1700 &3.3444 &3.4929  &3.6221 &3.7365  &3.8392  \\
 \hline
 $1 E^{\{n\}}_2 (0) $ &1 &1  & 1 & 1 & 1&1 &1  &1 &1  &1  \\
 \hline
$2 E^{\{n\}}_3 (0) $ & 1.1547 & 1.0425 & 1.0199 & 1.0115 & 1.0075 & 1.0053 & 1.0039 & 1.0030 & 1.0024 & 1.0020 \\
 \hline
 $3 E^{\{n\}}_4(0) $ &1 &1  & 1 & 1 & 1&1 &1  &1 &1  &1  \\
 \hline
 $4 E^{\{n\}}_5 (0) $ & 0.7698 & 0.9910 & 0.9982 & 0.9994 & 0.9998 & 0.9999 & 0.9999 & 1.0000 & 1.0000 & 1.0000 \\
 \hline
\end{tabular}
\end{center}
\caption{ Numerical values of the exponential-integral functions, $E^{\{n\}}_q(\tau)$ of (\ref{pa16}) 
at $\tau = 0$ for $q=1,2,3,4,5$ and $n=1,2,3,\ldots, 10$.
\label{table1}}
\end{table}
The exponential integral functions of (\ref{pa18}) are discussed in Appendix 1 of Chandrasekhar\cite{Chandrasekhar}, or {\bf 5.1.4} of  Abramowitz and Stegun\cite{Abramowitz}.  In Eq. (31) of their paper {\it Dependence of Earth's Thermal Radiation on the Five Most Abundant Greenhouse Gases}\cite{WH0},  van Wijngaarden and Happer used the Green's function (\ref{sf18}) with $E_{2}^{\{n\}}(\tau)\to E_{2}(\tau)$.  Fig. \ref{Ex5} shows that even for a modest number of stream pairs, $n=5$, one can hardly tell the difference between $E_{2}^{\{n\}}(\tau)$ and $E_{2}(\tau)$. Many familiar identities of the exponential integral functions (\ref{pa18}) have exactly analogous forms for the $n$-stream approximations (\ref{pa16}) or  are  limiting expressions for large $n$. For example, one can readily verify that
\begin{eqnarray}
\frac{d}{d\tau}E_{q}^{\{n\}}(\tau)&=&-E_{q-1}^{\{n\}}(\tau). \label{pa20}\\
\int_0^{\tau} d\tau' E_{q}^{\{n\}}(\tau')&=&E_{q+1}^{\{n\}}(0)-E_{q+1}^{\{n\}}(\tau),  \label{pa22}\\
E_{q}^{\{n\}}(0)&=&\frac{1}{q-1},\quad\hbox{for} \quad q=0,2,4,...\label{pa24}\\
\lim_{n\to \infty}E_{q}^{\{n\}}(0)&=&\frac{1}{q-1},\quad\hbox{for} \quad q=3,5,...\label{pa26}\\
\int_0^{\infty}d\tau' \,\tau' E_{q}^{\{n\}}(\tau')&=&\frac{1}{q+1},\quad\hbox{for} \quad q=0,2,4,...\label{pa28}\\
\lim_{n\to \infty}\int_0^{\infty}d\tau' \,\tau' E_{q}^{\{n\}}(\tau')&=&\frac{1}{q+1},\quad\hbox{for} \quad q=1,3,5,...\label{pa30}\\
\end{eqnarray}
Some representative values of $E^{\{n\}}_q(0)$ versus $n$ are shown in Table \ref{table1}.
The function  $E^{\{n\}}_1(0)$ increases slowly with $n$. 
The exact exponential function $E_1(\tau)$ diverges as $-\ln(\tau)$ as $\tau\to 0$.  
For $\tau = 0$  we find the  useful special case of (\ref{pa16}) 
\begin{eqnarray}
\lvec 0|\hat\mu_{\bf u}^{q}|0)=\frac{1}{2}E_{q+2}^{\{n\}}(0).
\label{pa32}
\end{eqnarray}
\begin{figure}\centering
\includegraphics[height=80mm,width=1\columnwidth]{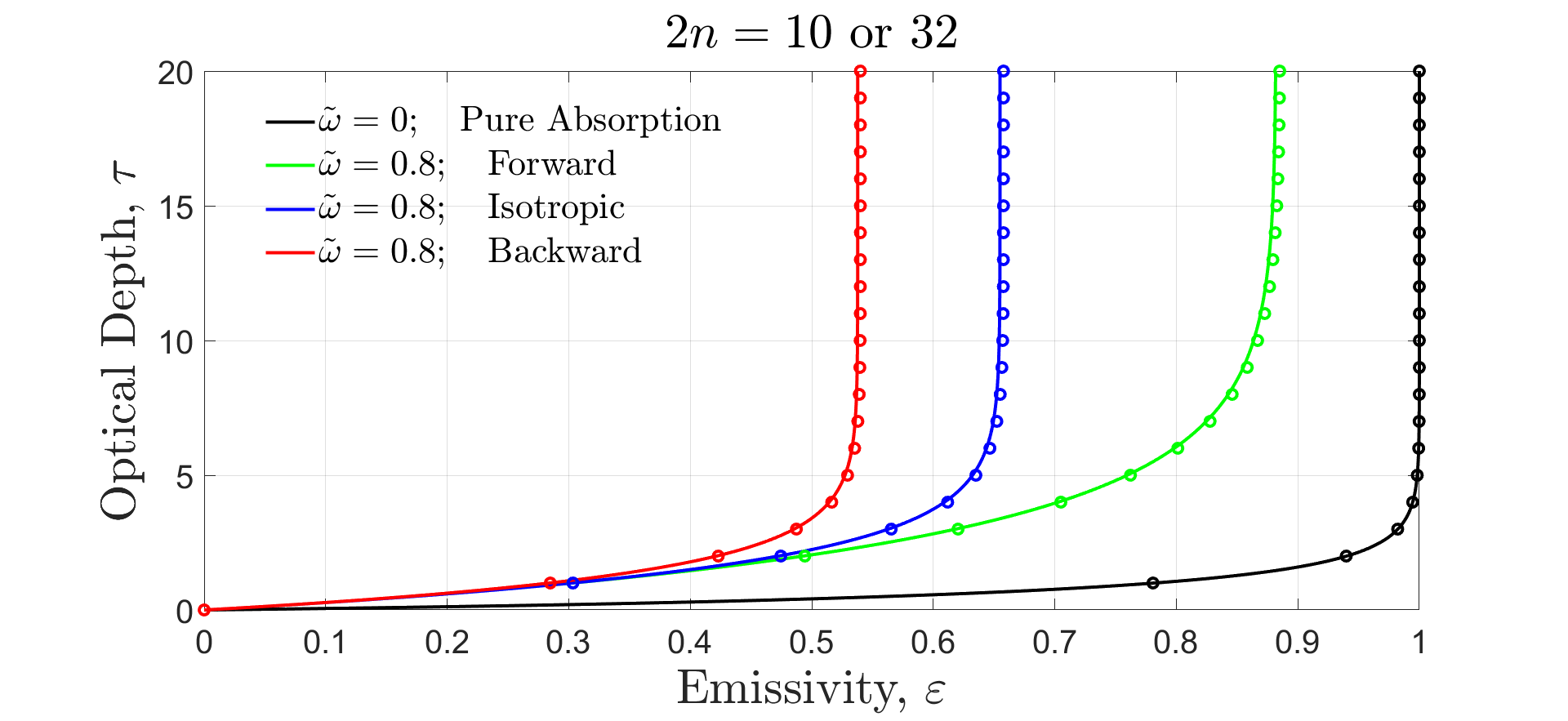}
\caption{ The thermal emissivity $\varepsilon$ of (\ref{se6}) as a function of the optical thickness $\tau_c$ of clouds. The  scattering parameters of the clouds are the same as those of Fig. \ref{Gr4}.  Evaluations for $2n=10$ are shown as smooth lines and evaluations for $2n=32$ are shown as small circles.  With respect to isotropic scattering, forward scattering increases the emissivity and backward scattering decreases it. The emissivity values for $2n = 10$,  $\tau_c=1$ and $\tau_c = 20$ were  cited in Fig. \ref{em1} and Fig. \ref{em20}.
\label{epsilon}}
\end{figure}

%
\subsection{Scalar emissivity $\varepsilon$ of an isothermal cloud\label{em}}
For an isothermal cloud, we can use (\ref{ex12}) and  (\ref{eic6})  to write the outgoing flux that is generated by thermal emission of the cloud particulates as
\begin{eqnarray}
\dot Z^{\{\rm out\}} &=&\lvec 0|\dot Z^{\{\rm out\}}\}\nonumber\\
&=&4\pi\lvec 0|(\hat\mu_{\bf u}-\hat\mu_{\bf d})|\dot I^{\{\rm out\}}\}\nonumber\\
&=&4\pi B\lvec 0|(\hat\mu_{\bf u}-\hat\mu_{\bf d})\mathcal{E}|0).
\label{se2}
\end{eqnarray}
Of special interest is a black cloud, for which the emissivity operator becomes the identity operator, $\mathcal{E}=1$, in accordance with (\ref{ot20}). Then we can use (\ref{pa32}) to write (\ref{se2}) as
\begin{eqnarray}
\dot Z^{\{\rm out\}} 
&=&4\pi B\lvec 0|(\hat\mu_{\bf u}-\hat\mu_{\bf d})|0)\nonumber\\
&=&8\pi B\lvec 0|\hat\mu_{\bf u}|0)\nonumber\\
&=&4\pi B E^{\{n\}}_3(0).
\label{se4}
\end{eqnarray}
We define the emissivity of an isothermal cloud as the ratio of the output flux (\ref{se2}) to the output flux of an ideal black cloud (\ref{se4}),
\begin{equation}
\varepsilon=\frac{Z^{\{\rm out\}}}{4\pi B E^{\{n\}}_3(0)}=\frac{\lvec 0|(\hat\mu_{\bf u}-\hat\mu_{\bf d})\mathcal{E}|0)}{E^{\{n\}}_3(0)}.
\label{se6}
\end{equation}
For a purely absorptive cloud, with $\tilde\omega =0$, we can can use the expression (\ref{eic28})  for $\mathcal{E}$  to  write (\ref{se6})  as
\begin{eqnarray}
\varepsilon&=&\lvec 0|(\hat\mu_{\bf u}-\hat\mu_{\bf d})[\hat 1-e^{-\hat\varsigma_{\bf u}\tau_c}-e^{\hat\varsigma_{\bf d}\tau_c}]|0)/E^{\{n\}}_3(0)\nonumber\\
&=&2\lvec 0|[\hat\mu_{\bf u} e^{-\hat\varsigma_{\bf u}0}-\hat\mu_{\bf u} e^{-\hat\varsigma_{\bf u}\tau_c}]|0)/E^{\{n\}}_3(0)\nonumber\\
&=&1-E_3^{\{n\}}(\tau_c)/E^{\{n\}}_3(0).
\label{em6}
\end{eqnarray}
Representative scalar emissivities $\varepsilon$ of (\ref{se6})  are shown in Fig. \ref{epsilon} as a function of the optical thicknesses $\tau_c$ of isothermal  clouds.
For clouds with relatively large single scattering albedos, $\tilde\omega = 0.8$, the limiting emissivities for optically thick clouds are substantially less than unity, especially for clouds with isotropic scattering or backscattering particulates.  
Fig. \ref{epsilon} also shows that the number $n$ of stream pairs used to model the cloud makes little difference.  One can barely notice the differnce between $n=5$ and $n=16$. 
\subsection{Flux from clouds with hot interiors\label{fch}}
We use (\ref{ex12}),  (\ref{chi14}) and (\ref{sf8}) to write the scalar flux from a cloud with a hot interior as
\begin{eqnarray}
\dot Z^{\{\rm out\}}=\lvec 0|\dot Z^{\{\rm out\}}\} &=&4\pi\lvec 0|(\hat\mu_{\bf u}-\hat\mu_{\bf d})|\dot I^{\{\rm out\}} \}\nonumber \\
 &=&4\pi\lvec 0|\hat\mu_{\bf u}(\mathcal{M}_{\bf u}-\mathcal{S}\mathcal{M}_{\bf d})(\mathcal{L}_{\bf u}-f\hat\lambda_{\bf u}/\lambda_{2n})^{-1}|0)B_0\nonumber\\
&&-4\pi\lvec 0|\hat\mu_{\bf d}(\mathcal{M}_{\bf d}-\mathcal{S}\mathcal{M}_{\bf u})
(\mathcal{L}_{\bf d}+f\hat\lambda_{\bf d}/\lambda_{2n})^{-1}|0)B_0.
 \label{fch2}
\end{eqnarray}
An interesting special case of (\ref{fch2}) is an optically thick, purely absorptive (black) cloud with a vanishing scattering matrix, $\mathcal{S}= \breve 0$
\begin{eqnarray}
\dot Z^{\{\rm out\}} 
 &=&4\pi\lvec 0|\hat\mu_{\bf u}(\mathcal{M}_{\bf u}-f\hat\mu_{\bf u}/\mu_{2n})^{-1}
 -\hat\mu_{\bf d}(\mathcal{M}_{\bf d}+f\hat\mu_{\bf d}/\mu_{2n})^{-1}|0)B_0.
 \label{fch4}
\end{eqnarray}
To order $f$, the pseudoinverse matrices  of (\ref{fch4}) become
\begin{eqnarray}
(\mathcal{M}_{\bf u}-f\hat\mu_{\bf u}/\mu_{2n})^{-1}&=&\mathcal{M}_{\bf u}+f\hat\mu_{\bf u}/\mu_{2n}\nonumber\\
(\mathcal{M}_{\bf d}+f\hat\mu_{\bf d}/\mu_{2n})^{-1}&=&\mathcal{M}_{\bf d}-f\hat\mu_{\bf d}/\mu_{2n}.
 \label{fch6}
\end{eqnarray}
The two terms of (\ref{fch4}) give equal amounts of outgoing flux from the top and bottom of the cloud. So we can use (\ref{fch6}) and (\ref{pa32}) to write (\ref{fch4}), to order $f$, as
\begin{eqnarray}
\dot Z^{\{\rm out\}} 
 &\approx&8\pi\lvec 0|\hat\mu_{\bf u}+f\hat\mu_{\bf u}^2/\mu_{2n}|0)B_0\nonumber\\
 &=&4\pi\left[E^{\{n\}}_3(0)+\frac{ f E^{\{n\}}_4(0)}{\mu_{2n}}\right]B_0\nonumber\\
 &\approx&2\pi\left[1+\frac{ 2 f}{3\mu_{2n}}\right]B_0.
 \label{fch8}
\end{eqnarray}
To get the last line of (\ref{fch8}),  we noted  from Table \ref{table1} that $E^{\{n\}}_3(0)=1.0075/2\approx 1/2$ and $E^{\{n\}}_4(0) =1/3$. For small values of $f \tau'/\mu_{2n} \ll 1$, we can write the Planck brightness (\ref{chi2}) near the bottom of the black cloud as
\begin{eqnarray}
B(\tau') &=&B_0 e^{f\tau'/\mu_{2n}} \nonumber\\
&\approx&\left[1+\frac{\tau' f}{\mu_{2n}}\right ]B_0.
\label{fch10}
\end{eqnarray}
We can therefore use (\ref{fch8}) and (\ref{fch10}) to write the radiative flux coming out of the top or bottom of a purely absorptive cloud with a hot interior as
\begin{eqnarray}
\frac{1}{2}\dot Z^{\{\rm out\}} \approx \pi B(2/3).
 \label{fch12}
\end{eqnarray}
The flux from the top or bottom of the  cloud is the same as the flux from a blackbody with the Planck intensity $B(\tau')=B(2/3)$ at an optical depth $\tau'=2/3$ into the cloud from the effectively emitting surface. Eq. (\ref{fch12}) is called the  Eddington-Barbier relation for the emergent flux.  For example, see Eq. (7.28) of Owocki\,\cite{Eddington}.  As we discussed in Section \ref{chi}, the angular distribution of radiation from purely absorptive clouds with hot interiors is not Lambertian, but is limb darkened.
\section{Heat Conduction\label{hc}}
For very optically thick clouds, with  $\tau_c\gg 1$, the incremental scalar radiative flux $Z(\nu, z)d\nu$ carried by radiation with spatial frequencies between $\nu$ and $\nu+d\nu$,  at the altitude $z$ above the bottom of the cloud, is well approximated by Fourier's law of heat conduction,
\begin{equation}
Z d\nu=-d\nu q\frac{dT}{dz}.
\label{hf2}
\end{equation}
Here  $q=q(\nu, z)$ is the monochromatic conductivity coefficient.  
The temperature gradient at the altitude $z$ is $dT/dz=dT(z)/dz$. The total heat flux $J=J(z)$ carried by all frequencies $\nu$ is 
\begin{equation}
J=\int_0^{\infty} Z(\nu)d\nu=- k\frac{dT}{dz},
\label{hf4}
\end{equation}
where the frequency-integrated conductivity coefficient is
\begin{equation}
 k =k(z)=\int_0^{\infty}q(\nu, z)d\nu.
\label{hf6}
\end{equation}
In the next section, we discuss  moments of the Green's function $\infF(\tau)$. It turns out that the monochromatic conductivity coefficient $q$ is proportional to the first moment $\langle \tau\rangle$ of $\infF(\tau)$.
\begin{figure}\centering
\includegraphics[height=80mm,width=1\columnwidth]{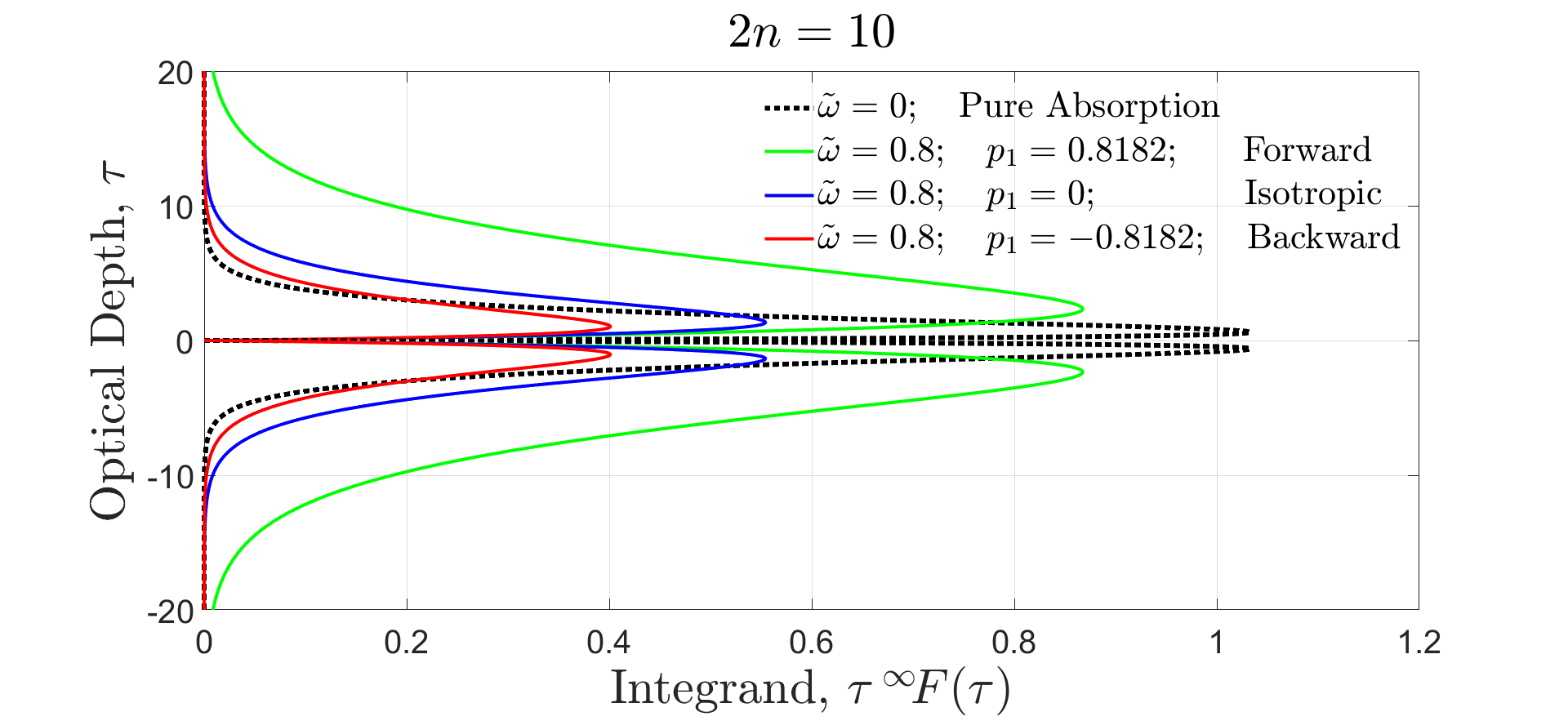}
\caption{The integrands, $\tau\,\infF(\tau)$, for the first moments $\langle \tau\rangle$ of (\ref{mo2}) for the Green's function $\infF(\tau)$ of Fig. \ref{Gr4}.  The areas of the integrands depend on the single scattering albedo $\tilde\omega$ and on the dipole moment $p_1$ of the scattering phase function, as given by (\ref{mo20}).  The radiative heat flux (\ref{hc6}) is proportional to $\langle\tau\rangle$. See the text for more detail.
\label{Gr5}}
\end{figure}

%
\subsection{Moments of $ \infF(\tau)$\label{mo}}
We define the $q$th moment of the Green's function (\ref{sf14}) for the scalar flux in an infinite cloud as
\begin{equation}
\langle \tau^q\rangle =\int_{-\infty}^{\infty}d\tau\, \tau^q \, \infF(\tau).
\label{mo2}
\end{equation}
Because $ \infF(\tau)$ is antisymmetric in $\tau$, in accordance with (\ref{sf16}), all the even moments vanish
\begin{equation}
\langle \tau^q\rangle =0,\quad\hbox{for}\quad q=0,2,4,\ldots.
\label{mo4}
\end{equation}
For odd $q=1,3,5,\ldots$, we can use $\infF(\tau)$ of (\ref {sf18}) to write (\ref{mo2}) as
\begin{eqnarray}
\langle \tau^q\rangle &=& 2\int_{0}^{\infty}d\tau\, \tau^q \,\infF(\tau)\nonumber\\
&=&8\pi\eta_0\lvec 0|\hat\mu\int_{0}^{\infty}d\tau\, \tau^q \,
\sum_k |\lambda_k)\lvec \lambda_k|e^{-\kappa_k\tau}\hat\varsigma|0)\nonumber\\
&=&8\pi\eta_0\lvec 0|\hat\mu
\sum_k q!\lambda_k^{q+1} |\lambda_k)\lvec \lambda_k|\hat\varsigma|0)\nonumber\\
&=&4\pi \eta_0 q!\lvec 0|\hat\mu\bigg[
\sum_k\lambda_k^{q+1} |\lambda_k)\lvec \lambda_k|+\sum_j\lambda_k^{q+1} \hat r|\lambda_j)\lvec \lambda_j|\hat r \bigg]\hat\varsigma|0).
\label{mo6}
\end{eqnarray}
In the last line of (\ref{mo6}) we used the reflection symmetry, $\hat r|\lambda_j)=|\lambda_k)$ and $\lvec\lambda_j|\hat r =\lvec \lambda_k|$,  of (\ref{lam6}).
The indices of downward modes, with $j=1,2,3,\ldots,n$ are related to the indices $k=n+1,n+2,\ldots,2n$ of the upward modes by the index reflection function of (\ref{int8}),
\begin{equation}
j=r(k)=2n+1-k,\quad\hbox{and}\quad k=r(j)=2n+1-j.
\label{mo8}
\end{equation}
The penetration lengths and basis vectors are related by
\begin{equation}
\lambda_j=-\lambda_k,\quad\hat r|\lambda_j)=|\lambda_k),\quad\hbox{and}\quad
\lvec \lambda_j|\hat r = \lvec \lambda_k|.
\label{mo10}
\end{equation}
Since $q$ is an odd integer, we can set
\begin{equation}
\lambda_k^{q+1}=(-\lambda_j)^{q+1}=\lambda_j^{q+1},
\label{mo12}
\end{equation}
and write (\ref{mo6}) as
\begin{equation}
\langle \tau^q\rangle  
=4\pi\eta_0 q!\lvec 0|\hat\mu\bigg[
\sum_k\lambda_k^{q+1} |\lambda_k)\lvec \lambda_k|+\sum_j\lambda_j^{q+1} \hat r|\lambda_j)\lvec \lambda_j|\hat r \bigg]\hat\varsigma|0).
\label{mo14}
\end{equation}
Since $\hat\mu\hat r=-\hat r\hat\mu$, $\hat r\hat \varsigma = -\hat\varsigma\hat r$, $\hat r|0)=|0)$ and $\lvec 0|\hat r =\lvec 0|$, we can write the last factor of (\ref{mo14}) as
\begin{equation}
\lvec0|\hat\mu\sum_j\lambda_j^{q+1} \hat r|\lambda_j)\lvec \lambda_j|\hat r \hat\varsigma|0)=\lvec0|\hat\mu\sum_j\lambda_j^{q+1} |\lambda_j)\lvec \lambda_j| \hat\varsigma|0),
\label{mo16}
\end{equation}
so that (\ref{mo14}) becomes

\begin{eqnarray}
\langle \tau^q\rangle  
&=&4\pi\eta_0q!\lvec 0|\hat\mu\bigg[
\sum_k\lambda_k^{q+1} |\lambda_k)\lvec \lambda_k|+\sum_j\lambda_j^{q+1} |\lambda_j)\lvec \lambda_j| \bigg]\hat\varsigma|0)\nonumber\\
&=&4\pi\eta_0 q!\lvec 0|\hat\mu\hat\lambda^{q+1}\hat\varsigma|0).
\label{mo18}
\end{eqnarray}

For our purposes, the most important moment is the first, with
$q=1$.  In this case we note
that $\hat \lambda =\hat\kappa^{-1}=\hat\eta^{-1}\hat\mu$, and $\lvec l| \hat\eta^{-1}=\lvec l|\eta_l^{-1}$ or $\hat\eta^{-1}|l)=\eta_l^{-1}|l)$,
and  we  use the identity (\ref{mm20}) to write (\ref{mo18}) as
\begin{eqnarray}
\langle \tau\rangle  
&=&4\pi\eta_0\lvec 0|\hat\mu\hat\lambda^2\hat\varsigma|0)\nonumber\\
&=&4\pi\eta_0\lvec 1|\hat\eta^{-1}\hat\mu\hat\eta^{-1}|0)\nonumber\\
&=&\frac{4\pi}{3\eta_1}\nonumber\\
&=&\frac{4\pi}{3(1-\tilde\omega  p_1)}.
\label{mo20}
\end{eqnarray}
Examples of first moment integrands from (\ref{mo2}) are shown in Fig. \ref{Gr5}.  The areas between the vertical axis and  the curves $\tau\infF(\tau)$ shown in Fig.  \ref{Gr5}  give the same numerical values as the algebraic formula (\ref{mo20}).

Algebraic formulas for higher odd moments can be found in analogous ways. For example, if $q=3$  we can use (\ref{mm20}) and (\ref{mm22}) write (\ref{mo18}) as
\begin{eqnarray}
\langle \tau^3\rangle  
&=&4\pi\eta_0 3!\lvec 0|\hat\mu\hat\lambda^4\hat\varsigma|0)\nonumber\\
&=&\frac{8\pi}{\eta_1^2}\lvec 1|\hat\mu\,\hat\eta^{-1}\hat\mu|1)\nonumber\\
&=&\frac{8\pi}{3\eta_1^2}\left(\frac{1}{\eta_0}+\frac{4}{5\eta_2}\right).
\label{mo22}
\end{eqnarray}
\subsection{Thermal conductivity}
Inside a very optically thick cloud, with total optical depth $\tau_c\gg 1$, we can ignore contributions  $\Delta F(\tau,\tau')$ of the cloud boundaries to the Green's function of (\ref{sf12}), and approximate  the flux (\ref{sf8}) by
\begin{equation}
Z(\tau)
=\int_0^{\tau_c}d\tau' {\infF}(\tau-\tau')B(\tau').
\label{hc3}
\end{equation}
As shown in Fig. \ref{Gr4},  $\infF(\tau-\tau')$ has a sharp, positive peak for $\tau-\tau'\to 0^+ $ and a sharp negative peak for $\tau-\tau'\to 0^-$. For $|\tau-\tau'|\gg 1$,  $\infF(\tau-\tau')$ rapidly goes to zero.
We can therefore  approximate the Planck intensity $B(\tau')$ in the integrand of (\ref{sf8})  by the first two terms of a Taylor series
\begin{equation}
B(\tau')=B+(\tau'-\tau)\frac{dB}{d\tau}.
\label{hc4}
\end{equation}
Here $B$ and $dB/d\tau$ are the values of the Planck intensity and its first derivative at the optical depth $\tau$. 
Then we can use (\ref{mo2}) and (\ref{mo20})  to write the scalar flux (\ref{hc3}) as
\begin{eqnarray}
Z(\tau)&=& \int_{-\infty}^{\infty}d\tau' \, \infF(\tau-\tau')
\left[B+(\tau'-\tau)\frac{dB}{d\tau}\right]\nonumber\\
&=&-\langle \tau\rangle \frac{dB}{d\tau}.
\label{hc6}
\end{eqnarray}
Here we noted from (\ref{sf16}) that $\infF(\tau-\tau')=-\infF(\tau'-\tau)$. 
 Setting
\begin{equation}
\frac{dB}{d\tau}=\left(\frac{\partial B}{\partial T}\right)\left(\frac{dT}{dz}\right)\frac{dz}{d\tau}.
\label{hc8}
\end{equation}
in (\ref{hc6}) and comparing to (\ref{hf2}), we can use (\ref{mo20}) to write  the monnochromatic heat conductivity coefficient $q$  as
\begin{eqnarray}
q&=&\frac{\langle \tau\rangle }{\alpha}\left(\frac{\partial B}{\partial T}\right)\nonumber\\
&=&\frac{4\pi}{3(1-\tilde\omega p_1)\alpha}\left(\frac{\partial B}{\partial T}\right).
\label{hc12}
\end{eqnarray}
The spatial attenuation rate $\alpha$, defined by  (\ref{int0}), is
\begin{equation}
\alpha = \frac{d\tau}{dz}.
\label{hc14}
\end{equation}
\begin{figure}\centering
\includegraphics[height=80mm,width=1\columnwidth]{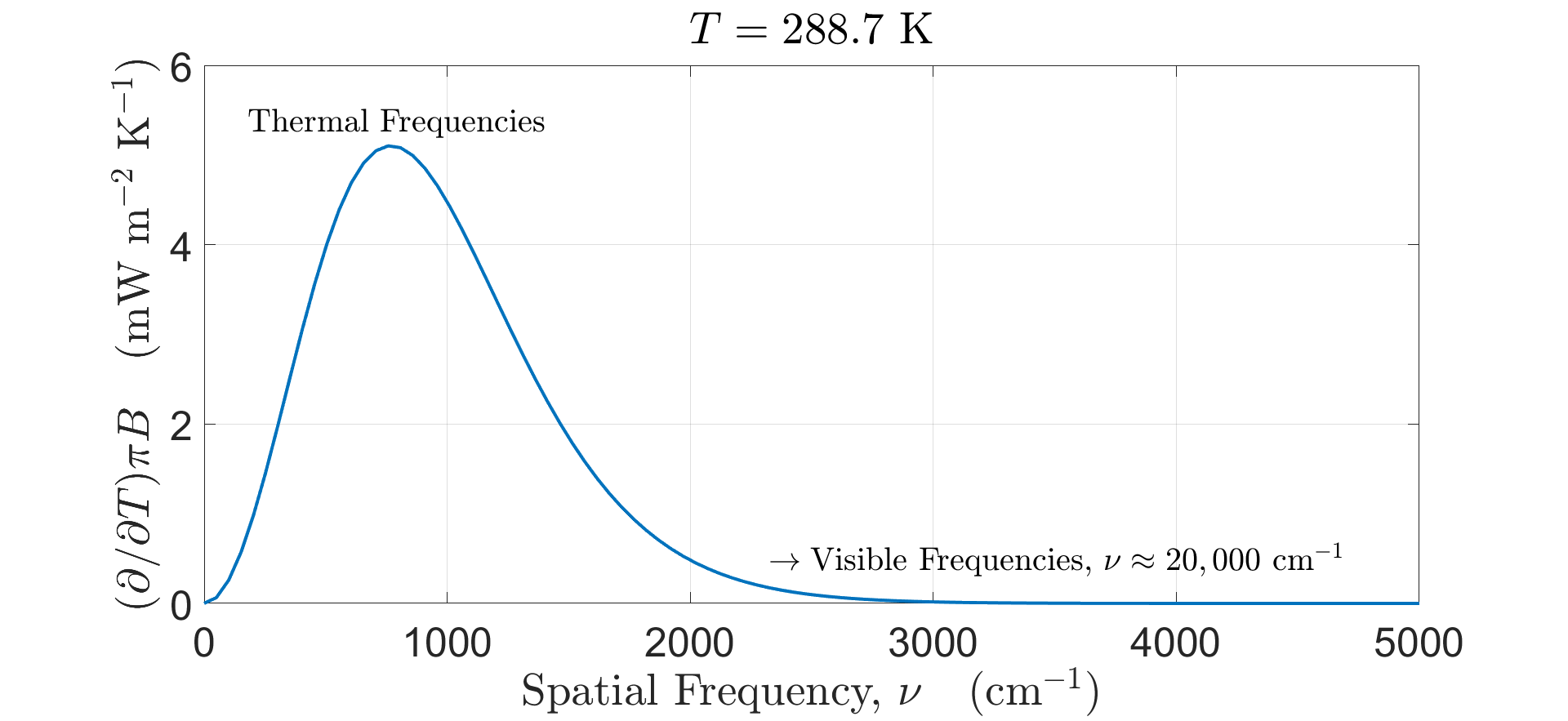}
\caption{$ (\partial /\partial T) \pi B$, the rate of change of the Planck surface flux,  $\pi B$, with temperature, $T$, 
from (\ref{hc16}).  The spatial frequency (in cm$^{-1}$) of the radiation is $\nu$. The cloud interior has a temperature $T=288.7$ K, a representative temperature of Earth's surface.
\label{dBdT}}
\end{figure}

%
The rate of change of the Planck intensity (\ref{et10}) with temperature $T$ is
\begin{eqnarray}
\frac{\partial B}{\partial T}
=\left(\frac{2k_{\rm B}^3T^2}{h_{\rm P}^2c}\right)\frac{x^4e^x}{\left(e^x-1\right)^2}.
\label{hc16}
\end{eqnarray}
The ratio $x$ of the photon energy $h_{\rm P}\nu c$ to the characteristic thermal energy $k_{\rm B}T$ is 
\begin{equation}
x=\frac{  h_{\rm P}\nu c}{k_{\rm B}T}.
\label{hc17}
\end{equation}
The function $\partial B/\partial T$ is plotted versus the radiation frequency $\nu$ in Fig. \ref{dBdT} for a representative surface temperature of $T=288.7$ K. There is negligible thermal heat conduction for frequencies above $\nu = 3000$ cm$^{-1}$, including the dominant frequencies of sunlight. Clouds glow in the dark and transport heat with thermal infrared radiation, not visible light.

\begin{figure}\centering
\includegraphics[height=80mm,width=1\columnwidth]{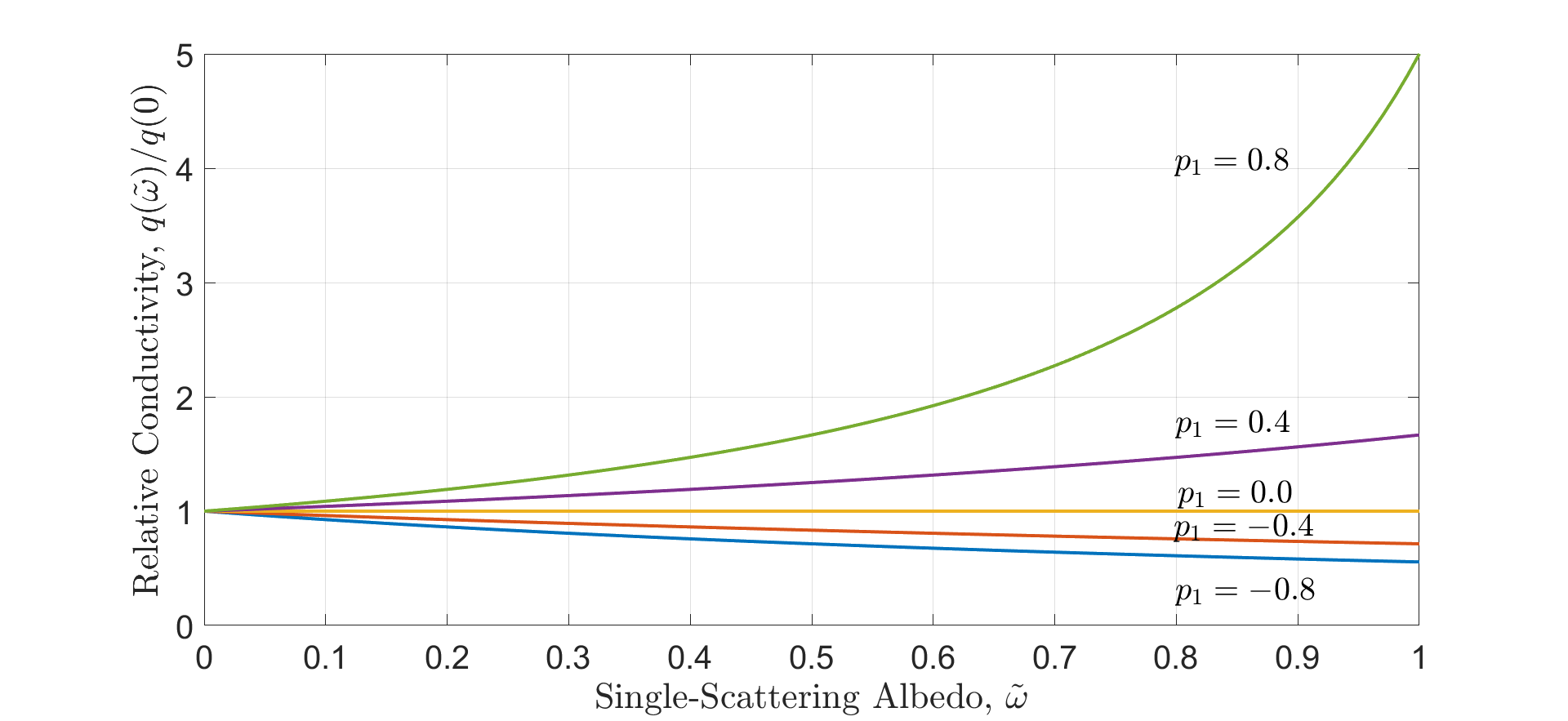}
\caption{The dependence of the relative heat conductivity,  $q(\tilde\omega)/q(0) =1/\eta_1=1/(1-\tilde\omega p_1)$,  on the single scattering albedo $\tilde\omega$ and on the dipole coefficient $p_1$ of the scattering phase function.  The absolute heat conductivity, $q=q(\tilde\omega)$, is given by (\ref{hc12}) and (\ref{hc18}). See the text for more details.
\label{cond0}}
\end{figure}

%
It is instructive to define the  relative (dimensionless) monochromatic heat conductivity  
\begin{equation}
\frac{q(\tilde\omega)}{q(0)} =\frac{1}{1-\tilde\omega p_1}.
\label{hc18}
\end{equation}
Eq. (\ref{hc18}) is the ratio of the absolute conductivity (\ref{hc12})  for a finite scattering probability, 
$\tilde\omega\ge0$, to the value for a purely absorptive cloud with $\tilde\omega = 0$.
As shown in Fig. \ref{cond0}, for a positive dipole coefficient $p_1$ of the scattering phase function, which implies predominantly forward scattering, the heat conductivity of (\ref{hc12}) will increase with an increasing  fraction $\tilde\omega$ of scattering, even though the emission per particle decreases as $1-\tilde\omega$.  For a negative dipole coefficient $p_1$ of the scattering phase function, which implies predominently backward scattering, the heat conductivity of (\ref{hc12}) will decrease as the scattering fraction increases. For strong forward scattering, with $\tilde \omega$ not much less than its maximum possible value of 1, and for $p_1$ close to its maximum possible value of
\begin{equation}
p_1\le  \frac{2n-1}{2n+1}<1,
\label{hc20}
\end{equation}
for a $2n$-stream model,  the heat conductivity is significantly enhanced compared to that of a cloud with pure absorption, $\tilde\omega =0$, or compared to particulates with $p_1=0$ and equal forward and backward scattering fluxes. The higher multipole coefficients $p_l$  of the scattering phase, with $l\ge 2$ have no influence on the heat conductivity.
\subsection{Gray clouds}
Substituting (\ref{hc12}) into (\ref{hf6}) we find that the frequency-integrated heat conductivity  is
\begin{equation}
k=\int_0^{\infty} d\nu\,\frac{4\pi}{3(1-\tilde\omega p_1)\alpha}\left(\frac{\partial B}{\partial T}\right).
\label{agc6}
\end{equation}
For a homogeneous grey cloud, where  none of the scattering parameters $\alpha$, $\tilde \omega$ or $p_1$ depend on frequency $\nu$,  the only factor in the integrand of (\ref{agc6}) that depends on $\nu$ is
 $\partial B/\partial T$. Differentiation both sides of (\ref{et11a}) with respect to temperature we find
\begin{eqnarray}
\frac{\partial}{\partial T}\int_0^{\infty} d\nu\,\pi B(\nu,T)
&=&4\sigma_{\rm SB}T^3\nonumber\\
&=&5.46\hbox{ W m$^{-2}$ K$^{-1}$}.
\label{agc7}
\end{eqnarray}
The numerical value on second line, for a representative temperature $T=288.7$ K of Earth's surface, is important. It says that a surface warming of 1 K for a blackbody Earth would increase the flux to space by 5.46 W m$^{-2}$.

Using (\ref{agc7}) in (\ref{agc6}) we see that 
 the conductivity coefficient becomes 
\begin{equation}
k=k_{\rm rad}
=\frac{16\sigma_{\rm SB}T^3}{3(1-\tilde\omega p_1)\alpha}.
\label{agc8}
\end{equation}
 At a representative atmospheric temperature of $T=288.7$ K and a cloud attenuation rate of $\alpha = 0.01 $ m$^{-1}$, a single scattering albedo, $\tilde\omega = 0.5$ and a dipole coefficient $p_1 = 0.8$ for anisotropic scattering, the radiative heat conduction coefficient (\ref{agc8}) becomes
\begin{equation}
k_{\rm rad}=1.2\times 10^3 \hbox{ W m$^{-1}$ K$^{-1}$}.
\label{agc10}
\end{equation}
For comparison, the molecular heat conductivity of air at a temperature of 300 K is more than four orders of magnitude smaller.  A representative value \cite{conductivity} is
\begin{equation}
k_{\rm air}=2.5 \times 10^{-2}\hbox{ W m$^{-1}$ K$^{-1}$}.
\label{agc12}
\end{equation}

For the conductivity coefficient (\ref{agc10}), and for
the radiative flux for a typical moist adiabatic lapse rate of $dT/dz=-6.5\times 10^{-3}$ K m$^{-1}$, the total radiative flux (\ref{hf4}) is
\begin{equation}
J= -k_{\rm rad}\frac{dT}{dz}= 8 \hbox{ W m}^{-2},
\label{agc14}
\end{equation}
about 2.7\% of the typical flux, $Z\approx 300$ W m$^{-2}$, at the top of the cloud-free atmosphere.

We can also write the heat flux (\ref{hf4}) in the form familiar from gas kinetic theory
\begin{equation}
J=- \,\frac{\lambda_{\rm rad}c}{3}\left(\frac{d u}{dz}\right).
\label{agc16}
\end{equation}
Here  $\lambda_{\rm rad}$ is the effective mean free path for heat transport by photons in the gray cloud. The mean energy density, $u$, of the photons can be written in terms of the Planck intensity $B$ of (\ref{int32}) as
\begin{eqnarray}
u&=&\frac{4\pi}{c}\int_0^{\infty}d\nu B\nonumber\\
&=&\frac{4\sigma_{\rm SB}T^4}{c}.
\label{agc18}
\end{eqnarray}
For (\ref{agc16}) to give the same flux $J$ as (\ref{hf4}) and (\ref{agc6}) we must have
\begin{equation}
\lambda_{\rm rad}=\frac{1}{\eta_1\alpha}=\frac{1}{(1-\tilde\omega p_1)\alpha}.
\label{agc20}
\end{equation}
For single scattering albedos $\tilde\omega$ not much less than 1, and for strong forward scattering, with dipole scattering-phase coefficients $p_1$ also not much less than 1, the effective mean free path $\lambda_{\rm rad}$ of (\ref{agc20}) for radiative energy transport can be much greater than the mean free path $1/\alpha$ for photon collisions with particulates. For strong forward scattering, most photons continue in approximately the same direction as before the collision.
\section{Summary}
We have given a quantitative overview  of the thermal emission of radiation by clouds.  In Section \ref{in}, we reviewed the vector and matrix methods for analyzing radiation transfer that were presented in an earlier paper,  {\it 2n-Stream Radiative Transfer}\cite{WH1}. We consider homogeneous clouds with negligible horizontal variation. The single scattering albedos $\tilde\omega$ of the cloud particulates, and the  phase functions $p(\mu,\mu')$, for scattering axially symmetric radiation of direction cosine angle $\mu'$ to  radiation of direction cosine $\mu$, are constant within the cloud. The cloud temperature can vary in the vertical direction. The radiation at the optical depth $\tau$ above the bottom of the cloud is characterized by the intensity values $I(\mu_i,\tau)$  at $2n$ Gauss-Legendre direction cosines $\mu_i$, the roots of the $2n$-th Legendre polynomal, $P_{2n}(\mu_i)=0$. The products of the sample intensities $I(\mu_i,\tau)$  and the Gauss-Legendre weights $w_i$ of (\ref{int12}) are equal to the
elements, $\lvec\mu_i|I(\tau)\}=w_iI(\mu_i,\tau)$, of $2n\times 1$ intensity vectors $|I(\tau)\}$.

As outlined in Section \ref{exin}, it is useful to write the intensity  $I(\mu_i,\tau)$ as a part $\dot I(\mu_i,\tau)$ that has been spontaneously emitted inside the cloud,  and a part $\ddot I(\mu_i,\tau)$ that comes from external radiation incident onto the top and bottom of the cloud. The first externally or internally generated photon to appear in the cloud may be scattered one or more times before escaping from the top or bottom, or being absorbed and converted to heat.   
Scattering operators $\mathcal{S}$ and their matrix representations $\lvec\mu_i|\mathcal{S}|\mu_{i'})$ are used to describe the fraction of external incoming intensity $\ddot I^{\{\rm in\}}(\mu_{i'})$  with direction cosine $\mu_{i'}$ that is scattered to outgoing intensity $\ddot I^{\{\rm out\}}(\mu_{i})$  along the direction cosine $\mu_{i}$.  

Green's-function operators $G(\tau')$ and their matrix representations $\lvec\mu_i|G(\tau')|\mu_{i'})$  are used to describe 
radiation  $\dot I(\mu_{i})$ that is thermally emitted by cloud particulates in  infinitesimally thin layers at the source optical depth $\tau'$. The analogous Green's functions for the electrostatic fields generated by point charges $q$ in the space between conductors can be written as the sum of the  central field $q\,{\bf r}/r^3$ generated by the point charge in conductor-free space at a vector displacement ${\bf r}$ from the charge, and a second part that accounts for the fields from surface charges  induced on the conductors. Similarly,  the Green's function $G$ for a cloud is the sum of a part $\infG$  for an infinite cloud and a part $\Delta G$ that accounts for modifications due to the boundaries at the top and bottom of the cloud. 

In Section \ref{eic} we derive Kirchhoff's law of emission and scattering of thermal radiation, the sum of the scattering matrix $\mathcal{S}$ and the thermal emission matrix $\mathcal{E}$ of an isothermal cloud is the identity matrix $\hat 1$, that is, $\mathcal{S}+\mathcal{E}= \hat 1$.

For isothermal clouds with small  optical thickness, $\tau_c=1$ or less, thermal emission is mainly determined by the single scattering absorption, $1-\tilde \omega$. As shown in Fig. \ref{em1} for $\tilde\omega = 0.8$, the emitted intensity  is less than half of  the blackbody limit, $\dot I(\mu_{i})=B$. The emitted intensity  is limb brightened and depends little on the  scattering phase function $p(\mu,\mu')$. There is much more emission for a purely absorbing cloud with $\tilde \omega = 0$, but there is still substantial limb brightening because the cloud is not optically thick.

Shown in Fig. \ref{em20} is the emission of isothermal clouds that are optically thick, with $\tau_c=20 \gg \lambda_{2n}$, where $\lambda_{2n}=-\lambda_{-1}$ is the maximum penetration length of radiation streams in the cloud.  Purely absorbing clouds with $\tilde\omega = 0$ emit at the blackbody limit,  $\dot I(\mu_{i})=B$.  Optically thick isothermal clouds with substantial scattering, for example, with single scattering albedos, $\tilde \omega = 0.8$, emit substantially less thermal radiation than the blackbody limit. The emission depends strongly on the anisotropy of the scattering phase function.  Clouds with strong forward scattering  are ``blackest" and emit the most thermal radiation. There is limb darkening, which is most pronounced for strongly forward-scattering phase functions. 

The limb darkening of isothermal clouds in Fig. \ref{em20} is not due to temperature gradients inside the cloud, as it is for the Sun.  Photodetaching collisions of visible photons with negative hydrogen ions H$^-$ dominate the opacity of the Sun's photosphere \,\cite{Photosphere}.  Scattering collisions are negligible in comparison. So there would be no limb darkening of  the Sun if the temperature did not increase substantially with depth below the surface of the photosphere.  In contrast,
for optically thick, isothermal, scattering clouds,  there is limb darkening because nearly vertically directed photons escape through the surface more efficiently than nearly horizontally directed ones.

The directional dependence of outgoing radiation from optically thick clouds with hot interiors is shown in Fig. \ref{em20h}. A purely absorbing cloud with a hot interior has limb darkening, like the Sun.  A hot interior increases the intensity of all emitted streams, with or without scattering, in  comparison to isothermal clouds with the same surface temperature.

Finally, in Section \ref{hc} we show that for typical optically thick clouds in Earth's atmosphere radiative conduction of heat is orders of magnitude faster than conduction due to molecular diffusion.

We have shown that
the  $2n$-stream matrix method of reference \cite{WH1} is a graphic, efficient and accurate mathematical tool for calculating thermal emission from isothermal or non-isothermal clouds with constant values of the single scattering albedo $\tilde\omega$ and the scattering phase function $p(\mu,\mu')$.
\section*{Acknowledgements}
 The Canadian Natural Science and Engineering Research  Council provided financial support of one of us.

\end{document}